%% file: Main.tex
\documentclass[journal]{IEEEtran}
\input{usepackage.tex}

\newcolumntype{P}[1]{>{\centering\arraybackslash}p{#1}}

\begin{document}

\title{Blockchain-Empowered Trustworthy Data Sharing: \\ Fundamentals, Applications, and Challenges}

\author{Linh T. Nguyen,
        Lam Duc Nguyen,~\IEEEmembership{Member, IEEE,}
        Thong Hoang,~\IEEEmembership{Member, IEEE,}  
        Dilum Bandara, \\
        Qin Wang,~\IEEEmembership{Member, IEEE,}       
        Qinghua Lu,~\IEEEmembership{Senior Member, IEEE,} 
        Xiwei Xu, Liming Zhu,\\
        Petar Popovski,~\IEEEmembership{Fellow, IEEE,} and
        Shiping Chen,~\IEEEmembership{Senior Member, IEEE}
\thanks{Linh T. Nguyen was with Posts and Telecommunications Institute of Technology, Hanoi 10000, Vietnam. Email: linhnt.B17VT215@stu.ptit.edu.vn}
\thanks{Lam Nguyen, Dilum Bandara, Qin Wang, James Hoang, Qinghua Lu, Sherry Xu, Liming Zhu, and Shiping Chen are with CSIRO Data61, Eveleight 2015, Sydney, Australia. Email: \{lam.nguyen,  james.hoang, dilum.bandara, qinghua.lu, Xiwei.Xu, liming.zhu, shiping.chen\}@data61.csiro.au }
\thanks{Petar Popovski is with the Connectivity Section, Department of Electronic Systems, Aalborg University, Aalborg 9220, Denmark. Email: petar@es.aau.dk }
\thanks{\textit{Corresponding Author: Lam Duc Nguyen}}
}


\maketitle

\begin{abstract}

Various data-sharing platforms have emerged with the growing public demand for open data and legislation mandating certain data to remain open. Most of these platforms remain opaque, leading to many questions about data accuracy, provenance and lineage, privacy implications, consent management, and the lack of fair incentives for data providers. With their transparency, immutability, non-repudiation, and decentralization properties, blockchains could not be more apt to answer these questions and enhance trust in a data-sharing platform. However, blockchains are not good at handling the four \emph{Vs} of big data (i.e., volume, variety, velocity, and veracity) due to their limited performance, scalability, and high cost. Given many related works proposes blockchain-based trustworthy data-sharing solutions, there is increasing confusion and difficulties in understanding and selecting these technologies and platforms in terms of their sharing mechanisms, sharing services, quality of services, and applications. 
In this paper, we conduct a comprehensive survey on blockchain-based data-sharing architectures and applications to fill the gap. First, we present the foundations of blockchains and discuss the challenges of current data-sharing techniques. Second, we focus on the convergence of blockchain and data sharing to give a clear picture of this landscape and propose a reference architecture for blockchain-based data sharing. Third, we discuss the industrial applications of blockchain-based data sharing, ranging from healthcare and smart grid to transportation and decarbonization. For each application, we provide lessons learned for the deployment of Blockchain-based data sharing. 
Finally, we discuss research challenges and open research directions. 

\end{abstract}





\input{Sections/1-Introduction.tex}
\input{Sections/2-Data_Sharing.tex}

\input{Sections/3-Background_knowledge.tex}

\input{Sections/4-Convergence-Blockchain-Data-Sharing.tex}
\input{Sections/5-Applications-Blockchain-Data-Sharing.tex}
\input{Sections/6-Open-Research-Issues.tex}
\input{Sections/7-Conclusion.tex}

\bibliography{Refs}

\end{document}

%% file: usepackage.tex
\usepackage{dsfont}
\usepackage{hyperref}
\usepackage{url}
\usepackage{cite}
\usepackage{xcolor}
\usepackage[nolist]{acronym}
\usepackage{graphicx} 
\usepackage{float}
\usepackage{threeparttable}
\usepackage{cancel}
\usepackage[official]{eurosym}
\usepackage{fancyhdr}
\usepackage{comment}
\usepackage{caption}
\usepackage{subcaption}
\usepackage{chemformula}
\usepackage{setspace}
\usepackage[T1]{fontenc} 
\usepackage{amsmath}
\usepackage[cmintegrals]{newtxmath}
\usepackage{bm} 
\usepackage[colorinlistoftodos,prependcaption,textsize=tiny]{todonotes}
\usepackage{amsmath,amsfonts}
\usepackage{algorithmic}
\usepackage{array}
\usepackage{textcomp}
\usepackage{stfloats}
\usepackage{url}
\usepackage{verbatim}
\def\BibTeX{{\rm B\kern-.05em{\sc i\kern-.025em b}\kern-.08em
    T\kern-.1667em\lower.7ex\hbox{E}\kern-.125emX}}
\usepackage{balance}
\usepackage{diagbox}

\usepackage{pdfrender}
\usepackage{xcolor}
\usepackage{glossaries}
\usepackage{makecell}
\usepackage{caption}
\usepackage[usestackEOL]{stackengine}
\usepackage{array, multirow, cellspace}
\usepackage{pifont}
\usepackage{lipsum, tabularx, ragged2e, multirow}
\usepackage{hyperref}
\usepackage[utf8]{inputenc}
\usepackage{enumitem}
\usepackage{wrapfig}
\usepackage{colortbl,hhline}
\usepackage{flushend}
\usepackage{balance}
\usepackage{booktabs}

\usepackage[switch, columnwise]{lineno}
\usepackage{blindtext}

\def\({\left(}
\def\){\right)}

\def\[{\left[}
\def\]{\right]}

\usepackage{glossaries}

\hypersetup{
     colorlinks = true,
     linkcolor = blue,
     anchorcolor = black,
     citecolor = blue,
     filecolor = blue,
     urlcolor = black
}

%% file: Sections/1-Introduction.tex
\section{Introduction}
\subsection{Motivation}

\textit{``Data is the new oil''}, a tremendously valuable and untapped asset in the 21\textsuperscript{st} century~\cite{humby2006data}. 
Today, every individual, business, and government needs data to offer, and improve their services and applications regardless of the discipline. 
However, data silos held by individuals and organizations are not diverse and significant enough to serve the needs of complex tasks such as training machine learning (ML) models \cite{alpaydin2020introduction, lecun2015deep} or supply chain management (SCM) \cite{hugos2018essentials}. 
Therefore, \emph{data sharing} has emerged to make data available to one or more individuals or organizations to maximize the value of the data and provide more robust data pools. 
The more data organizations hold via data sharing, the higher the chance they could maximize their business and social value.
There is a growing public demand for shareable data to reach such broader business and societal benefits. Governments also support this demand by mandating certain data types remain open and shareable through legislation.

Data sharing motivates parties to connect, collaborate, and provide more significant insights leading to better decision-making processes~\cite{gewin2016data}. For example, in healthcare, data sharing can reduce emergency department admissions, repeated doctor visits, and patient worries~\cite{hulsen2020sharing, kim2015comparison, shen2019medchain}. Sharing medical and health data helps doctors gain more information about a patient's medical and drug histories, thereby reducing diagnosis errors, medication errors, duplicate testing, and complicated document administration~\cite{zhang2018towards}. 

\textcolor{black}{Cloud service providers (CSPs) have emerged as a scalable and cost-effective platform for generating, collecting, storing, and sharing data that possess the four Vs of big data, namely volume, variety, velocity, and veracity~\cite{abadi2009data}. The use of CSPs, however, leads to several limitations and security/privacy concerns. First, due to a general lack of trust in sharing data with others and data sovereignty concerns, most users hesitate to store their sensitive data in the cloud~\cite{shao2015fine}. Second, due to the multi-tenancy paradigm, multiple customers are simultaneously vulnerable to attacks~\cite{odun2017cloud}. Finally, customers' data may be leaked, manipulated, and illegally exploited for profit by insiders~\cite{gupta2019layer}}.
The challenges in managing data sharing stem from the reliance on centralization for storage, sharing, and access to the data. However, this approach is only effective if the central storage, networking infrastructure, and data access control (DAC) services are functioning and well-equipped. Previous research~\cite{contreras2015sharing} has shown that data custodians face difficulties setting up central data access management. Moreover, the manual nature of traditional data sharing and access makes it difficult to monitor and enforce data consent and data access agreements. Furthermore, centralized platforms do not allow for the active involvement of various stakeholders, including individuals and organizations, in the management of data sharing.

With the emerging need for distributed data sharing, a spectrum of blockchain-based solutions is being proposed to build decentralized, trustworthy, and transparent data-sharing environments~\cite{soret2021learning}. 
On one end of the spectrum, all collected data are stored in the blockchain, and all access to the data is recorded as transactions in the distributed ledger in a redundant manner, and each miner verifies these transactions. This enhances the immutability, transparency, non-repudiation, and decentralization of a data-sharing platform, making it challenging for malicious parties to attack and manipulate the data for their advantage. Besides, blockchains help solve the problem of sharing data in heterogeneous systems by having a unified data model and interface~\cite{nguyen2022bdsp}. However, realizing the limited performance, scalability, privacy concerns, and high cost of the current blockchain technology, the other end of the spectrum uses blockchains only to facilitate the data-sharing business process without storing data on-chain. Therefore, exploring the emerging spectrum of blockchain-based data-sharing platforms is imperative. 

\subsection{Related Works} 
\label{subsec-rw}
Driven by the rapid innovations in blockchains, a plethora of studies have been conducted to review and survey related topics. However, a survey and taxonomy on data-sharing topics are still missing. For instance, a line of works covers key concepts like blockchain architecture~\cite{zheng2018blockchain, monrat2019survey, gao2018survey, belotti2019vademecum}, system components~\cite{wang2022sok,chatzigiannis2022sok,karantias2020sok,li2022sok}, properties~\cite{feng2019survey,sanka2021survey,li2020survey,wang2023exploring}, consensus algorithms~\cite{bano2019sok,garay2020sok, bach2018comparative}, and further analyzes blockchain applications in various domains from healthcare, smart grid to supply chains~\cite{ali2018applications, reyna2018blockchain, panarello2018blockchain,mollah2020blockchain,wang2021non,agbo2019blockchain,hewa2021survey}. In~\cite{belotti2019vademecum}, authors present distributed ledger technology (DLT)\footnote{In the scope of this paper, we use Distributed Ledger Technologies (DLTs) and Blockchain interchangeably.} and blockchain properties, architecture, etc., and provide instruments to decide \textit{when}, \textit{which}, and \textit{how} to use and deploy blockchains. 

For example, Sanka et al.~\cite{sanka2021survey} focused on related concepts in cryptography and applications like the Internet of Things (IoT) and banking with various quantitative surveys and analyses. Meanwhile, the survey in~\cite{hewa2021survey} presented blockchain-based smart contracts, underlying technologies, and atop applications. Besides, security and privacy are two major concerns needing to be investigated, and blockchain is of no exception. In~\cite{li2020survey}, the authors conducted the first systematic examination and survey on security risks and attacks on blockchain platforms, with potential research directions in this field. 

Many related works explore the applicability of blockchains in real-world applications (e.g., IoT and healthcare). In~\cite{ali2018applications, reyna2018blockchain, panarello2018blockchain}, motivations for integrating blockchains and IoT under different schemes are analyzed, as well as potential advantages, challenges, and applications are discussed. Mollah et al.~\cite{mollah2020blockchain} discussed blockchain's contribution to smart grid applications, e.g., applicability in advanced metering infrastructure. The authors also discussed the features and drawbacks of multiple blockchain-based industrial solutions for smart grids. Agbo et al.~\cite{agbo2019blockchain} conducted a systematic review in which they discussed questions about blockchain use cases in healthcare and presented blockchain-based healthcare applications and their challenges and limitations. From the application perspective, the use of blockchain combined with secure storage and access, the Internet of Medical Things, and Federated Learning (FL) for medical data sharing is studied in~\cite{xi2022review}. Moreover, Wan et al.~\cite{wan2020blockchain} identified the influences for the use of blockchains for data sharing by reviewing a large number of literature.

\begin{figure} [t!]
    \centering
    \includegraphics[width=0.8\linewidth]{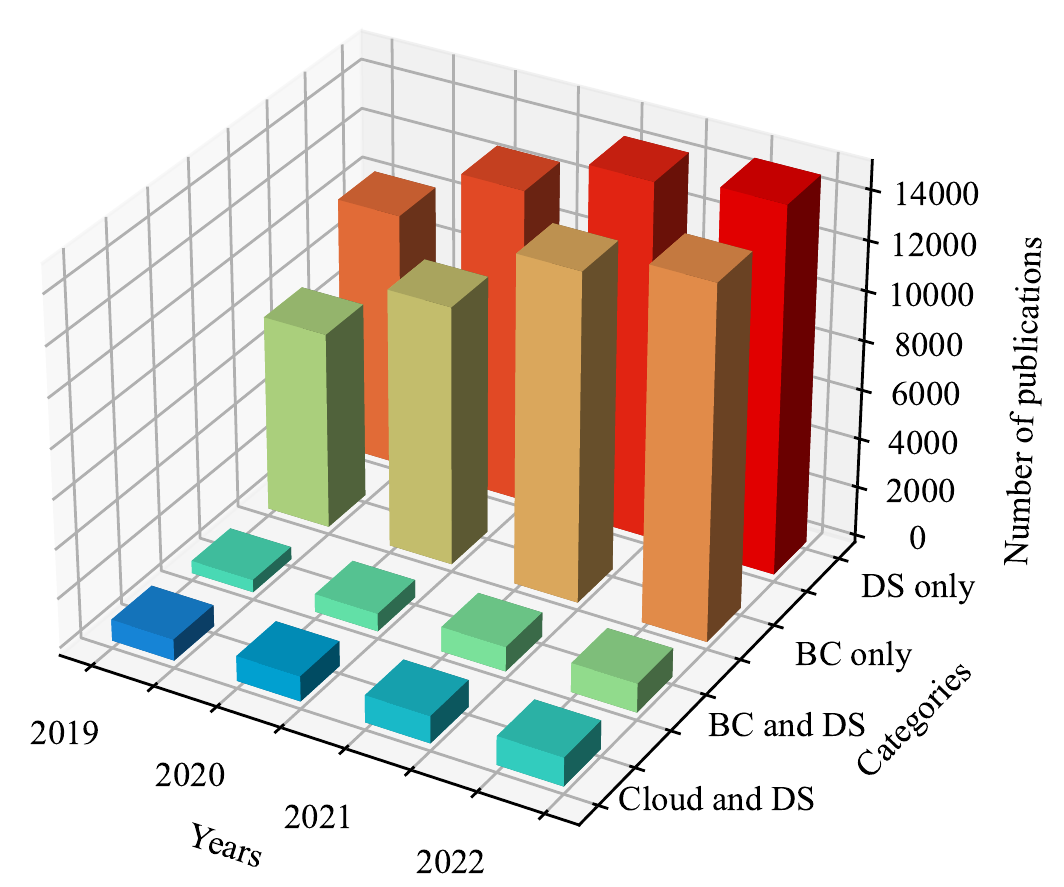}
    \caption{Number of annual publications on blockchains (BC) \& data sharing (DS)}
    \label{fig:statistic}
\end{figure}

In Figure~\ref{fig:statistic}, we summarize the number of related research papers that appeared between 2019 and 2022 as reported by the Dimensions AI software~\cite{dimentions}. The number of publications on blockchains and blockchain-based data sharing increases each year. This indicates the trend of blockchain and data-sharing technologies, as well as the need and significance of this survey and taxonomy study.
We conducted research on the number of publications with four specific keywords, namely, \emph{"Cloud and Data Sharing", "Blockchain and Data Sharing", "Blockchain", and "Data Sharing"}. 
 The fact is that the studies have related keywords of data sharing via cloud techniques and using Blockchain is just over \emph{1000} publications. Meanwhile, there are a tremendous number of research about Blockchain, and the \emph{"Data Sharing"} keyword is mentioned around \emph{12000-14000} publications.
As suggested by much literature above, blockchains are projected to bring new opportunities to mitigate risks and threats in terms of privacy and security, i.e., availability and integrity, for data sharing compared to centralized approaches. Table~\ref{tab:related-works} presents a detailed comparison between our survey and previous studies. 

\input{Tables/Related_works.tex}

\subsection{Contributions}

Even though emerging blockchain technology has accelerated the data-sharing deployment process, many significant challenges still exist. This manuscript takes a closer look at the design and applications of blockchain-based data-sharing deployments. To the best of our knowledge, even though several surveys and literature reviews have been proposed (Section \ref{subsec-rw}), they mainly discuss general blockchain constructions or the specific data-sharing scenario. Any work focusing on the status of combination, usage, and applications between blockchain and data sharing is still absent. It is beneficial to draw a clear roadmap to help learn the ways of adopting blockchain technologies, with appropriate nature and properties, for data sharing in environments with low or no trust. Moreover, the applicability of blockchain for data sharing in applications such as transportation, data marketplace, decarbonization, and digital industries has not been explored~\cite{xi2022review, wan2020blockchain}. To fill this gap, we conduct a survey and taxonomy on blockchain-based data sharing. The main contributions of this paper are highlighted as follows:

\begin{enumerate}
    \item[$\triangleright$] First, we \emph{present an overview of data sharing}, including concepts, types, and limitations. Then, \emph{we provide an overview of DLTs and blockchains}, including their classifications (types and properties), fundamental components (smart contracts and storage), and related techniques (e.g., privacy-preserving techniques). 

    \item[$\triangleright$] Second, we \emph{focus on exploring the need for applying blockchain technology in data-sharing solutions} and accordingly \emph{design a reference architecture for blockchain-based data sharing, called BlockDaSh}. The proposed \emph{BlockDaSh} provides a structural construction for a data-sharing solution and includes key building blocks and layers (majorly on data processing, sharing, and storage). We also discuss how the reference architecture can be adapted for different application scenarios. 
    
    \item[$\triangleright$] Third, we \emph{delineate an in-depth taxonomy of blockchain-based data-sharing applications in multiple use cases.} Specifically, we emphasize the cases as a set of scenarios that closely relate to users' daily needs (measured by frequency, convenience, and importance), including healthcare, supply chain, transportation, smart grid, data markets, education, government, and decarbonization.

    \item[$\triangleright$] Specifically, \emph{we provide detailed lessons learned from deploying Blockchain-based data-sharing schemes to various domains}. The \textit{lessons learned} provide a guideline for deployments of Blockchain-based data-sharing schemes in these applications. 

    \item[$\triangleright$] We \emph{discuss open research issues and potential research directions} on these approaches. We dive into the analysis and discussions of the several critical properties that contribute to data sharing and usage, covering connectivity, security, privacy, incentive/punishment, scalability, and editability.  
    
\end{enumerate}

\begin{figure*}[t!]
    \centering
    \includegraphics[width=0.84\linewidth]{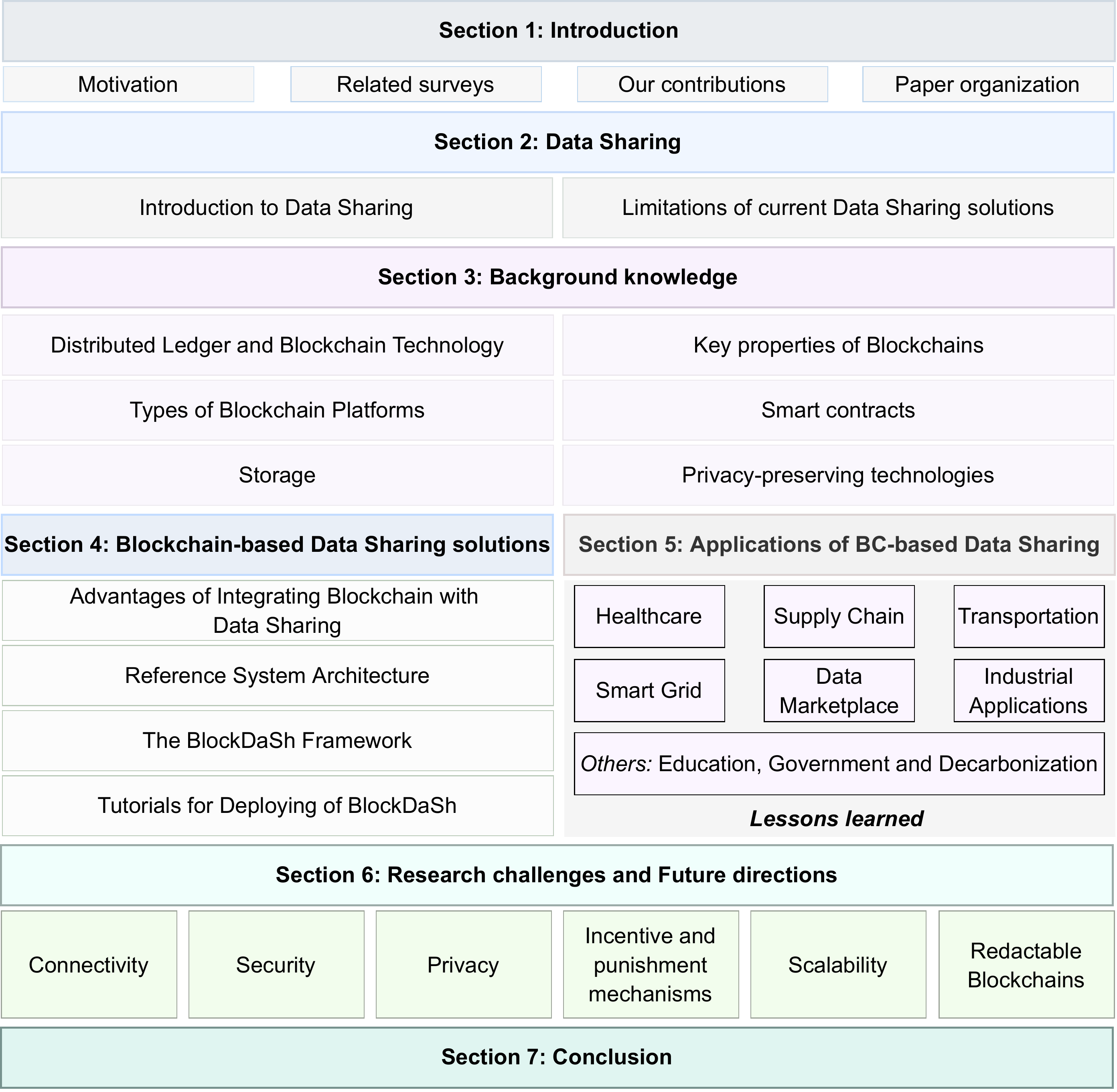}
    \caption{The organization for the paper's content }
    \label{fig:PaperOrganization}
\end{figure*}

\subsection{Paper Organization.}
The organization of this paper is depicted in Figure~\ref{fig:PaperOrganization}. First, we present the background on data sharing and blockchain technologies in Section~\ref{sec-datasharing} and~\ref{sec-background}, respectively. In Section~\ref{sec-BlockDaSh}, we present a reference architecture for blockchain-based data sharing, which can be applied in various application domains. Next, in Section~\ref{sec-taxonomyApp}, we present selected examples from these application domains while discussing their designs and crucially evaluating their abilities to achieve data-sharing goals. Section~\ref{sec-issuesDirections} discusses research challenges and future directions. Finally, we conclude the paper in Section~\ref{sec-conclusion}.

%% file: Tables/Related_works.tex
\begin{table*}[t!]
\small
 \renewcommand{\arraystretch}{1.3}
\centering
\caption{Existing surveys on blockchain-related topics and our research contributions.}
\label{tab:related-works}
\begin{tabular}{|p{0.11\linewidth} | P{0.05\linewidth} | p{0.42\linewidth} |p{0.29\linewidth}|}
\hline
\rowcolor[HTML]{EFEFEF} 
\textbf{Topics} & \textbf{Ref} & \textbf{Key Contributions} & \textbf{Differences From Our Research} \\ \hline

{\multirow{8}{*}{\parbox{2cm}{Blockchain concepts}}}& {\multirow{3}{*}{\parbox{1cm}{\centering \cite{zheng2018blockchain, monrat2019survey, gao2018survey}}}} & {A survey on blockchain design, characteristics, consensus algorithms, applications, research challenges, and future directions}. & {Lacks deep analysis on applications and does not cover blockchain-based data-sharing applications.} \\ \cline{2-4} 

{} & \multirow{3}{*}{\parbox{0.7cm}{\centering{\cite{belotti2019vademecum}}}} & {A survey on DLT and blockchain designs, consensus algorithms, and answering three questions when which, and how to apply blockchain.} & {Lacks application guidelines and does not cover blockchain-based data-sharing applications.} \\ \cline{2-4} 
  
{} & \multirow{2}{*}{\parbox{0.7cm}{\centering{\cite{hewa2021survey}}}} & {An extensive survey on specific applications, research challenges, and future works on smart contracts.} &
  {Focuses on smart contract technology.} \\ \hline

\multirow{5}{*}{\parbox{2cm}{Blockchain \\ \& security}} &  \multirow{3}{*}{\parbox{0.7cm}{\centering{\cite{li2020survey}}}} & {A systematic survey on security threats and attacks on blockchain platforms, along with potential solutions to enhance security.} & {Focuses on security aspects of blockchains.}\\ \cline{2-4} 

{} & \multirow{3}{*}{\parbox{0.7cm}{\centering{\cite{leng2020blockchain}}}} & A survey on blockchain security at process, data, and infrastructure levels from the perspective of information systems. &
Concentrates on the security aspects of blockchains. \\ \hline

\multirow{3}{*}{\parbox{2cm}{Blockchain \\ \& privacy}} & \multirow{3}{*}{\parbox{0.7cm}{\centering {\cite{feng2019survey}}}} & {A survey on privacy issues in blockchain platforms, along with possible cryptographic defense mechanisms and future research directions.} & {Focuses on the privacy aspects of blockchains.} \\ \hline

{\multirow{4}{*}{\parbox{2cm}{Blockchain \\ \& IoT}}} & \multirow{2}{*}{\parbox{1cm}{\centering{{\cite{ali2018applications, reyna2018blockchain, panarello2018blockchain}}}}} & {The applicability of integrating blockchain and IoT, issues, advantages, and applications are discussed.} & {Focuses on the applications of blockchain in IoT in general.} \\ \cline{2-4} 

{} & \multirow{2}{*}{\parbox{0.7cm}{\centering{\cite{mollah2020blockchain}}}} & {A survey on using blockchain for future smart grid: requirements, opportunities, and challenges.} & {Mainly discusses the applications of blockchain in smart grids in general.} \\ \hline  

\multirow{4}{*}{\parbox{2cm}{Blockchain \\ \& healthcare}} & \multirow{2}{*}{\parbox{0.7cm}{\centering{\cite{agbo2019blockchain}}}} & {A systematic review on blockchain use cases in healthcare, applications, and research challenges.} & {Mainly discusses the role of blockchain in healthcare.} \\ \cline{2-4} 

{} & \multirow{2}{*}{\parbox{0.7cm}{\centering{\cite{xi2022review}}}} &
  {A review of the development of blockchain-based medical data sharing} & {Lacks detailed architecture and techniques for data sharing in healthcare.} \\ \hline

{\multirow{2.5}{*}{\parbox{2cm}{Blockchain, data sharing, \\\& supply chain}}} & \multirow{3}{*}{\parbox{0.7cm}{\centering{\cite{wan2020blockchain}}}} & {A systematic review on impacts, potential challenges, and future work of applying blockchain for data sharing in supply chains.} & {Lacks data-sharing architecture and applications, and only discusses aspects of data sharing in the supply chain.} \\ \hline

\multirow{11}{*}{\parbox{2cm}{Blockchain-based data sharing}} & \multirow{11}{*}{\parbox{1cm}{\centering {\emph{Our survey}}}} & An extensive survey on blockchain-based data sharing:
  \begin{itemize}[noitemsep, topsep=0pt, leftmargin=15pt]
  
  \item We present an overview of the data-sharing topic, including fundamental concepts, diversity, and limitations, and We discuss the need for leveraging blockchains in data sharing. Then, we introduce a reference architecture called \emph{BLockDaSh}, which can be used widely for data-sharing applications.
  
  \item To our best knowledge, this is the first survey and taxonomy on leveraging blockchain/smart contract technology for sharing data in diverse applications, i.e., healthcare, supply chain, transportation, smart grid, data marketplace, and industry.
  
  \item Fine-grained tables of classification and key lessons learned are presented to provide insights into blockchain-based data-sharing solutions. 
\end{itemize} 
More details of contribution are given in the text.
&  \\ \hline
 
\end{tabular}
\end{table*}

%% file: Sections/2-Data_Sharing.tex
\section{Data Sharing}
\label{sec-datasharing}

 \begin{figure*}[t!]
	\centering
	\includegraphics[width=0.8\linewidth]{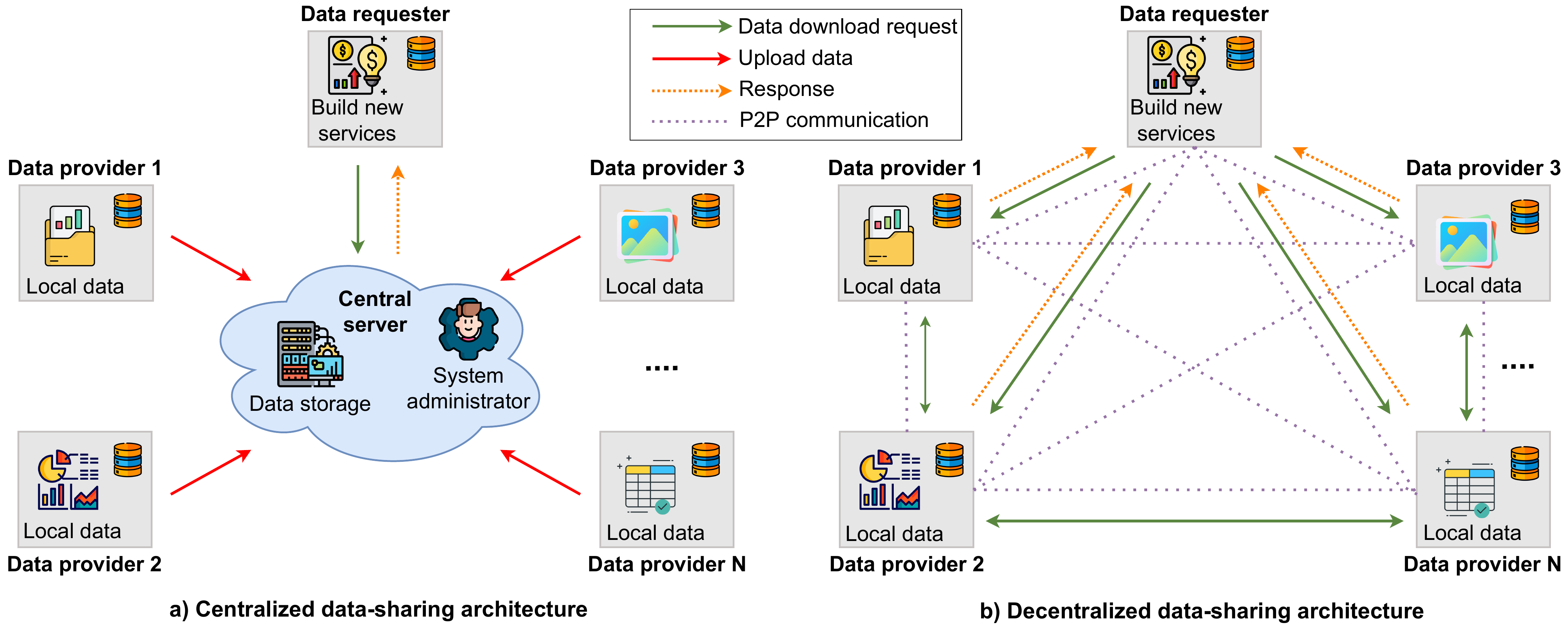}
	\caption{Types of data-sharing architectures}
	\label{fig:datasharingarchitecture}
\end{figure*}

This section introduces the basic concepts of data sharing and outlines the limitations of existing data-sharing solutions.

\subsection{Introduction to Data Sharing}

Data sharing is the practice of making data accessible to enhance its quality and value. It involves the exchange of data among individuals, organizations, and governments through various means, such as email, file-sharing software, and disk drives. Advancements in technology have made data sharing faster, more efficient, and more widespread. The growth of IoT devices and services has further increased the number of potential data sources, reinforcing the trend toward data sharing. By sharing data, organizations can gain valuable insights, improve decision-making, and foster innovation. However, data sharing also presents challenges related to privacy, security, and ethical considerations that must be carefully addressed to ensure the protection of sensitive information~\cite{datasharing}.

The data-sharing process involves two main entities: data providers (DPs) and data requesters (DRs). DPs and DRs (Table~\ref{tab:abbreviation-list} presents a list of abbreviations used in the paper) can be individuals, businesses, or organizations. DPs own meaningful data or don’t necessarily hold the data but they have the right to access it and share it with third parties. Besides, they are willing to share their data for non-profit purposes or monetize it. DRs aim to exchange data with DPs to derive business insights, build services, or create products. There are two main approaches to data sharing: 1) \emph{direct data sharing} between DPs and DRs and 2) \emph{platform-based data sharing} where DPs store data in a local or third-party platform/repository (e.g., cloud storage) and set rules and restrictions DRs' accessing on the shared data. 
The first approach is not scalable and raises serious security and privacy concerns. In addition, with this kind of sharing, DPs lose control over the usage of shared data, which poses potential risks for DPs, especially when sharing sensitive data.
Therefore, the latter approach has been considered an effective solution and is widely applied in a variety of applications.

\input{Tables/Glossary.tex}

Data-sharing software architectures (cf. Figure~\ref{fig:datasharingarchitecture}) can be classified into two main categories: centralized (e.g., cloud-based solutions) and decentralized (e.g., FL,  multi-party computation (MPC), or blockchain-based solutions).
In the centralized approach, DPs are limited to the scope of an individual, an enterprise, or a small group of parties. All shared data are kept on a local or a cloud server and are visible to the storage administrator. For example, consider a company (i.e., a DR) wanting to develop an ML-based tourist destination recommendation service. To train the ML model, it needs data from sources ranging from travel agencies to local tourism authorities (i.e., DPs). If the company can convince such parties to provide data, it can ask them to upload their data to its cloud-based Dropbox~\cite{drago2012inside} account (without losing generality). Alternatively, DPs may share data via their Dropbox, allowing the company to download their data. 
While most security issues can be overcome by setting appropriate access controls for DPs and DRs, this approach is vulnerable to a single point of failure, not scalable (e.g., high data volume, velocity, and requires too many integrations at the central server), and limits privacy (e.g., the administrator has access to all data and DPs lose control of their data as soon as they are shared). 

Conversely, the decentralized data-sharing approach can support many DPs (e.g., thousands of IoT devices and parties) to join the data exchange process. Shared data can be stored in a decentralized manner, e.g., in each DP's local storage, reducing system-wide data leakage and data breaches. For example, FL has emerged as a decentralized and collaborative data-sharing solution for training ML models while improving privacy and security~\cite{dayan2021federated, mcmahan2017communication,yu2023ironforge}. The core idea of FL is that DPs (aka FL clients), such as IoT sensors or institutions, collaboratively support the model training process with the federated server without revealing their raw datasets~\cite{nguyen2021federated, sheller2020federated}. To begin a training phase, the federated server (e.g., a DR in Figure~\ref{fig:datasharingarchitecture}) initiates a naive model and distributes it to a selected set of DPs for partial training tasks. DPs then conduct local training with the initial ML model and their private data. Next, DPs send updated model weights or gradients to the federated server. The server aggregates local updates and sends the upgraded model back to a new-selected set of DPs for the next round. The training process continues iteratively until desired performance metrics are met and/or the ML model reaches convergence.

MPC is another potential approach for decentralized inter-organizational data sharing using cryptographic techniques while maintaining the privacy of inputs~\cite{yao1982protocols, agahari2022not}. MPC typically makes use of the idea of homomorphic encryption (HE), which allows computing on encrypted data without having to decrypt them~\cite{rivest1978data}. MPC decreases the reliance on third parties as the computation is performed locally. Unlike FL which shares model weights after each iteration, MPC only shares the final result with DRs after the complete calculation round. Moreover, the final result is composed of encoded original data fragments. Therefore, MPC is deemed to have a higher level of privacy and security than FL.

\subsection{Limitations of Current Data-sharing Solutions}
\label{sec:limit_data_sharing}

Although existing data-sharing solutions offer salient features, they pose several challenges as follows:
\begin{itemize}
    \item \emph{Centralization.} When shared data are collected, stored, and processed in a central server managed by a single party, it leads to security, privacy, and trust issues.
    
    \item \emph{Heterogeneity and interoperability.} As shared data are often collected and gathered from various data sources in different data types and communication protocols~\cite{Stuart2018}, it is challenging to process and exchange data between vertical domains and organizations.
    
    \item \emph{Resource consumption.} In the centralized design, the central server needs sufficient bandwidth and storage to handle aggregated data from multiple DPs. The same concern exists, albeit to a lesser extent, also for the aggregated server in FL, even though DPs do not provide raw data.
    In addition, as MPC must produce a massive number of random numbers to prevent raw data from being leaked, it requires significant computational resources, potentially lowering the system's performance. Moreover, transferring many messages among MPC parties distributed across multiple locations increases communication delay and costs.

    \item \emph{Security.} Due to the decentralization and heterogeneity of data sources, and the central management and storage approach, current data-sharing systems are vulnerable to breaches and attacks. Moreover, there are known cases of shared data being modified, manipulated, or illegally exploited for profit by untrustworthy central server operators, and malicious clients, who submitted modified or misleading parameters, reducing the utility of the global model in FL \cite{gupta2019layer, tolpegin2020data}.

    \item \emph{Privacy.} During the data exchange process, shared data containing personally identifiable information (PII) may be intentionally or inadvertently leaked. Furthermore, once data are shared, DPs lose control over their shared data because they do not know what forms of data which DRs intend to collect, the purpose of their data collection, and where and with whom the data will be further shared. It leads to a high information asymmetry (IA) between DPs and DRs and threatens the privacy of DPs. Regulations on data sharing have been imposed to overcome some of these concerns, e.g., the Personal Data Protection Act~\cite{chik2013singapore}, the General Data Protection Regulation~\cite{gdpr}, and the Cyber Security Law of the People’s Republic~\cite{clpr}. 

    \item \emph{Trust.} 
    In an untrustworthy environment like the Internet, current solutions bring uncertainty and threats to customers, especially due to the aforementioned security and privacy issues. Consequently, there is a general lack of trust across the entire ecosystem, ranging from DPs, DRs, data subjects, and data aggregators to platform providers, inhibiting the contribution and use of data. 
\end{itemize}
  
Recent advancements in blockchain technology provide potential solutions to overcome many of these problems such as centralization, heterogeneity, poor interoperability, privacy, and security (e.g., integrity, availability, and non-reputation)~\cite{li2022smart,xiao2020survey}. Furthermore, blockchains can increase the provenance of data and the traceability of data-sharing systems. After providing a briefing on background knowledge about DLTs, blockchains, and privacy-preserving methods in Section~\ref{sec-background}, we then discuss how blockchain-based data-sharing solutions leverage these properties.

%% file: Tables/Glossary.tex
\begin{table*}[t!]
\small
\centering
\renewcommand{\arraystretch}{1.2}
\caption{List of abbreviations used in the paper}
\label{tab:abbreviation-list}

\begin{tabular}{|m{1.5cm}|m{6.5cm}|m{1.5cm}|m{6.5cm}|}
\hline
\rowcolor[HTML]{EFEFEF} 
\textbf{Acronym} & \textbf{Definition} & \textbf{Acronym} & \textbf{Definition}  \\ \hline

API & Application Programming Interface & IIoT & Industrial Internet of Things \\
ABE & Attribute-Based Encryption & IPFS & InterPlanetary File System \\

B2B & Business-to-Business & ITS & Intelligent Transportation System \\

CA & Certificate Authority & IV & Intelligent Vehicle \\

CHF & Chameleon Hash Function & KSI & Keyless Signature Infrastructure \\

CID & Content Identifier & ML & Machine Learning \\
CP-ABE & Ciphertext-Policy Attribute-Based Encryption & MPC & Multi-party Computation \\

CSP & Cloud Service Provider & OBU & Onboard Unit \\

dApp & Decentralized Application & P2P & Peer-to-Peer\\

DAC & Data Access Control & PBFT & Practical Byzantine Fault Tolerance \\ 
DAG & Directed Acyclic Graph & PET & Privacy-Enhancing Technology \\

DDB & Distributed Database & PII & Personally Identifiable Information \\
DHT & Distributed Hash Table & PoA & Proof of Authority \\ 

DLT & Distributed Ledger Technology & PoW & Proof of Work \\

DP & Data Provider & PPT & Privacy-Preserving Technique\\ 
DR & Data Requester & PRE & Proxy Re-encryption\\ 

DSP & Data Service Provide & RSU & Roadside unit\\ 
DSV & Data Shapley Value & SCM & Supply Chain Management\\
DUC & Data Usage Control & SHR & Surgery Health Record \\

EHR & Electronic Health Record & SPOF & Single Point of Failure \\
EMR & Electronic Medical Record & SCSSM & Supply Chain Social Sustainability Management \\
ERP & Enterprise Resource Planning & TEE & Trusted Execution Environment \\

FDI & False Data Injection & TET & Transparency-Enhancing Technology \\
FL & Federated Learning & TPS & Transactions per Second \\
FSC & Food Supply Chain & TSSS & Threshold Secret Sharing Scheme \\


HACCP & Hazard Analysis and Critical Control Point & V2V & Vehicle to Vehicle  \\
HE & Homomorphic Encryption & VANET & Vehicular ad-hoc Network \\ 

IA & Information Asymmetry & VECON & Vehicular edge Computing and network\\

IoT & Internet of Things & zk-SNARK & Zero-Knowledge Succinct Non-Interactive Arguments of Knowledge  \\

\hline
\end{tabular}
\end{table*}

%% file: Sections/3-Background_knowledge.tex
\section{Background Knowledge}
\label{sec-background}

In this section, we present some background knowledge about blockchain technology and privacy-preserving techniques (PPTS). 

\subsection{Distributed Ledger and Blockchain Technology}

{\textit{Distributed Ledger Technologies.}} 
In recent years, we have heard various buzzwords appearing in every media hype like Metaverse \cite{dionisio20133d}, cryptocurrencies \cite{mukhopadhyay2016brief}, or non-fungible tokens\cite{nadini2021mapping}. Their appearance has piqued the interest of researchers from various disciplines, e.g. finance \cite{treleaven2017Blockchain}, technology \cite{yang2022fusing}, economics \cite{catalini2020some} and management \cite{berdik2021survey} to participate in the research and study. And, DLT is the underlying underpinning of the explosion and rapid development of these technologies. A distributed ledger is an immutable, transparent, and decentralized data storage and transfer protocol that is maintained by a peer-to-peer (P2P) network of participants. Besides, the network participants use a consensus protocol to determine which participant has the right to update the ledger and in which order. The ledger is updated using transactions that contain a set of attributes like sender and receiver addresses, the amount of cryptocurrency to transfer, a smart contract function to invoke and its input parameters, and a digital signature. A distributed ledger can be structured as a blockchain (i.e., a linked chain of blocks), directed acyclic graph (DAG) \cite{wang2022sok}, HashGraph, HoloChain, or Tempo \cite{panwar2020distributed}. 


In the first type of DLT, blockchain organizes transactions into blocks, which are linked together by consensus mechanisms \cite{nakamoto2008bitcoin}. Differently, DAG technology does not require miners to confirm and validate transactions like Proof of Work consensus (PoW) does \cite{cullen2020resilience}. The DAG transactions are confirmed in the sequential flow by nodes, which validated at least two of the previous transactions on the ledger. Furthermore, a node's transactions on the distributed ledger database become more valid when this node validated more preceding transactions. DAG is considered as an alternative approach to the blockchain with greater improvements in terms of scalability and fee-less nano-transactions. Next, for the hashgraph structure, transactions are stored in a parallel structure on the same timestamp. With the support of gossip and virtual voting protocols, the transaction is verified by nodes in the network before being recorded in the ledger \cite{akhtar2019Blockchain}. This structure stands out for its small storage unit requirement and the fact that each node will be aware when a consensus has been reached that the blockchain's nodes are uncertain. Especially, a much more decentralized-driven architecture, named Holochain, intends to avoid using any global consensus protocol. Therefore, it provides every node with its own chain and becomes an agent-centric structure \cite{zaman2022thinking}. The last one, called Tempo, uses a technique known as sharding to divide the ledger, which properly orders all of the network events that occurred. In essence, transactions are recorded in the ledger according to the sequence of events rather than the time stamp \cite{panwar2020distributed}. Until now, since BC is the distributed ledger technology that is most commonly used, we mainly focus on BC technology in this work.

{\textit{Blockchain.}} Satoshi Nakamoto introduced the first blockchain, Bitcoin, in 2008 \cite{nakamoto2008bitcoin}, and it was released in 2009 as the sparking platform for more than 10000 cryptocurrencies that have evolved since then \cite{crypto}. blockchain is made up of records of transactions or blocks that are chained together to form a tamper-resistant ledger, and it allows for secure decentralized data storage in distributed networks. As an element of the blockchain, each block consists of a header and a body containing a batch of validated transactions. The header of a block, in particular (except for the first, known as the genesis block), contains an inverse reference pointing to a previous block. The hashed value of the previous block is used as the inverse reference. This prevents fraudulent actions because a change in any block in historical records invalidates all blocks because all subsequent hashes change, and nodes in the blockchain detect it. A block header also contains other fields such as Version, Timestamp, MerkleRoot, or Nonce, counting on specific purposes. Besides, different blockchains have different block size limits. For instance, while the block size of Bitcoin is limited to 1 MB and is able to store around 2000 transactions, each Ethereum block has a target size of 15 million gas, expandable to 30 million based on demand \cite{nakamoto2008bitcoin, buterin2014next, ether}.

To form a distributed consensus between nodes, blocks and their transactions are transmitted and verified across the network. When a transaction is created, it is signed by the owner of the source assets, who possesses a pair of public and private keys. Then, the transaction is broadcast to the entire blockchain network, where each network node validates and propagates it based on a list of criteria. Every node, known as a miner or validator, that receives transactions from various sources, only forwards transactions to its peers when it has already verified these transactions. This ensures invalid transactions are immediately dropped. They will have a memory pool of verified transactions for building blocks at the time. In addition, if a miner wants to create a block, the distributed consensus problem must be solved. Therefore, only miners who solve the consensus problem can broadcast their new blocks across the network. Now, other nodes will participate in the block validation process \cite{miller2015discovering, wang2019survey}. When a block is proved to be eligible by any node who saw it, it is appended to miners' local chains where the miners have not solved the consensus problem yet, via the inverse reference pointing to the parent block. A block's generator may claim a certain number of new coins as well as fees collected from all enclosed transactions.

\begin{figure*}[t!]
 	\centering
 	\includegraphics[width=0.7\linewidth]{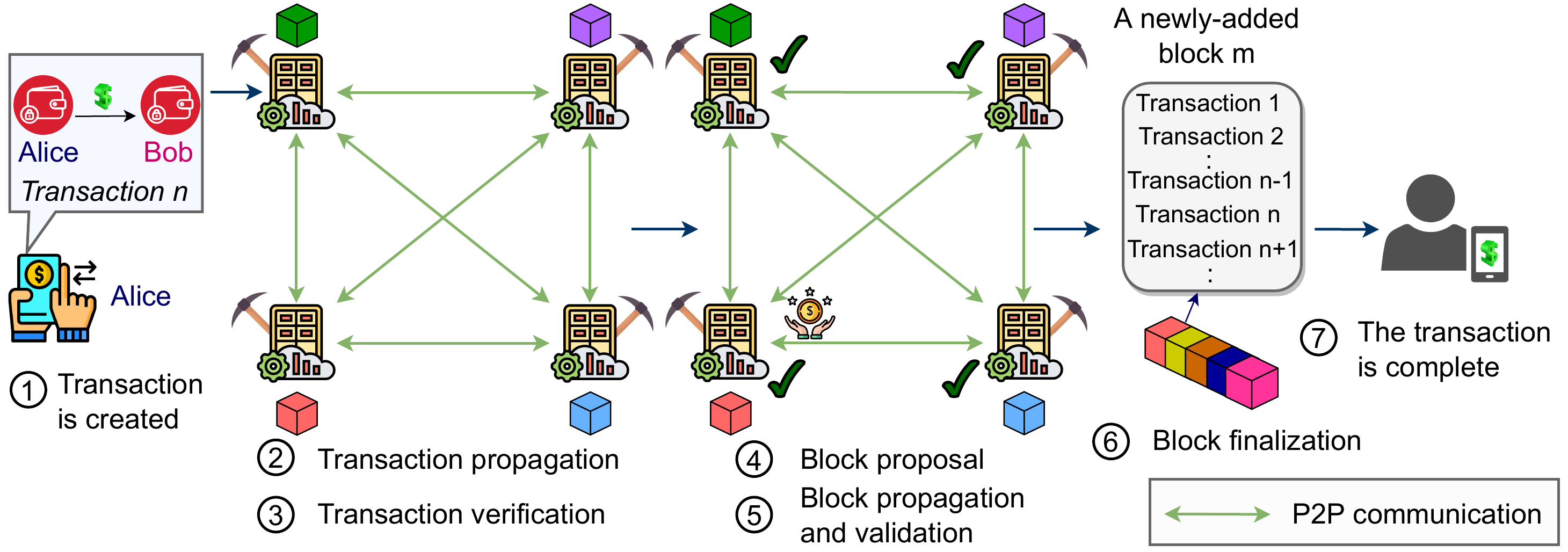}
 	\caption{The general working flow of a blockchain network.}
 	\label{tab:Fig2}
 \end{figure*}

For example, in the case of Bitcoin, Alice and Bob create their own public and private keys. They are able to share public keys with each other in the entire network, while private keys are kept secretly. Alice wants to send two bitcoins to Bob. To accomplish this, Alice must sign a transaction with her private key containing the main information, such as transaction inputs and outputs. This transaction is now broadcast to the Bitcoin network to wait for verification. Then, miners in the network try to verify Alice's transaction by validating a list of criteria such as the signature in this one. If all requirements are valid, the transaction is grouped into blocks with other verified transactions by miners. Each node or miner works on solving a difficult puzzle for its block in the PoW consensus algorithm. The first miner who successfully completes PoW is permitted to propagate its newly-created block to neighbors. After that, other miners will double checks this block under shared rules. For instance, miners ensure that all transactions in the proposed block are valid (e.g., the block size is within acceptable limits and the first (and only) transaction is a coinbase generation transaction). Next, they also check whether the new block points to its predecessor in the longest chain. If it does, this block would be assembled into miners' local chains. Different miners will build upon it by referring to it as the prior block. Eventually, as blocks are added to the longest chain, the transaction is completed, and Bob receives Alice's coins. The miner who creates the accepted block receives a number of coins and transaction fees for their effort. The entire process of a transaction working is illustrated in {Figure~\ref{tab:Fig2}}.

\subsection{Key Properties of Blockchain}


Blockchains have several unique properties that make them suitable for new trustworthy and efficient ways of employing them~\cite{monrat2019survey, dai2019blockchain}. Following are some of the key properties: 

\emph{Decentralization.} 
Blockchain is maintained and operated by a P2P network of nodes rather than a single authority. With decentralization, everyone has access to the blockchain network, can send transactions, and build blocks. Moreover, users can securely store and share their assets on the blockchain, and they can exercise full control over their assets using their private keys. It eliminates the dependence on a trusted third party that is likely to lead to a single point of failure, performance concerns, and censoring. For example, in traditional payment systems, clients must rely on a bank to hold their funds and wait for the bank's authentication and validation before completing transactions. 
If the bank network faces a large-scale attack (e.g., distributed denial-of-service) it could result in a single point of failure and loss of funds. 

\emph{Immutability.} On blockchain platforms, any manipulation or attempt to tamper with transactions would be detected and aborted by peers. The main reason is that blocks are linked together, and each block header contains the hash of the predecessor block. Specifically, the hash value of a block is computed from its header content (e.g., its parent hash value and Merkle root hash that act as a fingerprint of all transactions in the block) using a hash algorithm like SHA-256. Therefore, if malicious entities attempt to modify any data in a transaction, they must regenerate the new block hash to match the modified transaction(s). Also, if any successor blocks are generated, attackers need to regenerate each of them to match the predecessor's new block hash. However, depending on the consensus protocol adopted by the blockchain modifying a chain of blocks is either computationally impossible or would require a supermajority (typically more than 2/3) of votes or cryptocurrency held by the blockchain peers. Therefore, a blockchain can be used to store data with high a level of immutability.

\emph{Transparency.} The content of transactions and the ledger state are visible to all network nodes that validate transactions, build blocks, and store a copy of the ledger. Such a high level of transparency helps blockchain-based applications instantly identify non-compliant actions, operational mistakes, and execution errors.

\emph{Traceability.} Because all transactions and ledger state changes are immutably recorded, timestamped, and transparent, a blockchain provides an audit trail of activities on an asset or by a party~\cite{yu2018decentralized, monrat2019survey}. For example, SCM systems use blockchains to track the provenance and chain-of-custody of goods~\cite{kshetri20181, agrawal2021blockchain}.
 \emph{Non-repudiation.} A blockchain transaction must be cryptographically signed by the sender and the signature is validated by all network nodes before including in the leader. Hence, a party cannot repudiate later that the transaction did not happen.

 \emph{Pseudonymity.} Blockchain transactions are recorded based on sender and receiver addresses derived from the public keys. Hence, blockchain addresses do not reflect the individuals or organizations behind a transaction. However, through other means such as tracking IP addresses, analyzing payments and withdrawals at centralized exchanges, and social media posts, it has been possible to deanonymize transacting parties, particularly if addresses are reused in multiple transactions. Hence, blockchain provides pseudonymity~\cite{al2019privacy}. 


\subsection{Types of Blockchain Platforms}
When developing a blockchain-based solution, it is necessary to determine which type of blockchain is best suited for the project. Therefore, it is critical to have a thorough understanding of blockchain structure classifications.

Blockchain platforms can be broadly classified across two dimensions: 1) based on the technical excludability of users (i.e., public vs. private) and 2) the ability to perform activities such as validating transactions and updating the protocol (i.e., permissioned vs. permissionless)~\cite{iso,ahmadjee2022study, zheng2018blockchain, makhdoom2019blockchain}. 
A \emph{public blockchain} is accessible to the public for use, and provides incentives for users to join, contribute computing power, and charge fees for the use. No single entity owns or controls the entire blockchain~\cite{swan2015blockchain}.
A \emph{private blockchain} is accessible only to a limited group of users for use and typically sits within a physical or virtual private network. The network peers are authorised by some centralized party or consortium and are not offered explicit incentive mechanisms like paying transaction fees. Any blockchain platform can be used as a private blockchain. 
In a \emph{permissioned blockchain}, pre-authorisation is required to perform a particular activity, such as issuing transactions, validating them, and building a block. They may also offer fine-grained permissions such as permission to create a particular asset/data and execute a function in a smart contract. Conversely, in a \emph{permissionless blockchain}, no such authorization is required to perform any activity.

Based on these two dimensions, there can be four combinations of blockchain platforms: public-permissionless, public-permissioned, private-permissionless, and private-permissioned.
\emph{Public-permissionless blockchains}, such as Bitcoin and Ethereum, can be used by anyone to transact, validate transactions, and build blocks. While they have the highest level of transparency and security, they typically have lower transaction throughput and higher latency to finalize a transaction. 
\emph{Public-permissioned blockchains}, such as Ripple, Avalanche, and Algorand, allow anyone to transact but only a pre-authorised set of network peers can validate transactions. They still charge transaction fees. These blockchains tend to have better transaction throughput, lower latency to finalize a transaction, and lower costs compared to public-permissioned blockchains due to the lower number of preselected peers involved in transaction validation. Conversely, the lower number of peers and their selection by a centralized authority also make them more vulnerable to traditional attacks, hence, are considered less secure than public-permissionless blockchains~\cite{ahmadjee2022study, li2019information}.
While a \emph{private-permissionless blockchain} has pre-authorised network peers, those peers do not need any further permissions to issue transactions. For example, Ethereum, which is permissionless, can be used in a private network. Private networks tend to have higher transaction throughput and lower latency to finalize a transaction compared to public networks~\cite{mazzoni2022performance}.
\emph{Private-permissioned blockchains}, such as Hyperledger Fabric, have pre-authorised network peers and users need to satisfy fine-grained permissions to transact. They are more suitable for regulated industries (e.g., finance and healthcare industries) where it is essential to know your customers. Hence, an approval process is used before giving access to the blockchain, and further authorization is required to issue transactions. Also, they offer the greatest level of privacy~\cite{mazzoni2022performance}. This privacy and transparency trade-off needs to be balanced when sharing data using blockchains~\cite{ahmadjee2022study}.

When a group of organizations use a private blockchain it is typically referred to as a \emph{consortium blockchain}. Such blockchains are more accessible than private blockchains, as they are partially decentralized. Moreover, most consortium blockchains are private-permissioned; hence, have better privacy properties. Because of the decentralization and better privacy private-permissioned blockchains are more attractive for data-sharing use cases. 

Hybrid blockchains combine properties of both private and public blockchains. They are essentially sidechains that anchor to public-permissionless blockchains to enhance data integrity~\cite{wu2017democratic}. This is achieved using approaches like storing the Merkel Root hash of the ledger state or block hash of the private blockchain on a public blockchain. Hybrid blockchains can be widely applied for sharing data in application domains such as SCM, banking, and governments.

\subsection{Smart Contract}

A traditional contract is an agreement between individuals or organizations to carry out a set of activities. Proposed in the 1990s by Nick Szabo, smart contracts are digital programs where traditional contractual stipulations are embedded into hardware or software, which automatically executes pre-agreed contract terms under different constraints~\cite{smartcontract}. With the advent of blockchain technology, smart contracts are being applied in a wide range of applications (e.g., SCM and healthcare). A \emph{smart contract} is a set of executable codes that represents an agreement between two untrustworthy entities without reliance on a trusted third party~\cite{buterin2014next, syed2019comparative}. It is built above the ledger and consensus layers of a blockchain architecture~\cite{xie2019survey}. Applications that are built while using smart contracts and the underlying business logic and data storage are known as \emph{decentralized applications} (dApp)~\cite{khan2021Blockchain, wu2021first}.

Once deployed a smart contract is recorded immutably on the blockchain. Its functions could be invoked by submitting a transaction to its address~\cite{hewa2021survey}. Distinct access controls can also be defined for each function in a smart contract. When conditions in the contract are met, the corresponding actions are enforced and executed automatically~\cite{koulu2016Blockchains}. The updated states and recorded on the ledger and the results are sent to the transaction issuer. Every peer in the blockchain network keeps a copy of the smart contract in its local ledger and executes the smart contract based on the transaction inputs. Like data immutability, this hinders tampering attempts from malicious attackers. By utilizing smart contracts, actions can be immediately enforced and executed in a deterministic manner when certain pre-defined conditions are satisfied. 
Therefore, smart contracts add programmability to data and assets managed by a blockchain. Moreover, the smart contract is transparent to all users as per the blockchain's permission setting; thus, increasing trust and fairness.


\subsection{Storage}


One of the most vital aspects of blockchains is ledger storage, where data are stored, uploaded, and retrieved. As of October 2022, the sizes of the Bitcoin~\cite{bitcoinsize} and Ethereum~\cite{ethersize} ledgers were approximately 433 and 975~GB, respectively. These numbers continue to increase as more transactions are processed. Therefore, recording large volume data is a challenge not only for cryptocurrencies but also for IoT~\cite{hajjaji2021big} and data-sharing applications.

A blockchain-based application can store data on the blockchain (on-chain) or outside (off-chain)~\cite{hepp2018chain}. With on-chain solutions, data can be sent to the blockchain via transactions and can be stored in the ledger state or on the transaction log. In public blockchains, a transaction fee must be paid to store data on-chain. Also, transactions with large data payloads reduce the number of transactions that can be included in a block, reducing the transaction throughput of the network.


Alternatively, off-chain solutions store data off the blockchain but record metadata about data on the blockchain. This provides an integrity check for data stored off-chain. Generally, the hash of data and a reference to access data (e.g., URL) are stored on the blockchain. This establishes a link between off-chain data and on-chain metadata, which can be used to validate the integrity of off-chain data. Because metadata tends to be smaller, the corresponding storage cost will be lower and the impact on the transaction throughput is minimal. Therefore, off-chain storage is popular among blockchain-based data-sharing solutions. However, this approach provides only an integrity check as data cannot be recovered if off-chain data are modified or lost. 

InterPlanetary File System (IPFS) is a distributed hash table (DHT) based P2P storage network solution to share and retrieve data in a distributed system~\cite{benet2014ipfs, bernardini2019Blockchains}. IPFS introduces the concept of content addressing when content can be retrieved based on a content-based routing mechanism rather than its location like in traditional protocols. When a client imports a file to the IPFS network, the file is split into small chunks~\cite{tang2020research}. The hash of a chunk's data is used as its content identifier (CID)~\cite{ipfs_addressing}. These CIDs form a Merkle-directed acyclic graph to link chunks together. IPFS uses a DHT to allocate chunks to IPFS nodes and later locate them using their CID value~\cite{athanere2022Blockchain}.
When retrieving chunks, they are hashed again and matched against their CID to verify the integrity. IPFS is a popular off-chain storage solution for dApps and is also adopted by data-sharing constructions ~\cite{huang2020secure, klems2017trustless, khatal2021fileshare, chen2022fine}.

\subsection{Privacy-Preserving Technologies}

Due to growing data privacy concerns and privacy policies restricting storing, exchanging, and utilizing data, especially PII data, PPTs have been developed to safeguard data from unintentional leaks and deliberate disclosure efforts~\cite{kaissis2020secure, 7163220, wang2020preserving}. The PPTs are commonly categorized into two sub-classes: privacy-enhancing technology (PET)~\cite{heurix2015taxonomy} and transparency-enhancing technology~(TET) \cite{tet} with various protection purposes~\cite{SDMMetho12}. While PETs focus on data minimization, confidentiality, and unlinkability, TETs tend to reinforce transparency and intervenability.

PETs can be implemented in hardware and software, and enable individuals and organizations to benefit from a significant volume of data without disclosing it. PETs also ensure that sensitive data are gathered as infrequently as feasible (i.e., data minimization) and only authorized parties have access to and use the data (i.e., data confidentiality). Moreover, the data are processed and analyzed exclusively for the purpose for which they were acquired (i.e., data unlinkability). There are three approaches to protect data privacy in PETs~\cite{Part3Pri51}. 
The first approach uses specialized hardware and cryptographic techniques to isolate input data and computations performed on them, and then share only the output data with DRs. Typical techniques include the use of a trusted execution environment (TEE) like Intel SGX, MPC, and HE. For example, with HE~\cite{acar2018survey}, DPs can achieve a higher standard of data security and privacy by performing computations on encrypted data and sharing outputs with DRs in encrypted form.
Theoretically, for a given HE function $E$ with respect to a function $f$, the encrypted of $f$ can be computed by computing a function $g$, over encrypted variables of $X = \{x_1, x_2, ... x_n\}$.
\begin{equation}
    E(f(x_1, x_2, ..., x_n)) \equiv g(E(x_1), E(x_2), ..., E(x_n))
\end{equation}
Each participant can encrypt his private $x_i$ and transmit $E(x_i)$ to another. Next, that party calculates the function $g$ on encrypted data via homomorphism of the encryption method, and the party gets the encrypted value of function $f$. However, it is still a challenge to apply HE in real applications \cite{sen2013homomorphic} as HE requires either application modifications or specialized client-server applications to make it work.

\input{Tables/privacy-preserving-techniques}

Another possible solution to enhance data when making sensitive private data securely available is differential privacy. The definition of differential privacy is defined as follows: a randomized algorithm $\mathcal{A}: \mathcal{D} \rightarrow \mathcal{R}$ with domain $\mathcal{D}$ and range $\mathcal{R}$ is $(\epsilon, \delta)$ - differential privacy if for any two adjacent training dataset $D_1, D_2 \subseteq \mathcal{D}$, in which the data points in these two datasets are difference, and any subset of output $S\in \mathcal{R}$, satisfies the condition: 
\begin{equation}
    Pr[\mathcal{A}(D_1) \in S ] \leq e^\epsilon Pr[\mathcal{A}(D_2) \in S] + \delta,
\end{equation}
where $\epsilon$ and $\delta$ are called privacy budget and failure rate, respectively. A smaller $\epsilon$, a stronger privacy guarantee. 
However, DP also has some serious flaws \cite{domingo2021limits}. Some DP queries leak a small amount of data, hence if an attacker is able to repeat similar queries, the total loss could be catastrophic. 
A quantitative comparison of privacy-preserving techniques is described in Table \ref{tab:table-comparision}

To date, Internet users have given up their personal information and privacy in order to use services and applications in the Internet environment. On the one hand, the more information users supply, the better the service providers' experience. When providers understand their customers through shared data, they may customize and improve their services and applications to adapt to customers' demands. Customers, on the other hand, have a difficult time recalling and determining what data providers collect and use for what purposes. The TET was created to give users a clear picture of what and how their data are gathered, analyzed, and utilized to promote information transparency between users and services \cite{zimmermann2015categorization} without aiming at data minimization. With this mission, the TET is a complementary tool to PET to bring a comprehensive privacy-preserving solution. 

Integrating privacy protection solutions, i.e., PPTs, helps businesses and organizations solve the worry of adopting blockchain technology in data-sharing applications. This is because the risk of revealing personal and sensitive information is high when sharing it via transactions stored in a publicly accessible ledger, even if encryption algorithms and pseudonyms are used \cite{herrera2014research, 8888155}. Therefore, the combination of PPTs and blockchain technology is urgently needed to develop secure and privacy-preserving data-sharing applications.

%% file: Tables/privacy-preserving-techniques.tex
\begin{table}[!h]
\centering
\caption{Quantitative comparison of privacy-preserving data sharing strategies 
}
\begin{threeparttable}
\resizebox{\linewidth}{!}
{
\begin{tabular}{l c c c c c c} 

\toprule
\multicolumn{1}{c}{\diagbox{\textbf{PPT}}{\textbf{Attributes}}} & \begin{tabular}[c]{@{}c@{}}\textbf{Privacy}\end{tabular} & \textbf{Size~} & \textbf{Speed} & \textbf{Data Utility} & \textbf{Deployment}  \\ 
\midrule
\textit{Secure Multi-party Computing}    
&  $\star$$\star$$\star$$\star$$\star$     
& $\star$$\star$      
& $\star$$\star$    
& $\star$$\star$$\star$$\star$         
& $\star$ \\ 

\textit{Homomorphic Encryption}         
& $\star$$\star$$\star$$\star$$\star$     
& $\star$$\star$   
& $\star$$\star$    
& $\star$$\star$$\star$$\star$              
& $\star$  \\ 

\textit{Differential Privacy}           
& $\star$$\star$$\star$$\star$    
&  $\star$$\star$$\star$$\star$        
&  $\star$$\star$$\star$$\star$  
& $\star$$\star$$\star$               
& $\star$$\star$$\star$  \\ 
\textit{Synthetic Data}                 
& $\star$$\star$$\star$$\star$$\star$ 
&  $\star$$\star$$\star$       
& $\star$$\star$$\star$        
& $\star$$\star$$\star$              
& $\star$$\star$$\star$ \\ 

\textit{Federated Learning}             
& $\star$$\star$$\star$$\star$$\star$   
& $\star$$\star$$\star$       
& $\star$$\star$$\star$       
& $\star$          
& $\star$$\star$ \\ 
\bottomrule
\end{tabular}
}
\end{threeparttable}
\label{tab:table-comparision}
\end{table}

%% file: Sections/4-Convergence-Blockchain-Data-Sharing.tex
\section{Blockchain-based Data Sharing Solution}
\label{sec-BlockDaSh}

This section presents some approaches that take advantage of blockchain platforms to solve data-sharing problems. 

\subsection{Advantages of Integrating Blockchain with Data Sharing}

In the preceding Section~\ref{sec:limit_data_sharing}, we highlighted the research obstacles that data-sharing solutions face, such as centralization, security, and privacy concerns, and a lack of trust among parties. Blockchain can improve data sharing by enabling decentralization, increasing security and privacy levels, and providing incentive mechanisms. The primary benefits of integrating blockchain with data sharing are summarized here.

\emph{Decentralization of data-sharing systems}. Shared data and access to the data would be stored as transactions on distributed nodes for storage purposes. The blockchain platform is managed by participants rather than a central operator, ensuring that the entire system is not governed by a single entity.

\emph{Traceability of data-sharing systems}. Using a traceable and immutable distributed ledger, blockchain technology provides an audit trail for parties to retrieve information at various stages. In a supply chain, for example, all information on product provenance, manufacturing, distribution, and transportation until the product reaches customers would be shared and recorded in the blockchain managed by authorized stakeholders. They have the authority to inspect, test, and verify the product's dependability throughout the entire process.

\emph{Security of data-sharing systems}. Blockchain technology ensures the integrity and availability of data-sharing solutions. It decreases the risks of data modification and falsification without permission by eliminating the reliance on a single entity for data storage and management. However, to protect data confidentiality, data-sharing solutions must adopt permissioned blockchains, i.e., private or consortium ones, instead of public ones, or combine blockchains with cryptographic technologies, e.g., symmetric encryption and asymmetric encryption. Also, thanks to digital signature technology, and immutability and traceability properties, any malicious operations on the data in the blockchain network will be recognized, rejected, and traceable, which is not possible with other data-sharing systems. For example, it might appear out of control to identify and track malevolent clients who provide altered and deceptive learning parameters in FL-based decentralized data-sharing systems.

\emph{Privacy of data-sharing systems}. On the blockchain platform, an individual's identity is represented by a public address. In other words, the public address is a pseudonym, which is displayed in every transaction for the individual in possession of the accompanying public key. Although transactions made with this address could be traced back due to blockchain's transparency, blockchain technology adds a new layer of identity protection to protect the individual's identity rather than using the real identity like traditional approaches, i.e., using CSP's solutions. Moreover, because of cryptographic keys, users are capable of controlling who and how third parties access and use their data. 

\emph{Incentive and punishment mechanisms of data-sharing systems.} Integrating blockchain technology into data-sharing solutions offers both financial rewards and non-monetary incentives. In addition to making revenues by sharing quality data, individuals, businesses, and organizations can gain rewards by participating in consensus processes to validate and append blocks of transactions to blockchains. More importantly, with the reliability in terms of privacy and security that blockchain brings, it encourages parties to join the blockchain network to exchange their data with others, which promotes collaboration among them in vertical disciplines for building better user-oriented services and applications. Besides, malevolent parties, who try to corrupt the data-sharing system, will lose their deposit in the blockchain and be detected and banned from the system if seriously violate regulations.  
    
\begin{figure*}[t!]
	\centering
	\includegraphics[width=0.8\linewidth]{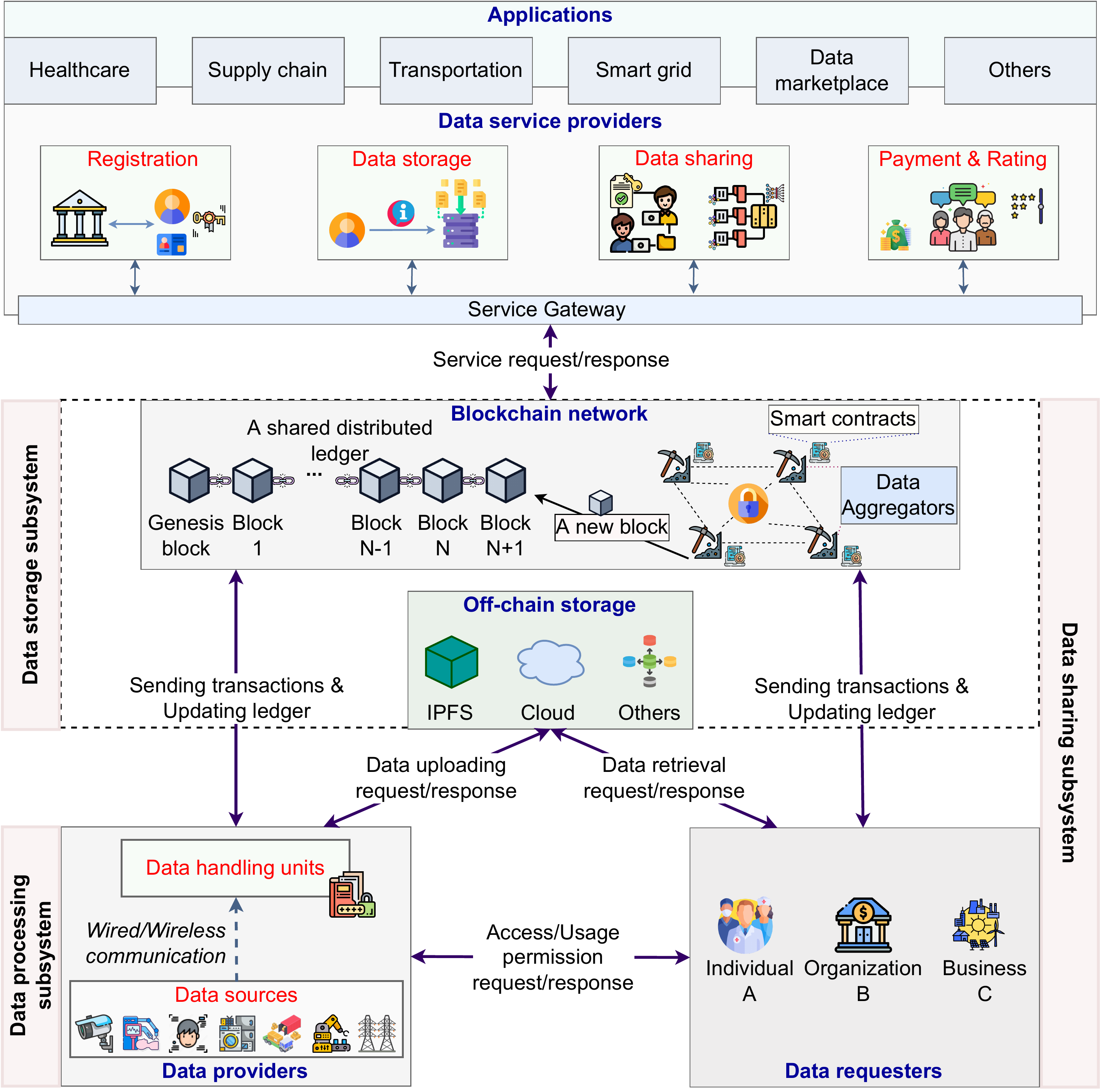}
	\caption{\emph{BlockDaSh} - our proposed blockchain-based data-sharing reference system architecture}
	\label{fig:BlockDaSh}
\end{figure*}

\subsection{Reference System Architecture}

In this section, we propose a permissioned (i.e., private or consortium) blockchain-based data-sharing system architecture called \emph{BlockDaSh}. There are several reasons that hinder integrating permissionless (i.e., public) blockchains with general data-sharing applications, especially the security and privacy paradox in the permissionless blockchain: 

\textit{Transparency.} In permissionless blockchain platforms, transactions are transparent and accessible to everyone. This leads to a situation in which malicious parties access, view, analyze, and use data (e.g., sensitive information from credit cards, and medical records) for illegal purposes without the data provider's consent. 

\textit{Immutability.} Also, because everyone is accessible to permissionless blockchain-based data-sharing systems, the risks of storing improper, sensitive, or illegal data are inevitable. Still, these data cannot be withdrawn and deleted due to the immutable characteristic \cite{ateniese2017redactable}. It takes up storage space, interferes with personal life, breaches intellectual rights, and violates users' rights to control their data. 

\textit{Traceability.} When there are hundreds, thousands, or millions of users in the blockchain platform, resulting in massive transactions, tracking information at milestones when disagreements arise becomes increasingly complex and difficult.


Therefore, we focus on utilizing permissioned blockchains in designing blockchain-based data-sharing architecture in this survey. The proposed \emph{BlockDaSh} consists of five main components: \textit{Data Providers} \textit{(DPs)}, \textit{Data Aggregators} (DAs), \textit{blockchain network}, \textit{ Data Requesters (DRs)} and \textit{Data Service Providers} (DSPs), and three subsystems: \textit{data processing},  \textit{data storage}, and \textit{data sharing}. The architecture is shown in {Figure~\ref{fig:BlockDaSh}}. The detail of the components is described below: 

\noindent\hangindent 1em\textit{DPs} are the owners of smart homes, smart grids, hospitals, or banks that gather, store, transform, integrate, and encrypt the huge amount of raw data generated from data sources (DSs) such as humans, services, transactions, appliances, manufacturing machines, IoT devices, and enterprise resource planning (ERP) systems by using their connected data handling units. Data handling units are devices or servers which have a stable network connection, medium storage capacity, and computational capability. Besides, DPs are willing to share their data ( i.e., their collected and processed data, or third parties' data under their consent) with other parties to monetize data or for public contributions. DPs acting as lightweight nodes are in charge of communicating and interacting with the blockchain network instead of using local resource-constrained devices to reduce storage and communication costs, and latency of the system. To protect data confidentiality, DPs always encrypt their data before uploading it to the outside.

\noindent\hangindent 1em\textit{DAs} play the role of full nodes, which maintain a copy of the entire ledger in the blockchain network and are responsible for fully validating transactions and blocks. Besides, DAs also assist lightweight clients in transmitting transactions to the blockchain network. DAs are pre-selected and distributed across the network and serve as edge computing nodes that manage a cluster of DPs \cite{wang2019novel, kang2018blockchain}. They have a stronger wireless connection and larger computing and storage capacities than DPs, and they provide services for DPs within their coverage. Especially in the second use case of the data-sharing subsystem, DAs also work as aggregators to aggregate local model updates sent from FL clients, i.e., DPs.  

\noindent\hangindent 1em\textit{Blockchain network} is maintained by a group of consensus nodes known as DAs that are pre-selected and ensure the data they hold is valid, secure, and accessible to authorized users. The indices of DPs' data returned from online repositories and any operations (e.g., access or usage) on shared data are recorded and stored on the blockchain. In the blockchain network, DPs and DRs play the role of lightweight nodes. Therefore, each DP and DR only downloads the block header instead of storing the whole blockchain and can interact with the blockchain network via full nodes. 


\noindent\hangindent 1em\textit{DRs} are individuals, businesses, or organizations who desire to access and use data shared by DPs for their own purposes. For example, to choose the best route to a location, drivers want to be informed of the weather and road conditions along particular paths. With shared information from other vehicles on different routes, drivers will make better decisions for their travel. Besides, DRs need to pay an amount of money or cryptocurrency for DPs to purchase the data they want, in addition to transaction fees. Furthermore, DRs are able to rate DPs based on the relevance and quality of shared data. 

\noindent\hangindent 1em\textit{DSPs} are data-sharing platforms built over the blockchain network and run by a single organization or groups of organizations to provide a secure place for data sharing between DPs and DRs. They offer a wide range of services for DPs and DRs to join, upload data, or request data in a variety of real-world applications such as healthcare, supply chain, or transportation. In addition, user registration, data storage, data sharing, and payment and reputation management services are exposed to DPs and DRs to make the entire data-sharing procedure more convenient and accessible to people who wish to exchange data in different application areas. Regarding storage, DSPs can use external online repositories, e.g., cloud servers or IPFS, which have almost infinite storage space for off-chain processing and storage, to reduce storage and networking burdens on the blockchain network.

\subsection{The BlockDaSh Framework}

As shown in {Figure~\ref{fig:BlockDaSh}}, our \emph{BlockDaSh} system comprises three subsystems, which are data processing, data storage, and data sharing, making trustworthy data sharing between DPs and DRs without relying on a central third party:

\subsubsection{Data processing}
The data processing subsystem provides two functions: data collection and data pre-processing. In the former function, DPs possess and collect raw data that is diverse, noisy, and heterogeneous by nature from a range of data sources (DSs). DSs transfer their raw data to the connected data handling units via wired and wireless communication. In the case of DPs directly sending the original data to the blockchain-based system infrastructure, it will be costly in terms of communication overhead and energy consumption throughout the transmission process. Besides, sending them still containing noise without pre-processing may affect the quality of the data, leading to low rewards. As a result, raw data must be pre-processed by using data handling units to clean noise (e.g., data duplication, incomplete data, sensitive data, or incorrect data), integrate data from multiple DSs, and transform them into meaningful information. For example, a hospital is capable of collecting patient data generated from EMRs uploaded by doctors and medical devices in various departments. The data are then cleaned, transformed, and classified into different collections based on the sickness, such as the data collection of lung scans of patients after contracting COVID-19 or based on each patient. At the time, because DPs own transformed data and find it valuable and useful, they are willing to share it to monetize it and assist each other, e.g., other hospitals.

\begin{figure}[t!]
	\centering
	\includegraphics[width=\linewidth]{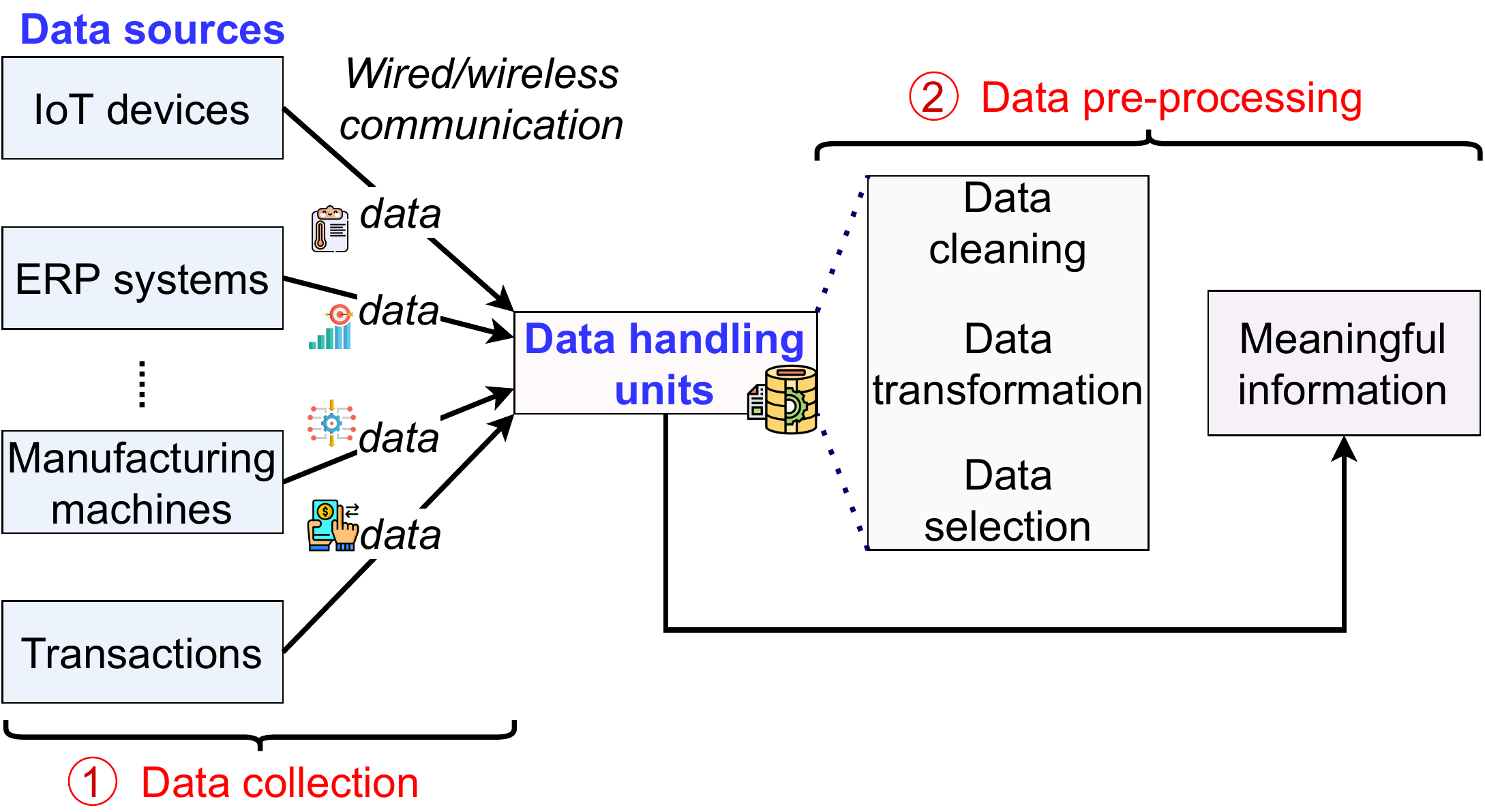}
	\caption{The flow chart of the data processing subsystem}
	\label{fig:DataProcessingSubsystem}
\end{figure}

\subsubsection{Data storage}
Shared data is an essential resource for offering more valuable insights, promoting collaboration, and developing services for different applications, so the confidentiality, availability, and integrity of the shared data are critical. For that reason, the storage of shared data for data discovery and retrieval must be taken into account to guarantee data security. Accordingly, a permissioned blockchain network is utilized to manage shared data to achieve decentralized, privacy-preserving, and secure storage and management. The processed data from DPs is safely stored using the permissioned blockchain network and the off-chain storage solution. The specific process of shared data storage is shown in {Figure~\ref{fig:DataStorageSubsystem}} and demonstrated through the following steps:

\begin{figure}[b!]
	\centering
	\includegraphics[width=\linewidth]{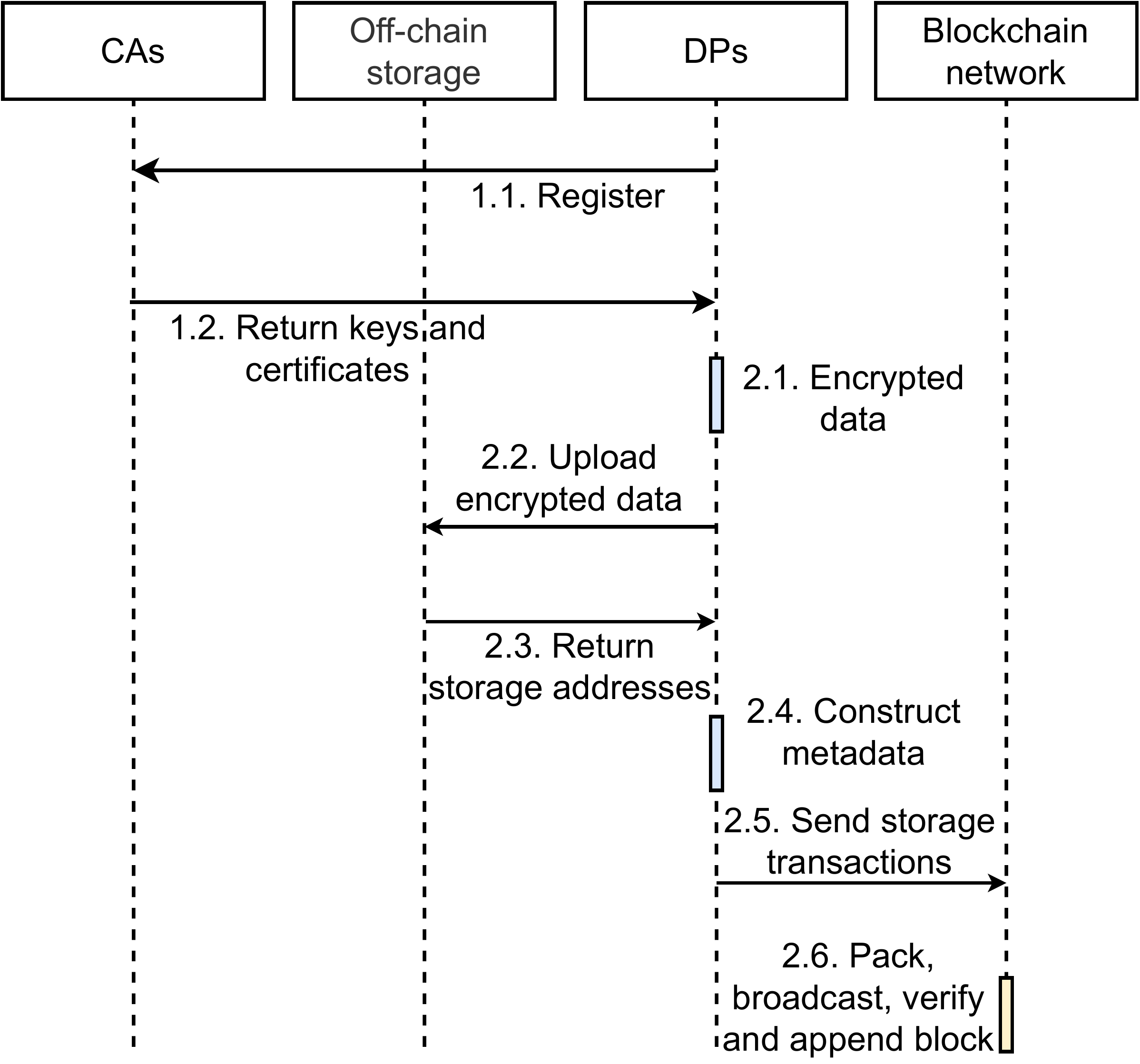}
	\caption{The workflow diagram of the data storage subsystem}
	\label{fig:DataStorageSubsystem}
\end{figure}

\noindent\hangindent 0.6em\textit{Step 1. (System initialization):} At first, DPs have to register with trusted certificate authorities (CAs), e.g., government departments of public health, to join the blockchain-based data-sharing system. When the authentication process is finished, registered DPs and DRs can join and connect to the blockchain network as trusted participants to exchange data. Each legitimate user holds a public key, a private key, and a corresponding digital certificate for authentication, authorization, and encryption tasks. After that, legitimate users download the latest block header from full nodes in the blockchain network to stay updated about the current state of the shared ledger.


\begin{figure*}[t!]
	\centering
	\includegraphics[width=0.85\linewidth]{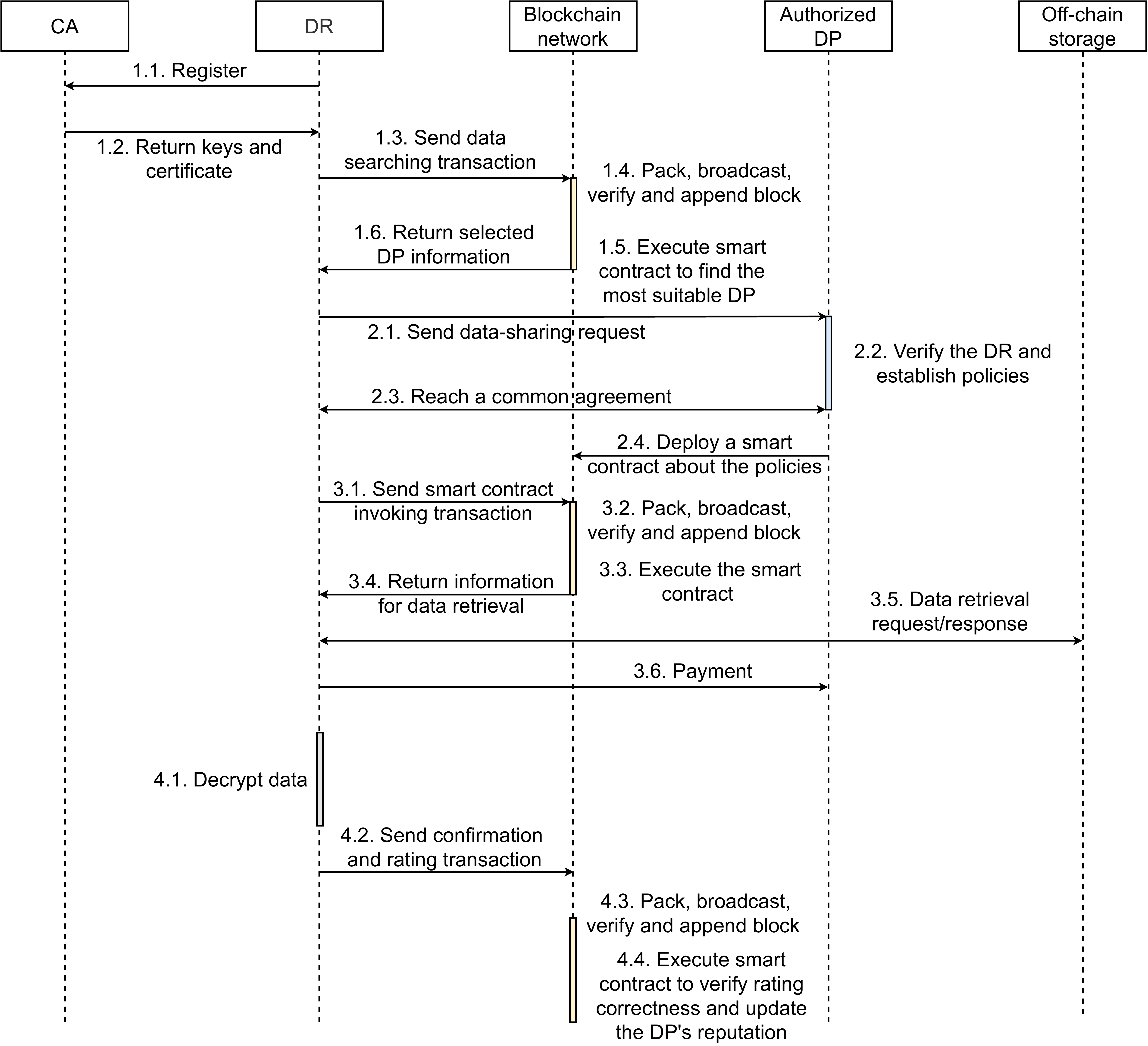}
	\caption{The workflow diagram of the data sharing subsystem}
	\label{fig:DataSharingSubsystem}
\end{figure*}

\noindent\hangindent 0.6em\textit{Step 2. (Data uploading):} As the volume of shared data in a wide range of applications is huge and may be sensitive, storing all data on the blockchain, which has limited space, is expensive and inefficient for a long period. Thus, storing metadata on-chain and shared data off-chain is adopted to achieve high scalability and security. Then, processed data will be encrypted by DPs before being sent to the off-chain storage source (i.e., IPFS or cloud servers) to improve scalability and protect the confidentiality of shared data. DPs extract searchable keywords and generate data descriptions from their local data. After the off-chain storage source returns a storage address for the ciphertext of shared data, a DP sends a transaction signed by its private key, in which the transaction contains information such as the timestamp of transaction creation and metadata, to the blockchain network for storage and sharing. The metadata may comprise information, e.g., shared data identification (i.e., hash digest), DPs identification, encrypted data storage address, DPs reputation, searchable keywords, and the data description.

\noindent\hangindent 0.6em\textit{Step 3. (Block generation and consensus process):} DAs gather transactions sent from DPs in their communication coverage. DAs compete with each other to create new blocks filled with valid transactions. The block production will generate a cryptographic hash link between the current block and its prior block to ensure authenticity, immutability, and traceability. Consensus algorithms are used to achieve common agreement and trustworthiness across DAs regarding the current state of the distributed ledger. The winner, i.e., a DA, of the consensus procedure will broadcast the new block to peer nodes or other DAs in the blockchain network for verification and audit. If other DAs verify the newly created block as valid, it will be added on-chain for storage.

\subsubsection{Data sharing} This is a core subsystem in the proposed \emph{BlockDaSh} system architecture where DPs and DRs can securely exchange data. Also, we provide a set of practical use cases to illustrate the workflow of the data-sharing subsystem.

\noindent \textbf{Use Case 1: Transferring Data among Parties.} \\
In the first use case, DPs and DRs can be matched based on DRs' requests, DPs' relevant data, and DPs' reputation, and then data, such as electronic and energy consumption information or traffic-related data, can be exchanged via smart contracts in the proposed \emph{BlockDaSh} architecture. We present the interactive process between DPs and DRs as follows:


\noindent\hangindent 0.6em\textit{Step 1. (Sending requests)} DRs having interests in exchanging data in the permissioned blockchain network need to pass the registration phase under the control of trusted CAs. DRs begin by sending data-searching transactions to the blockchain network for searching and discovering their data of interest based on metadata saved on-chain. When the blockchain finds suitable DPs who have the highest reputation and possess the most relevant data for DRs' needs, the blockchain network will return the result to DRs containing information for data retrieval. Now, DRs can submit data-sharing requests to DPs to request rights for accessing the shared data. 

\noindent\hangindent 0.6em\textit{Step 2. (Data sharing permission grant)} After a chosen DP receives a data-sharing request from a DR, it first verifies the identity of the DR via the DR's digital certificate to assure that the request comes from a trustworthy and legitimate user. After that, DP will establish payment, DAC, and data usage control (DUC) policies with the expiration time of these policies for their data. While DAC policies specify who can access what kind of data, DUC ones define how data are used for what purposes. When DP and DR reach a common agreement about DAC, DUC, and payment, DP will deploy a smart contract to permit transactions and agreements between them to be automatically carried out. When most consensus nodes verify the smart contract, it is irreversible, available, and accessible in all blockchain nodes. 

\noindent\hangindent 0.6em\textit{Step 3: (Data retrieval)} With the consent granted by a DP, a DR can submit a transaction containing the amount of cryptocurrency to deposit to invoke the deployed smart contract on the blockchain network. The DP's metadata (e.g., the DP's data storage address) and needed information for data retrieval will be sent back to the DR. Based on the data storage address of the data retrieved from the blockchain network, the DR sends a request to query data from the off-chain storage source. This process does not cost anything. When retrieval results are returned to the DR, it uses its cryptographic key to decrypt them, get the data it requires, and verify them to ensure the shared data has not been modified. At this time, DR can use the shared data for their work. 

\noindent\hangindent 0.6em\textit{Step 4. (Payment and rating management)} In the end, after a DR obtains shared data from a DP, it will pay the DP using cryptocurrencies. The DR will also initiate another transaction to inform peer nodes that shared data has been successfully obtained. Besides, if the DR finds the shared data from the DP to be useful, relevant, and proper to its requirements as well as useful for its tasks, it generates a rating transaction on the blockchain. The rating transaction will invoke a rating verification smart contract to verify the rating message to ensure its correctness. The high rating value would reflect the quality of shared data and affect the reputation of the DP. The reputation of the DP will be updated over the blockchain network. Besides, DRs who actively contribute accurate feedback will have a great chance to receive incentives for their contributions. 

\smallskip
\noindent \textbf{Use Case 2: Leveraging trustworthy data-sharing for collaborative learning.}

Blockchain-based FL refers to a decentralized approach for data sharing and analysis, where multiple parties can collaborate on developing ML models without the need to share their sensitive data with each other\cite{nguyen2021marketplace}. In this approach, the data remains local to the participating parties and is never transferred to a central location. 
Instead, a blockchain is used to store and verify the accuracy of the ML models. Each party contributes to the training of the model by computing local updates and sharing only a summary of these updates with the other parties via smart contracts. The use of blockchain technology provides transparency, immutability, and security to the machine learning process, ensuring that the shared models are trustworthy and can be audited at any time. This approach has the potential to enable data sharing in various domains, including healthcare, finance, and smart cities while preserving data privacy and security.

\noindent \textit{FL Concept}. Suppose that a DP or a client $i$-th has a private dataset $D_i$. In FL, this translates to finding an optimal global model parameter $w \in \mathbb{R}^d$ that minimizes the empirical risk on all of the  distributed training data samples\cite{pandey2022fedtoken}. The clients collaborate to solve the distributed optimization problem as follows:
\begin{equation}
    \underset{w \in \mathbb{R}^\textrm{d}}{\text{min}}F(w) := \frac{1}{N} \sum_{i=1}^{N} F_i(w)
\label{eq:objectivefunction}
\end{equation}
where $N$ is the number of clients (i.e., DPs). The objective function for the client $i$ is as follows:
\begin{equation}
    F_i(w) := L(w; D_i),
\end{equation}
where $L$ is a loss function \cite{chu2004bayesian}. We assume that the configuration ensures that if two FL clients have identical local datasets, then they have the same local models, e.g, $D_i = D_j$, so $F_i = F_j$. In order to solve the optimization problem of (\ref{eq:objectivefunction}), we can leverage the federated averaging (FedAvg) algorithm \cite{bonawitz2019towards}, which is widely used in FL. The FedAvg algorithm employs stochastic gradient descent (SGD)\cite{malinovskiy2020local} parallelly on a randomly sampled subset of clients and submits local model updates to a central server in each round. Specifically, $I=\{1,...,N\}$ is the set of FL clients in each training round $t$. We present the FedAvg as follows:

\begin{figure}[t!]
	\centering
	\includegraphics[width=0.95\linewidth]{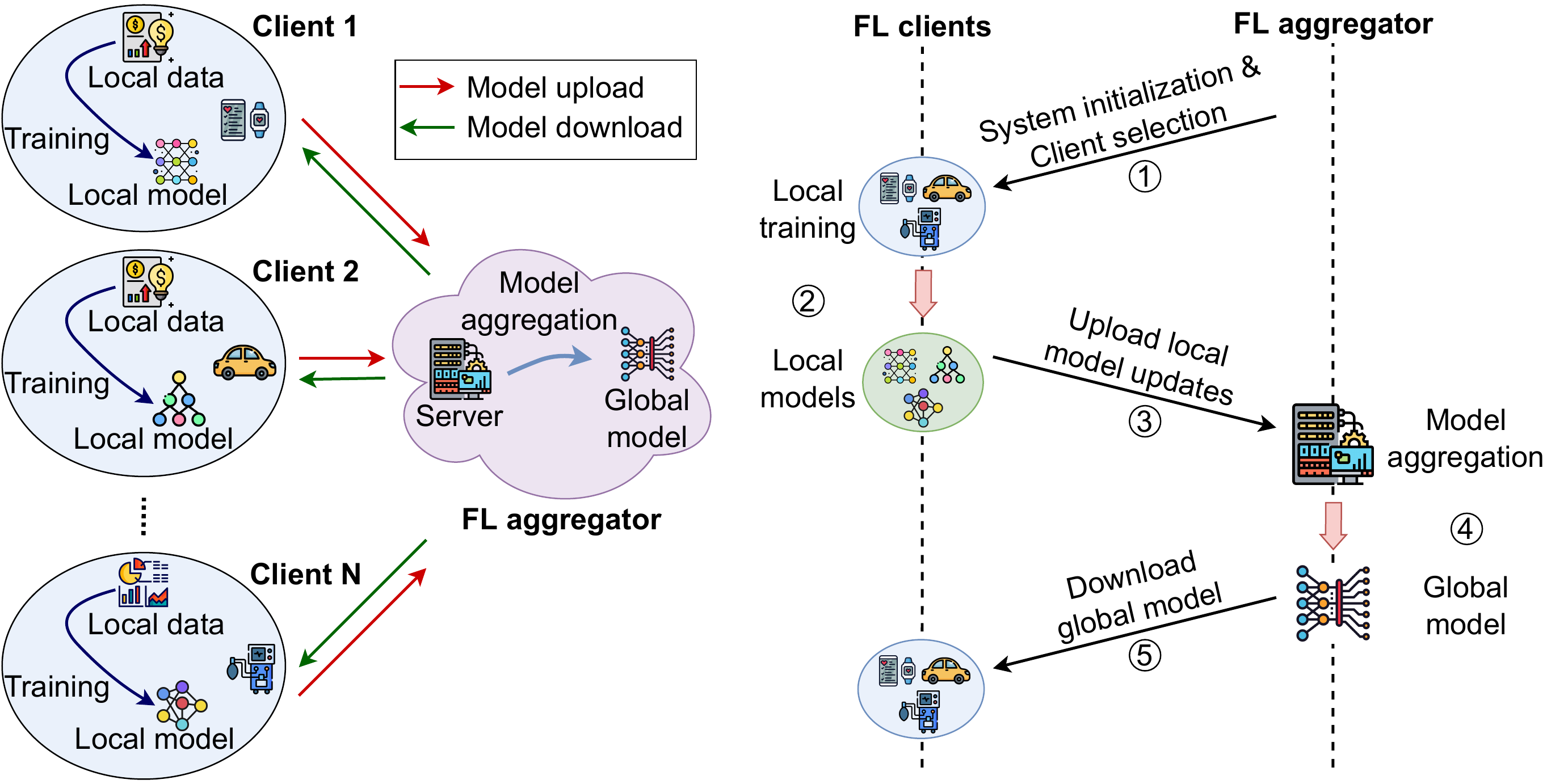}
	\caption{The standard FL architecture and workflow diagram \cite{lo2021flra}}
	\label{fig:StandardFL}
\end{figure}

\begin{enumerate}
    \item First, the centralized server broadcasts a last global model $w^t$ to all FL clients, say data providers. 
    \item Then, all data provider $i$ updates its local model by configuring  $w_i^t = w^t$ for all FL client $i$. 
    \begin{equation}
        w_i^{t_1} = w_i - \eta^t \nabla F_i(w_i^t), 
    \end{equation}
    where learning rate $n^t$ is used in round $t$-th. The default value of $n^t=0.1$
    \item Based on the client selection scheduling algorithm, a subset $I_t \subseteq I$ of FL clients is selected. The FedAvg algorithm randomly selects clients for local updates.
    \item Next, the central server aggregates the submitted local models to generate a new global model
    \begin{equation}
        w^{t+1} = \frac{1}{|I_t|} \sum_{i\in I_t} w^{t+1}
    \end{equation}
\end{enumerate}

The FL training process is iterated until the global loss function converges or achieves a desirable test evaluation metric, e.g., accuracy. The network architecture and communication process of standard FL are shown in Figure~\ref{fig:StandardFL}.

 \begin{figure}[t!]
	\centering
	\includegraphics[width=0.9\linewidth]{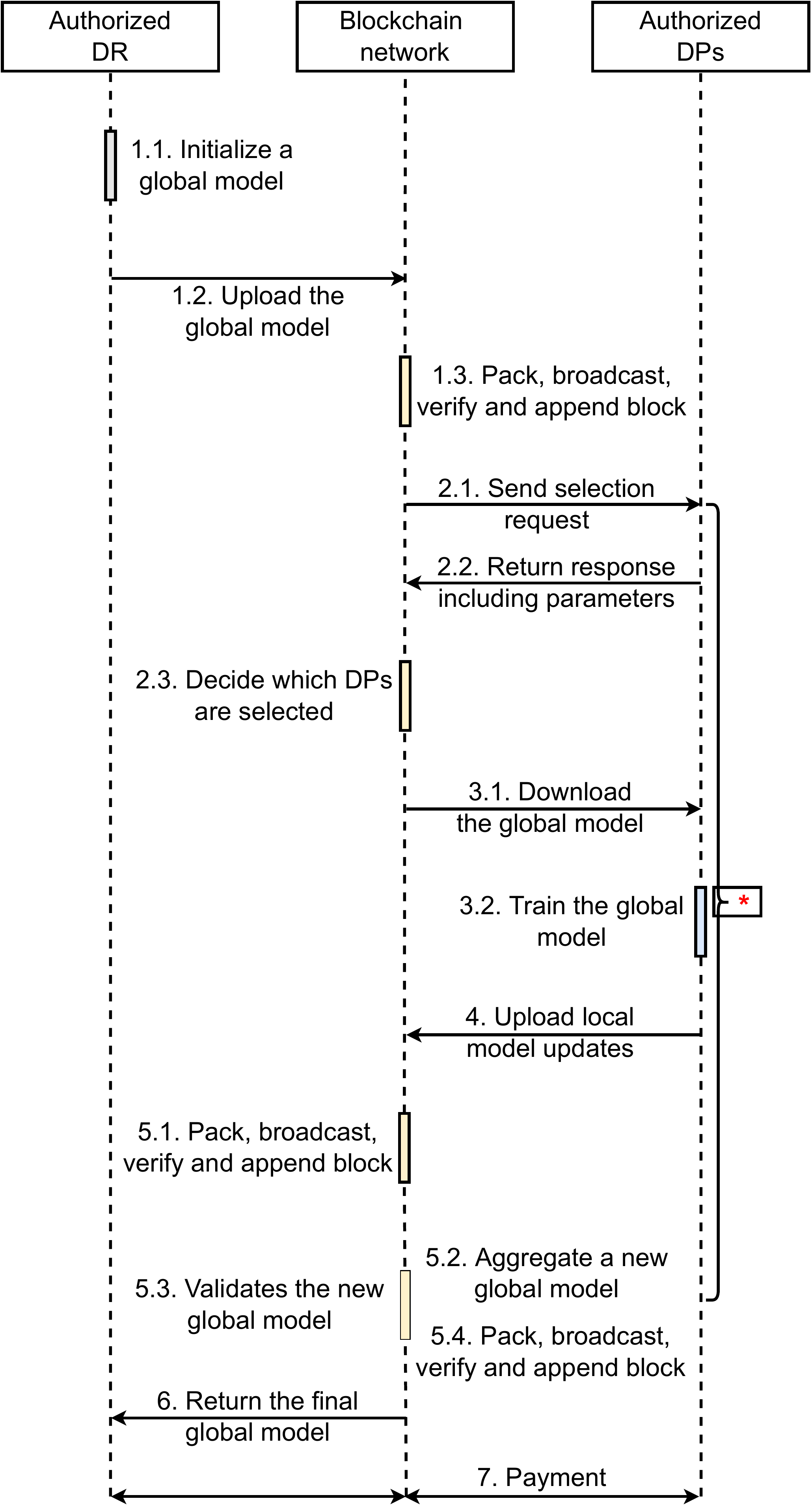}
 {\footnotesize \newline \emph{{\color{red} {*}} These steps will be iterated in multiple rounds if pre-defined criteria of the global model are not met.}\par}
	\caption{The workflow diagram of the trustworthy data sharing for collaborative learning}
	\label{fig:TrustedFL}
\end{figure}

\textit{Workflow}. This use case is different from the first one because it requires DPs to contribute not only their data but also their computational resources in multiple communication rounds. While the first service provides a one-by-one match between a DP and a DR for trading data, e.g., electronic medical records (EMRs), this service allows DRs, with the support of smart contracts, to train effective ML models that are hard for a single individual, organization, or business to achieve due to the shortage of data in terms of the amount, quality, and diversity. It is important to note that DRs can also buy pre-trained ML models if available from the first service, but they want to train their models from scratch in this case. Therefore, as shown in Figure~\ref{fig:TrustedFL}, a DR will be able to get its well-trained ML model via \emph{BlockDaSh} in the following steps: 

\noindent\hangindent 0.6em\textit{Step 1. (Initializing and Publishing global model):} At the early stage, the DR initializes a naive ML model (or global model) and publishes it to the blockchain following  the transaction format. Then, the transaction consisting of the global model is validated, verified, and appended to the blockchain.

\noindent\hangindent 0.6em\textit{Step 2. (Selecting DPs):} DAs playing the role of model aggregators send requests to random DPs to ask for their data size relating to the current training task, available computational resources, and communication channel states \cite{8761315}. Then, based on DPs' responses, DAs estimate the time required for downloading/uploading local model updates and the training time of each DP. As a result, DAs will have solid evidence to decide which clients are selected for the training task and ensure efficient utilization of DPs' resources.  

\noindent\hangindent 0.6em\textit{Step 3. (Downloading and Training global model):} The global ML model on the blockchain is downloaded by chosen DPs to train with their data with the goal of minimizing the loss function or optimizing the evaluation metric, e.g., accuracy or mean squared error. Also, the data of DPs will be kept locally and not exposed to the outside. 

\noindent\hangindent 0.6em\textit{Step 4. (Uploading local model updates):} Selected clients to upload transactions of their local model updates on-chain. 

\noindent\hangindent 0.6em\textit{Step 5. (Cross-verification and aggregation of local models):} These transactions are then broadcast, validated, put into blocks, and verified by other DAs on-chain. In the meantime, each DA validates the quality of the received local model updates by treating their data as a validation dataset. Only local model updates meeting pre-defined criteria of performance, e.g., a prediction accuracy of above 95\%, given by DR, are accepted as valid transactions and kept for aggregation. It is noteworthy that the local model update validation process is integrated into consensus protocols to address model/data poisoning attacks and save computational resources for DAs. After that, validated local model updates are aggregated into a new global model by DAs. The new global model will be published, propagated, validated, verified, and appended to the blockchain. Until the global model achieves a desirable evaluation metric value, all steps are iterated except for the first step \emph{Initializing and Publishing global model}.

\noindent\hangindent 0.6em\textit{Step 6. (Settlement):} When the global model reaches a certain level of performance, and the block containing the global model is included in the blockchain, a smart contract is triggered to settle the deal between DPs and the DR. At this time, DR is able to download the desired ML model from the blockchain, and rewards will be given to DPs.

\noindent\hangindent 0.6em\textit{Step 7. (Payment):} Finally, based on the efforts in training the global model and improving its performance, the smart contract calculates appropriate incentives to reward DPs.

\input{Tables/Section_4_Convergence_BC_Data_Sharing/BC_platforms.tex}

\subsection{Tutorials for Deploying of BlockDaSh}

Generally, integrating blockchain with data-sharing solutions is promising when blockchain technology brings trustworthiness and significant benefits to users. While users are able to monetize their data in a secure and privacy-preserving manner, they also benefit from enhanced services and applications that are customized, personalized, and upgraded using their sharing data. To realize such blockchain-based data-sharing systems, designing a detailed plan, choosing suitable technologies, and building a practical architecture are mandatory. In the following part, we provide recommendations for successfully deploying \emph{BlockDaSh} in a wide range of real-world applications in terms of designing blockchain networks.  Regarding the data storage system, we can refer to the storage part in \emph{the BlockDaSh framework section}. 

\subsubsection{Objectives}
It is crucial to first identify the problem, its scope, intended outcomes, and potential technologies before choosing to implement blockchain technology and the \emph{BlockDaSh} reference system architecture. Organizations can refer to a list of assessment criteria published by PwC\footnote{\url{https://www.pwc.com/}} to ensure that the final decision is the best fit for their project. The first questions to consider are the number of parties that need access, viewing, and retrieval of shared data and whether these actions must be recorded. Verification requirements for these actions among stakeholders must also be considered. Organizations must also assess costs and complexity when deciding to adopt \emph{BlockDaSh} in their data-sharing solutions, compared to traditional approaches such as relying on central providers like CSPs. The delay in interactions among stakeholders is crucial as it directly affects the quality of business services, so organizations must determine whether interactions need to be time-sensitive. Finally, it is essential to consider if transactions interact with each other in the data-sharing model.

\subsubsection{Design the Blockchain network}
\emph{BlockDaSh} is utilized as the reference system model for a blockchain-based data-sharing solution. The next step is to determine a suitable permissioned blockchain platform for storage and management. A wide range of permissioned blockchain platforms are being developed and offer distinct technical properties targeted at different use cases. The authors in \cite{nguyen2022analysis,nguyen2020trusted} comprehensively analyzed other distributed ledger technologies, e.g., Hyperledger Fabric\cite{androulaki2018hyperledger}, Ethereum, IoTA\cite{pervez2018comparative}, Solana\cite{yakovenko2018solana}, and Quorum\cite{baliga2018performance}, applied in industrial IoT applications. 

Selecting an appropriate Blockchain platform could be based on several factors, e.g., scalability, latency, throughput, security, and smart contract functionality. These factors are crucial for data-sharing applications because large volumes of data from DPs generated  millions of transactions daily~\cite{hampton2013big}, requiring a high-efficiency consensus mechanism. Latency is also essential, and confirmation times for Bitcoin and Ethereum may not be suitable for real-time monitoring. Transaction fees are another critical factor to consider, as they can significantly increase operational costs and negatively impact throughput \cite{kuzlu2019performance}. Ethereum requires fees and gas for each transaction, while Hyperledger Fabric and IOTA offer free solutions. Meanwhile, Solana is a high-performance, decentralized blockchain platform designed for building decentralized applications and facilitating fast, secure transactions. Solana is designed to scale linearly as the network grows. It can currently process over 65,000 transactions per second (TPS), making it one of the fastest blockchain platforms in existence.

Another critical factor for a secured data-sharing platform is support for permissioned and permissionless nodes. Both Ethereum and Hyperledger Fabric support public and private solutions, while Bitcoin and IOTA only provide public ones. Public networks may be more secure because the data is encrypted, verified, and stored on all devices, making it transparent. However, permissionless blockchains are not ideal for enterprise use, where companies deal with sensitive data and cannot allow anyone to join their network. Permissioned blockchains can be altered by their owners, making them more vulnerable to hacking, but they provide lower fees for validation and a faster consensus process~\cite{valenta2017comparison}.

Finally, smart contract technology is considered as the key innovation in the blockchain era. Smart contracts act as autonomous entities on the ledger and execute logic expressed as functions of the data. Smart contracts can automatically react to specific events, such as real-time policy enforcement for sharing data. Solana, Ethereum, and Hyperledger Fabric support smart contracts, while IOTA has a smart contract type called Quobic. Hyperledger Fabric also provides various features such as identity management, transaction integrity, and authorization with a trusted CA. These features are essential in a trusted IoT system. The comparison of DLTs in these areas is illustrated in Table~\ref{tab:BCplatforms}. 

%% file: Tables/Section_4_Convergence_BC_Data_Sharing/BC_platforms.tex
\begin{table*}[t!]
\small
\caption{Comparison permissioned blockchain platforms}
\renewcommand{\arraystretch}{1.3}
\centering
\small
\label{tab:BCplatforms}
\begin{tabular}
{P{0.8cm} 
 P{0.08\linewidth }
 P{1.8cm}
 P{0.08\linewidth}
 P{0.1\linewidth} 
 P{0.12\linewidth}
 P{0.14\linewidth}
 P{0.06\linewidth}
 P{0.05\linewidth}}
\toprule
\multicolumn{1}{c}{\textbf{Ref.}} &
  \multicolumn{1}{c}{\textbf{Platform}} &
  \multicolumn{1}{c}{\textbf{Governance}} &
  \multicolumn{1}{c}{\makecell{\textbf{Customer} \\ \textbf{target}}} &
  \multicolumn{1}{c}{\makecell{\textbf{Consensus} \\ \textbf{protocol}}} &
  \multicolumn{1}{c}{\makecell{\textbf{Smart} \\ \textbf{contract} \\ \textbf{language}}} &
  \multicolumn{1}{c}{\textbf{Performance}} &
  \multicolumn{1}{c}{\makecell{\textbf{Smart} \\ \textbf{contract}}} & 
  \multicolumn{1}{c}{\makecell{\textbf{Open} \\ \textbf{source}}}
\\ \midrule

\cite{wood2014ethereum} & Enterprise Ethereum & {Enterprise Ethereum Alliance} & Enterprise & PoW, PoS & Solidity, Python, Go, C++ & {-} & \checkmark & \checkmark\\\hline

\cite{androulaki2018hyperledger} & Hyperledger Fabric & Linux Foundation & Enterprise & Pluggable consensus & JavaScript, Go, Java & {Strong throughput and latency} & \checkmark & \checkmark\  \\ \hline

\cite{quorum} & Quorum & ConsenSys & Financial Enterprise & Raft, Istanbul BFT & Solidity, Vyper & {Strong throughput, poor latency} & \checkmark & \checkmark \\ \hline

\cite{corda} & Corda & R3 Consortium & Financial Enterprise & Pluggable consensus & Java, Kotlin & {Poor throughput, strong latency} & \checkmark & \checkmark\\ \hline

\cite{multichain} & Multichain & Coin Sciences & Enterprise & Distributed consensus & Javascript & {Strong throughput, decent latency} & \checkmark & \checkmark\\ \hline

\cite{hydrachain} & Hydrachain  & Brainbot Technologies & Enterprises & PBFT & Python & {-} & \checkmark & \checkmark \\ \hline

\cite{neo} & NEO & NEO Foundation & Enterprise & dBFT & Python, Java, Go, C\#& {-} & \checkmark & \checkmark\\ \hline

\end{tabular}

\end{table*}

%% file: Sections/5-Applications-Blockchain-Data-Sharing.tex
\section{Applications of Blockchain-based Data Sharing}
\label{sec-taxonomyApp}
With the rapid advancement and innovation driving today's society, blockchain is a technology that possesses all of the properties required to assist data-sharing solutions in overcoming present obstacles and progressing to the next level. In this section, we review existing blockchain-based data-sharing applications in a wide range of fields ranging from healthcare, supply chain, transportation, smart grid, and data marketplace to a bunch of industrial applications. 

\subsection{Healthcare}

Good health is required to completely enjoy life. The significance of it is far reinforced, especially now that the entire world has confronted the health repercussions of the COVID-19 pandemic \cite{fauci2020covid}. However, there are numerous hurdles that relate to authenticity, security, and privacy concerns for patients' medical data and negatively affect medical staff in the process of analyzing and diagnosing patients' health in the traditional healthcare domain \cite{gordon2018blockchain}. For example, patients' EMRs, which record the whole process of their medical treatment, are very sensitive, scattered, and not synchronized among various medical institutions. Moreover, patients also do not have permission to access their EMRs because they are privately stored in the local databases of these institutions. Fortunately, blockchain technology has emerged and solved the above-mentioned problems by providing distributed and immutable storage to facilitate the sharing of medical data in a secure and privacy-preserving way among healthcare organizations while patients keep control over their data. The relevant study on blockchain-based medical data-sharing solutions will be summarized in the following.

In \cite{zhang2018towards}, a scheme named BSPP was presented to improve diagnosis in e-Health systems. Authors used both private and consortium blockchains for storing and sharing EHRs in a secure and searchable way. While the private blockchain is used to store encrypted data, the consortium maintains records of EHR indexes. A keyword search protocol is introduced to allow authorized doctors to search historical EHRs while precluding them from searching for future records in order to improve data retrieval for treatment. Furthermore, the authors defined a Proof-of-Conformance consensus algorithm for broadcasting, validating, and appending new blocks to the chains. The proposed scheme was then implemented on JUICE platform for evaluation and achieved considerable results in terms of computation, storage, and time.

In \cite{chen2022fine}, Chen et al. proposed BFHS, a framework that provides a comprehensive DAC and data storage solution for electronic health record (EHR) sharing. To store EHRs in a decentralized and secure manner, and effectively retrieve them, BFHS used IPFS technology in the company of the consortium blockchain. In particular, while IPFS keeps encrypted patient's EHRs and returns indices of them, the consortium blockchain stores ciphertext of indices that are encrypted by patients using ciphertext-policy attribute-based encryption (CP-ABE) \cite{bethencourt2007ciphertext}. Furthermore, the smart contract was created to manage access control policies for patients, individuals, and organizations who desire to own EHRs. It also supports credit calculation for consensus nodes. The higher a node's reputation, the more likely it is to be picked to participate in the consensus process.

In \cite{7990130}, authors presented a blockchain-based system, named MeDShare, to reduce risks to patients' EMRs related to privacy issues when sharing them among CSPs. MeDShare, with the help of smart contracts, enabled capabilities like data traceability, data audit, and DAC over medical records during data-sharing processes. The system was designed with four layers: user, data query, data structuring and provenance, and existing database infrastructure. Smart contracts and an access control mechanism were utilized to trace actions performed on requested data to ensure that the data was used properly. If detecting any misbehavior on shared data, smart contracts revoke the access permission of the requester. Furthermore, these actions on the request and delivery of data would be stored in an immutable blockchain network maintained by consensus nodes. The authors evaluated and discussed the system performance by employing scenarios of two different threat levels (data and report swap). 

The work in \cite{ZOU2021102604} aimed to improve the efficiency and privacy of patient EMR sharing and retrieval in eHealth systems. The authors developed SPChain, a public or permissionless blockchain system that uses chameleon hash functions (CHFs) \cite{krawczyk1998chameleon} to create new kinds of block structure (keyblocks and microblocks) to securely store EMRs, as well as proxy re-encryption (PRE) to ensure data privacy while exchanging EMR. Furthermore, to incentivize medical institutions to participate in the data-sharing and mining process, a reputation-based consensus protocol was designed. To demonstrate the effectiveness and scalability of the proposed system, performance metrics such as storage cost, throughput, and time cost were assessed in a real-world implementation. Besides, the proposal was also resilient to different types of attacks. 

\begin{figure}[t!]
	\centering
	\includegraphics[width=0.9\linewidth]{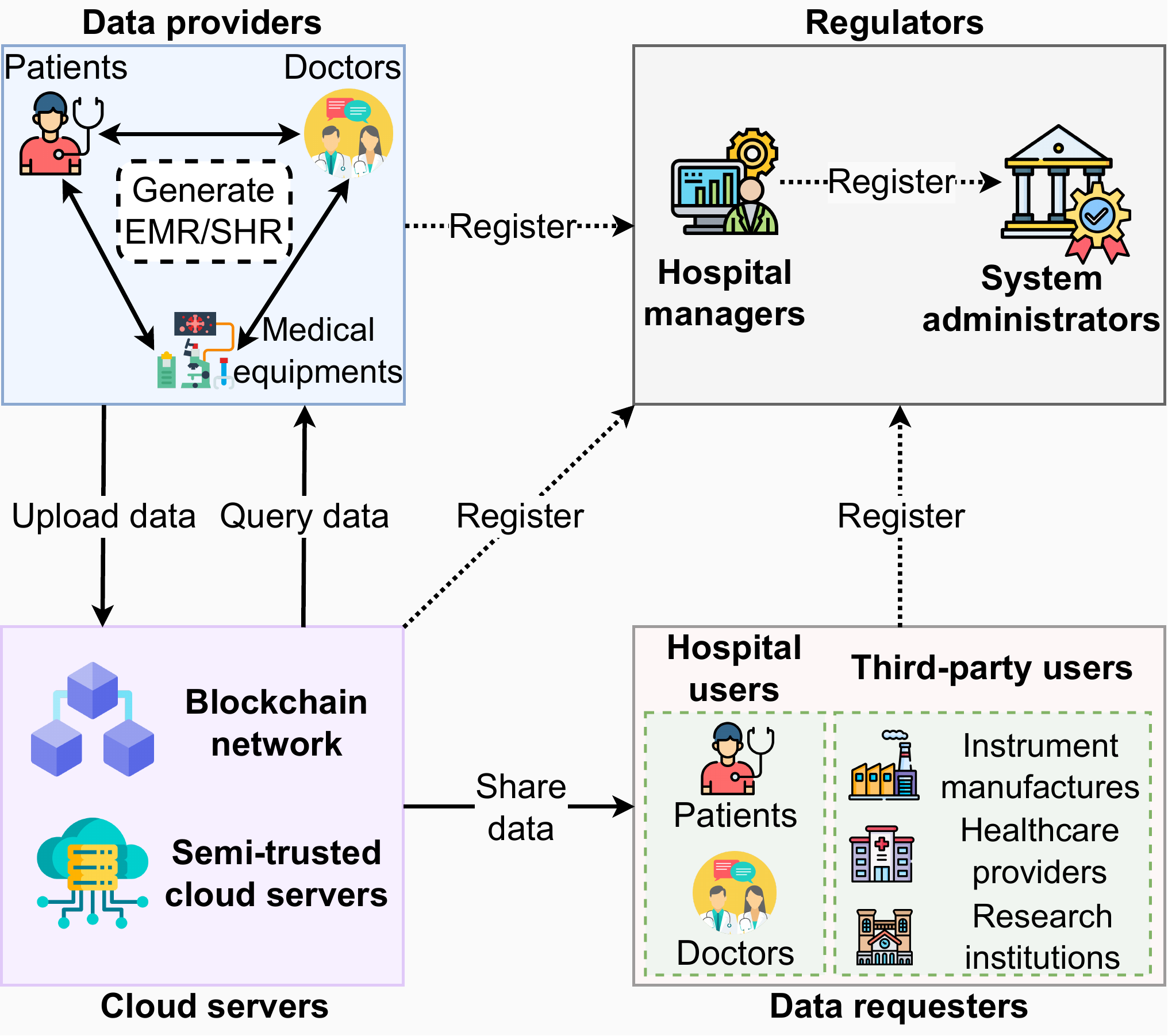}
	\caption{BC-based data-sharing system in healthcare \cite{CHEN2021338}}
	\label{fig:BC-healthcare}
\end{figure}

The authors in  \cite{fan2018medblock} proposed a blockchain-based system to share data between patients and hospitals with efficient EMR retrieval and access control mechanisms. The proposed system, named MeDBlock, uses local databases to store encrypted EMRs of patients among different hospitals, while their hash values are stored on the blockchain network. Because searching a historical patient's EMR over the entire ledgers across hospitals is inefficient and takes time, authors used a breadcrumbs mechanism that caches the hash of department-based patient-related blocks in hospitals per query for faster data retrieval. Moreover, for unauthorized users, information about signatures and encrypted summaries of EMRs is hidden before their identities are authenticated to ensure privacy for sensitive data.  

The authors in  \cite{DAGHER2018283} proposed a blockchain-based framework called Ancile to balance privacy and the necessity for access to EHRs. Ancile used smart contracts to manage access permissions for individuals who needed to access data for a specific reason. Furthermore, to improve data privacy, a distributed PRE method with blinding is given for data sharing between parties. This approach assures that messages and cryptographic key information cannot be fully decrypted if any hazards exist. To reduce storage costs, only transaction hashes were stored on the blockchain, while query link information was transmitted through the Hypertext Transfer Protocol Secure protocol. 

\input{Tables/1_Taxonomy_of_applications.tex}

Zhang et al. \cite{9831062} contributed a hierarchical access control mechanism for those who need to perform necessary operations on malicious content without affecting block hash values. Attribute authorities, DPs, data modifiers, and a blockchain network are the four components of the proposed system. The authors employed a decentralized attribute-based encryption (ABE) method with hierarchical authorization to avoid the single-point-of-failure problem and limit the power of modifiers. Furthermore, when employing redactable blockchain technology with the CHF, the hash value of a block is not changed if the transactions within it are amended. If no updating operations are required, the traditional hash function will be performed instead of the Chameleon one. 

In \cite{CHEN2021338}, authors focused on building a blockchain-based EMR/surgery health record (SHR) sharing system using IoT devices, Hyperledger Fabric, cloud servers, and the PRE algorithm. The proposed system architecture includes five layers: the system management layer, the data collection layer, the blockchain network layer, the cloud service layer, and the application layer. Thanks to doctors and medical IoT devices, real-time EMR/SHR data are collected and gathered for processing and storage. Regarding storage, encrypted EMR/SHR are kept on cloud servers, and the indexes with data usage records are stored on-chain. The authors proposed a PRE-based data sharing scheme to ensure secure data sharing on semi-trusted cloud servers, guaranteeing that those cloud servers cannot obtain any raw data and cryptographic keys. The performance has been discussed, and achieved significant results in terms of throughput and attack resistance. 

MedShare \cite{9547814} is a privacy-preserving consortium blockchain-based EHR sharing system that provides fine-grained access management and efficient data retrieval. MedShare's architecture is comprised of four components: health centers, users with different EHR access permissions, a blockchain system, and attribute authority. Encrypted EHRs were stored locally at healthcare institutions in this study, whereas encrypted searchable indices based on access permissions for distinct roles were put on the blockchain network. Specifically, the authors developed an on-chain constant-size ABE-based access control method to embed access control policies into search results without incurring high storage costs, as well as a multi-keyword conjunctive boolean search approach for efficiently performing an on-chain search.

The authors of \cite{9442539} tackled the problem of cross-hospital diagnosis by developing an EHR-sharing system called CPDS. CSPs and private hospital blockchains were used to build CPDS storage. To cut storage costs on the blockchain while also complying with data protection rules, researchers employed the former to store original EHRs and the latter just EHR indices. It is worth mentioning that each hospital owns a private blockchain, thereby creating a huge storage network, including blockchains between hospitals. CPDS leverages Wanchain technology for rapid EHR transfer among hospitals, resulting in a robust solution for scalability and variety for further development. With this solution, patients have the ability to delete EHRs when they desire. Finally, the security level of the proposal against multiple adversary attacks was discussed and analyzed. 

\smallskip
\noindent\textbf{Lessons learned.} Blockchain technology promotes EMR sharing among patients, hospitals, and third parties for better diagnosis, detailed treatment plans, and efficient data management.
\begin{itemize}
    \item DPs must encrypt their EMRs before uploading them to storage systems due to security and privacy concerns. Because the number of EMRs is massive and increases exponentially day by day, it is inefficient and costly to store all raw EMRs on blockchain network. Therefore, only hash values of EMRs are kept on the blockchain, while encrypted EMRs are stored on local hospital servers, cloud servers, or distributed storage systems such as IPFS and Swarm. 

    \item In blockchain-based healthcare applications, EMR/EHR access control also needs to be carefully investigated and developed because it allows DPs to monitor and govern actions performed on their shared data. By using smart contracts and cryptographic techniques (e.g., ABE), DPs are able to flexibly set access control policies to prevent EMRs from being accessible to unauthorized entities.

    \item Another issue that has arisen in blockchain-based healthcare applications is the retrieval or querying of historical EMRs. Although a patient's previous EMRs may be scattered across multiple medical institutions, applications must ensure that DRs, in general, can easily query them to provide quick and accurate diagnoses for a specific purpose. Because searching the entire ledger is inefficient, techniques and mechanisms (e.g., Bread crumbs \cite{fan2018medblock}) might be used to speed up data accessing and querying.

    \item  The majority of current blockchain-based healthcare applications have not considered using PPTs to protect the sensitive data privacy of patients. If DRs are nurses and doctors who are in charge of the patient's medical examination, data sharing without PPTs may be acceptable because they still need to see explainable raw data (e.g., X-ray images) for diagnosis. Regarding third-party requesters, not all patients want their raw EMRs leaked, viewed, and monetized. As a result, there is a need to combine PPTs and blockchain in order to build a secure and privacy-preserving blockchain-based healthcare application like \emph{BlockDaSh}. Especially, when the number of intelligent AI-based systems having large impacts on primary care and requiring a ton of medical data for training models increases at a rapid pace \cite{buch2018artificial, malik2019overview}, \emph{BlockDaSh} is able to support them in reducing the risks of patients' data leakage thanks to the trusted data sharing for collaborative learning service.  
\end{itemize}

\subsection{Supply Chain}

A supply chain is a series of processes in which multiple entities, suppliers, manufacturers, carriers, retailers, and customers take part in producing and distributing goods. Besides, SCM is a set of approaches to manage the entire supply chain to realize high performance with lower risks \cite{wieland2021dancing}. To achieve the smooth functioning of SCM, it requires active collaboration, engagement, and faithful commitment among entities in the supply chain. However, due to a lack of transparency, trust, and a strict binding force between supply chain members, existing SCM systems have faced various problems in product availability, product quality, product monitoring, product traceability and information synchronization. These flaws can be fixed with the solution of data sharing among entities by adopting blockchain technology. Blockchain technology provides supply chains with a decentralized, secure data repository to store and exchange information. Malicious entities and attackers cannot falsify cargo information during the process of goods' circulation. Therefore, information about products or business processes is transparent, traceable and symmetric for members of the supply chain, making SCM systems more effective and manageable. In the subsection, the related research on blockchain-based data-sharing solutions for supply chains will be summarized. 

The authors in \cite{agrawal2021blockchain} centered on traceability to address IA and poor visibility issues in the textile and clothing supply chain. All transactions were securely recorded and stored using blockchain technology, guaranteeing traceability. The proposed private blockchain-based framework was designed by the authors on two levels: organization and operation. While the organizational layer specified access control configurations between the blockchain network and partners, the operational layer described how smart contracts and transaction rules in supply chain traceability applications work. Afterward, an example of a supply chain for organic products was provided to demonstrate the proposed framework's data-sharing and traceability capabilities.

In \cite{8780161}, Wen et al. solved the IA by proposing a blockchain-based data-sharing scheme using Industrial IoT (IIoT) technology, the ABE method, and smart contracts. IIoT devices are in charge of collecting, gathering, and transmitting product-related data in real-time to the blockchain network. Only parties having attributes that match the access policies posed by DPs could execute transactions using ABE and smart contracts, eliminating the need for an intermediary party. With the help of IIoT and blockchain, data provenance, data integrity, and data transparency in SCM are guaranteed, thereby realizing information symmetry among parties. 

The authors of \cite{8010716} suggested a blockchain-based information-sharing system to handle the capacity risk problem in SCM caused by IA and demand forecasting capacities. To safeguard data privacy, HE technology was used, which allows computational processes to be conducted on encrypted data without disclosing any original information. The authors provided a Bitcoin-based modified block for sharing of "one data," which includes different sorts of transactions between suppliers, manufacturers, and retailers.

Dwivedi et al. \cite{DWIVEDI2020102554} used blockchain technology to develop a pharmaceutical SCM system that tracked data and ensured data integrity, participant authentication, and secure sharing. The suggested blockchain-based architecture has two layers: physical (end users, pharmacists, retailers, warehouses, manufacturers, and raw material suppliers) and consensus. With the help of CA servers and state machine model-based smart contracts, cryptographic keys were securely handed to organizations that then conducted actions based on converting states. Furthermore, in the presented consensus mechanism, a leader-validator node validates and creates transaction blocks. Performance metrics such as communication, calculation, execution, and storage costs have been analyzed and measured.

In \cite{casino2021blockchain}, authors proposed a decentralized blockchain-based information-sharing architecture for traceability in the food supply chain (FSC). Three smart contracts have been deployed to enable the sharing of information and the traceability of products and raw materials, stakeholders, and processes. All hashes of the product's data were stored in IPFS storage. To demonstrate the applicability of the suggested framework, a real-world situation involving dairy products was implemented and discussed. 

Authors employed IoT and blockchain technology to safeguard food safety by presenting a reliable and scalable real-time FSC traceability and sharing solution between stakeholders in \cite{7996119}. To manage and prevent hazards in FSCs, the Hazard Analysis and Critical Control Points (HACCP) approach was used in conjunction with a food chain model that included five processes: manufacturing, processing, warehousing, distribution, and retail. Furthermore, the suggested system used IoT devices to collect data, which was subsequently kept in BigChainDB \cite{mcconaghy2016bigchaindb} for scalability. Proposed smart contracts define a set of rules for supply chain members to follow when accessing, exchanging, and interacting with data and each other.

In \cite{baralla2018blockchain}, authors proposed an Ethereum-based FSC platform that provides transparency and sharing of product information between customers, local producers, and companies as well. All records of product-related data could be visible, integral, immutable, and verified by customers and members in the FSC by using Keyless Signature Infrastructure (KSI) and blockchain technology. In addition, two smart contracts were put in place to oversee the entire production and supply chain.

The authors in  \cite{venkatesh2020system} introduced a blockchain-based system architecture for transparency in supply chain social sustainability management (SCSSM). The proposed system was divided into five layers: intelligent objects, communication channels, data analysis, blockchain network, and applications. Data from each layer was transparent, visible, and authenticated by its typical users (i.e., inspectors and operators at the intelligent objects layer). Mainly, the blockchain network layer played a vital part in securely storing processed data from the data analysis layer with a fair incentive mechanism to encourage stakeholders to record and share their data.

\smallskip
\noindent\textbf{Lessons learned:} The following are key lessons learned via using blockchain-based data sharing in supply chains:
\begin{itemize}
    \item  IoT and IIoT technologies, when combined with the blockchain, facilitate traceability that allows for the real-time and nearly real-time capture and tracking of product quality, operation control, and logistical information throughout the supply chain. As a result, there is less chance of data manipulation, and the supply chain's operation runs smoothly.

    \item  Supply chain will always provide the latest information and have solid and accurate evidence to trust others and make better decisions during the process of circulation because the information is shared, readily available, and transparent to authorized stakeholders in the supply chain thanks to blockchain and cryptographic techniques. 

    \item  Although current blockchain-based data-sharing solutions for supply chains have solved most of the drawbacks of conventional approaches, such as product IA, product traceability, and transparency, there is anxiety among entities about sharing their (sensitive) data in the trend of outsourced manufacturing and distribution, and supply chain globalization. Access control mechanisms are not enough to safeguard the confidentiality of data when the number of malicious entities makes up the majority of a supply chain. Therefore, considering PPTs to create an additional security layer for data is needed for blockchain-based supply chain data-sharing applications. 
\end{itemize}

\subsection{Transportation}

With the brisk advancement of ML, Big data, wireless communication, IoT, and traffic infrastructure, the intelligent transportation system (ITS) has captivated attention in academia and industry in recent years \cite{zhu2018big, mollah2021blockchain, kaffash2021big}. ITSs assist in operating the entire transport system, minimizing travel delays, reducing fuel consumption, and improving road safety by providing transport network operators and drivers with accurate and timely information about traffic conditions.  In ITS, intelligent vehicles (IVs), which are equipped with onboard units (OBUs), play a crucial role in collecting and sharing a large amount of traffic-related data with the internet of other vehicles (IoVs). Nevertheless, sharing data in current transportation systems has security and privacy problems. The plans still rely on a centralized node to store data, creating a single failure point (SPOF) from large-scale attacks. Besides, in an open communication environment, the risk of data manipulation and eavesdropping is high, thereby threatening drivers' safety when they use tampered data to make decisions on the road. Blockchain technology provides a secure and trusted environment for vehicles to exchange their data without relying on an intermediary. We review recent studies on transportation data-sharing applications that apply blockchain technology in the following. 

In \cite{9042215}, Di et al. focused on developing a data-sharing system architecture containing essential components such as OBU, roadside unit (RSU), trusted authority (TA), and consortium blockchain for vehicular ad-hoc networks (VANETs). The consortium blockchain, in which RSUs serve as verification nodes, stored encrypted data sent from OBUs while reducing communication overhead and enhancing security. The searchable attribute-based proxy re-encryption (ABPRE) \cite{ABPRE} technique was applied to help ensure data privacy at OBUs and enable secure data sharing. Furthermore, smart contracts are deployed to provide different services for police, vehicle owners, and insurance companies.

\begin{figure}[t!]
	\centering
	\includegraphics[width=1.0\linewidth]{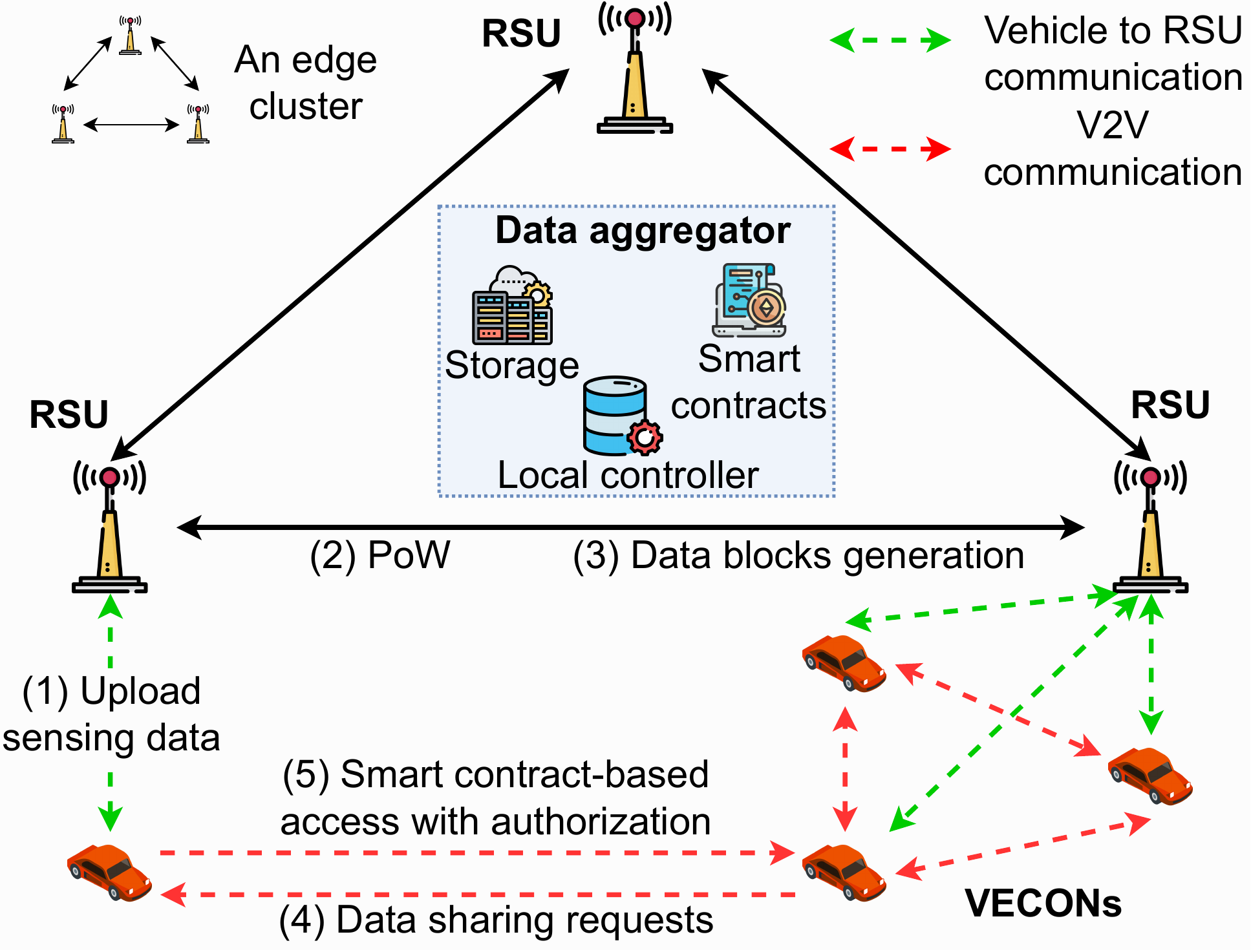}
	\caption{A consortium blockchain-based storage and data sharing model in VECONs \cite{kang2018blockchain}}
	\label{fig:BC-transport}
\end{figure}

The authors in  \cite{8746499} proposed DrivMan, a trust management and data-sharing solution for VANETs. To maintain the reliability of shared data, a physical unclonable function (PUF) is adopted to allocate each IV a unique cryptographic fingerprint to protect IVs' identities. In addition to serving as consensus nodes, in the proposed network model, RSUs also play a role as CAs for registering and revoking IV permissions, ensuring shared data comes from trusted sources. Two smart contracts are used to validate data, store data, and retrieve data from the blockchain network.

In \cite{zhang2019data}, authors introduced a consortium blockchain-based secure data sharing system among IVs and RSUs, named DSSCB, for VANETS. The consortium blockchain, in which RSUs are consensus nodes, was implemented to store shared data. Besides, to protect data provenance and integrity, shared data would be signed before transmission using a digital signature technique. Smart contracts are designed to determine a reasonable incentive for data contributors. Authors also imposed constraints on conditions to trigger smart contracts in order to regulate data usage and sharing better.

The authors in  \cite{su2020lvbs} demonstrated a lightweight blockchain-based data-sharing framework among vehicles and UAVs for rescuing in natural disasters where network infrastructures were damaged. A vehicular blockchain was established to record blocks of transactions between authorized UAVs and vehicles. In addition, the authors presented a credit-based DPoS consensus process with faster verification and lower energy consumption. Finally, reinforcement learning techniques were used to facilitate the sharing of high-quality data.

The authors of \cite{9457110} presented a blockchain-based high-speed data-sharing system for vehicle-to-vehicle (V2V) communications in the Internet of Vehicles (IoV) \cite{8746499, lu2018survey}. Some authorized IVs are considered data providers since they possess data collected and observed by their drivers in the presented system. 5\textsuperscript{th} generation base stations provide high-speed transmission for V2V communication and serve as data forwarders for vehicles. Data providers submit their data indexes to consortium blockchain for rewards. A smart contract was developed to verify and authenticate ratings from DRs after they receive providers' shared data, thereby facilitating the sharing of higher-quality data. To increase the performance of the system in terms of delay and throughput, the authors designed an enhanced delegated proof-of-stake (DPoS) consensus protocol.

The authors in  \cite{lu2020blockchain} integrated blockchain technology with FL to enable sharing of data among vehicles in the IoV. In this paper, a hybrid blockchain method for securely storing shared data events and ML model parameters was developed, which included a permissioned blockchain established by RSUs and a local DAG run by vehicles. With a DAG deployed on each vehicle, data sharing became more efficient and rapid, especially in time-sensitive scenarios in the IoV. Moreover, the shared data will be examined twice, promoting the exchange of qualified data while avoiding the sharing of fraudulent data.

The authors in  \cite{chen2019toward} focused on motivating data sharing among vehicles with a mechanism to evaluate data quality using blockchain technology in the IoV. The proposed architecture includes three main components: vehicles that collect, sell, or request data; cloud/edge servers; and RSUs, which are a communication bridge between vehicles and cloud/edge servers and run the consortium blockchain network. Cloud/edger servers receive requests, conduct checks on sellers' off-chain data using an expectation maximization (EM) algorithm, and make payments using a data quality-driven auction (DQDA). The data-sharing process is automatically executed by the smart contract between vehicles and edge/cloud servers.

In \cite{kang2018blockchain}, authors proposed a blockchain-based secure storage and data-sharing solution for vehicular edge computing and networks (VECONs). The solution utilized a consortium blockchain run by RSUs to store and retrieve data from data providers (vehicles) and DRs (vehicles), respectively. An edge node, e.g., a RSU, that uses the most storage space to keep shared data is recognized and receives rewards via a scheme managed by a data storage smart contract and Proof-of-Storage protocol. In terms of the sharing scheme, an ISSC smart contract, the PoW protocol, and a reputation-based mechanism with three-weight subjective logic (TWSL) are utilized to assist in selecting credible data providers and enable the exchange of high-quality data.

The work \cite{firdaus2021blockchain} utilized the consortium blockchain and smart contracts to present a data storage and sharing architecture. The rating of a vehicle's shared data is calculated by its neighboring vehicle and then reviewed by the nearest RSUs. This assures that shared data are trustworthy and high-quality, boosting driving safety. Authenticated data would be considered to be appended to the blockchain by employing a smart contract with the Practical Byzantine Fault Tolerance (PBFT) consensus protocol. Vehicles that share and rate data receive rewards for their contributions as an encouragement.

\smallskip
\noindent\textbf{Lessons learned.} The following are the principal lessons from the applications towards the transportation sector:

\begin{itemize}
    \item There are three main types of shared data in transport applications: 1) vehicle-generated data, such as weather conditions, road conditions, or engine status, collected by sensors installed on vehicles; 2) data imported by drivers and passengers, like nearby gas stations, charging stations, or ratings of services in restaurants and hotels; and 3) verified data that drivers found trustworthy and useful that is received from other vehicles and RSUs in other areas but might be helpful and suitable on this road. 

    \item The majority of current research prefers to use permissioned blockchain, to construct vehicular blockchain systems for data sharing due to resource-constrained vehicles and traffic infrastructure, security, and privacy issues. In order to store records of data sharing, including indexes of shared data, ratings of shared data, or reputation scores of IVs who shared data, RSUs or IVs are configured to act as consensus nodes. The vehicular blockchain network only allowed authenticated and authorized vehicles to join and share data, thus shared data are high-quality, traceable, and reliable. Additionally, smart contracts are used and deployed on these blockchain nodes to make it easier to store and share data by defining sets of rules for access control and rewards.

    \item Similar to healthcare and supply chain applications, data providers in transportation applications are also concerned about the risk of sensitive and personal data leakage, such as driver licenses and frequently traveled routes, and are reluctant to share their data, even when the reward is high. Not to mention that sharing data also costs them because they are resource-limited. Almost current solutions have not yet pondered utilizing PPTs to enhance the privacy of shared data for data providers. Although there have been a few attempts to use PPTs, they are limited and need to be widely adopted.
\end{itemize}

\input{Tables/2_Taxonomy_of_applications.tex}

\subsection{Smart Grid}

A smart grid is a self-healing power network that uses digital technologies to supply electricity to consumers via bidirectional communication \cite{5357331, dileep2020survey, bayindir2016smart}. This system enables monitoring, measuring, sharing, and controlling of power and information flow to increase efficiency and resilience, optimize energy consumption and costs, and reduce transmission losses. Smart grid systems can be achieved by implementing electronic power generators, transmission, distribution systems, control centers, renewable energy sources, and energy storage integration. Following this development, smart grid 2.0 has been developed to connect all energy sources anywhere, at any time, with the aid of the Internet technology as the number of power resources surged and the grid networks become more complex and distributed \cite{mollah2020blockchain, mahmud2020internet}. Although the trend toward moving to decentralized structures ameliorates the connectivity and integration problems of the smart grid, how to exchange information among entities in the grid system while complying with data privacy legislation and ensuring data security remains a major barrier for researchers and companies. Fortunately, blockchain technology offers excellent properties to not only create decentralized grid systems but also enhance trust, data security, and privacy for the information exchange of grid systems\cite{10061596}. Here, we review studies on smart grid applications  relating to our framework.

\begin{figure}[t!]
	\centering
	\includegraphics[width=0.9\linewidth]{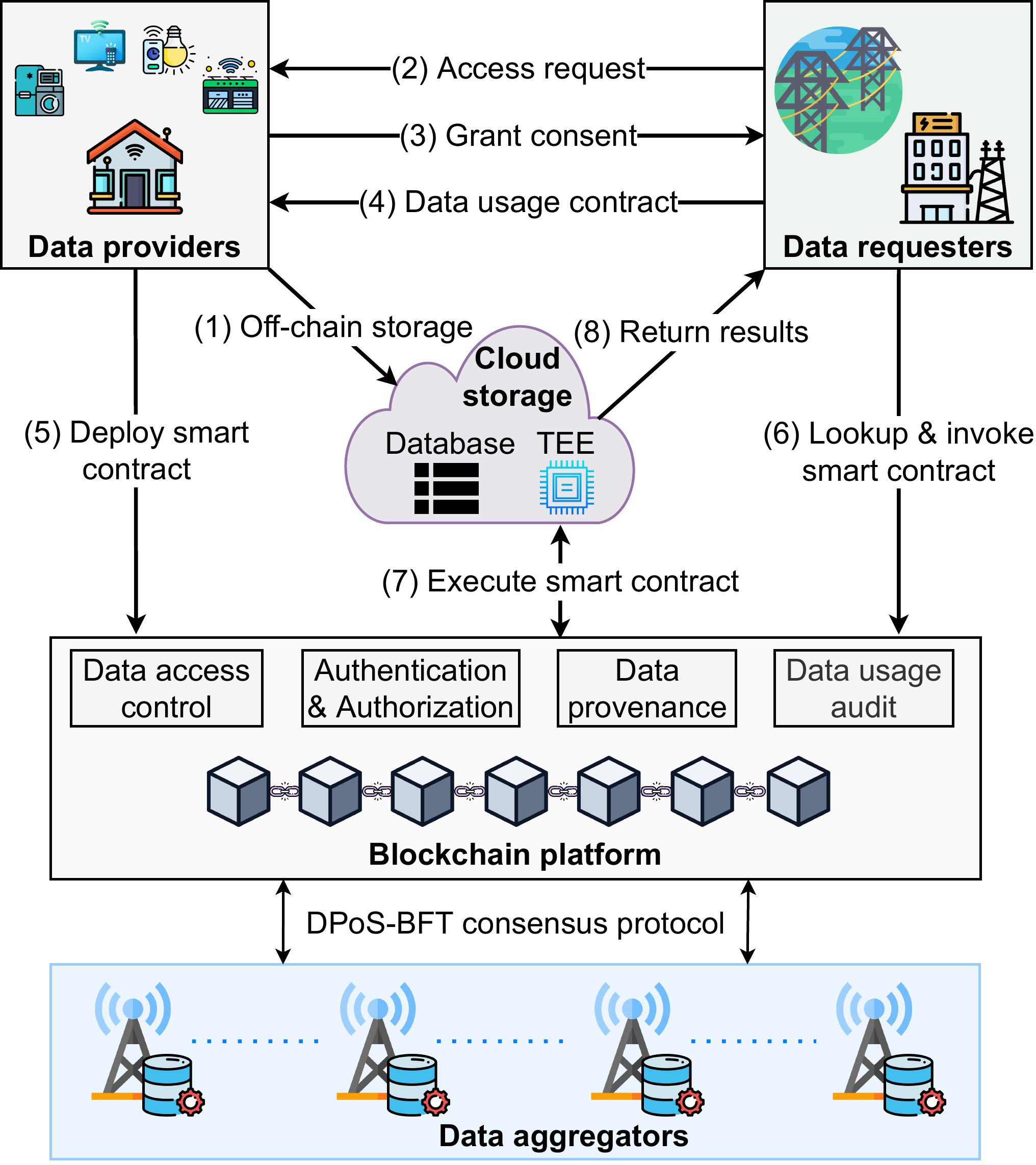}
	\caption{A blockchain-based privacy-preserving data sharing system in smart grid domain \cite{wang2020spds}}
	\label{fig:BC-smartgrid}
\end{figure}

In \cite{wang2020spds}, blockchain technology was used to present a data-sharing scheme with the management of private DAC and DUC. A framework, called SPDS, was proposed allowing DPs to establish access policies and track historical usage records over their shared data. Moreover, a smart contract-based data processing as a service (DPaaS) solution was developed to secure shared data privacy and reduce the computing demands on user devices.

In \cite{samuel2019blockchain}, authors proposed a blockchain-based data-sharing system that trades off privacy risks against rewards in deregulated smart grids. A Proof of Authority (PoA) protocol that enables a mining node with the highest reputation score to have a chance to append its block to the chain was presented to minimize computational and gas costs, and achieve reliability for the process of reaching a consensus in the blockchain network. Moreover, the authors designed a fair incentive mechanism based on the levels of risks related to privacy to motivate customers who have valuable data to share. Finally, to reduce the possibility of original information leakage, differential privacy technique was utilized. 

The authors in  \cite{yang2020secure} focused on designing an edge blockchain-enabled DAC framework using CP-ABE, \emph{(t, n)} threshold secret sharing scheme (TSSS), IoT devices and Hyperledger Fabric. CP-ABE was used to establish an access control attribute-based mechanism over shared data for DRs. Besides, both on-chain and off-chain models were proposed for flexible data sharing and data storage. To mitigate computational burdens on resource-constrained device nodes, cryptographic operations are outsourced to a consortium blockchain network. Unlike conventional centralized authority approaches, \emph{(t, n)} TSSS was given to build up a distributed authority that assists to solve the SPOF problem.

In \cite{li2021lightweight}, authors proposed a dual blockchain-based electricity data-sharing solution for power supply’s pricing center in smart grids. With the support of cloud infrastructure, encrypted shared data was securely stored, and it minimized storage costs for blockchain network. The private blockchain is presented to protect the user's identity by associating it with a pseudonym. To facilitate the sharing of data with the outside system, the public blockchain was used to expose shared data to authenticated entities. 

The authors in  \cite{reijsbergen2022securing} presented a blockchain-based data-sharing solution between operators to defend against False Data Injection (FDI) attacks. A blockchain network was availed to enable data sharing among operators in the power grid. Additionally, the authors proposed an incentive mechanism encouraging operators to give meaningful measurements and penalizing them if detecting any malicious ones. The entire data-sharing and incentivizing processes are managed and run by the smart contract.

The authors in \cite{guan2021blockchain} focused on protecting data privacy on both DPs' and requesters' sides of a blockchain-based data-sharing system between power transmission companies, electricity sales companies, and consumers. The consortium blockchain is utilized to store information that DPs and DRs exchange. Raw data collected from edge servers was collaboratively processed and calculated by running a smart contract. To send computing tasks and processed data without revealing the identities of senders and receivers and reducing information leakage, a data obfuscation method including main steps: data segmentation and data registration, ring signature, and multicast was used.

\smallskip
\noindent\textbf{Lessons learned.} Key lessons learned from the review of the blockchain-based solutions in the smart grid discussed above are summarized below:

\begin{itemize}
    \item In blockchain-based smart grid systems, blockchain technology ensures data sharing between DPs (energy customers) and DRs (energy service providers) is trustworthy by providing immutability, transparency, and traceability while complying with GDPR legislation. While blockchain network is commonly used to store metadata of raw data and records of data usage, smart contracts play a vital role which specifies incentive mechanisms, DAC and DUC policies. Besides, edge computing and cloud servers are utilized to handle computing outsourcing tasks, reducing the computational and communication burden on resource-constrained devices when these devices have to process (e.g., encryption) and send a huge amount of data over a long distance. 

    \item With blockchain-based data-sharing solutions, DPs are able to manage their energy usage plans and profiles, keep ownership and control over shared data, and be motivated to share their data. Meanwhile, DRs govern the production and distribution of energy effectively, have a better understanding of their clients, and provide DPs with efficient services. 

    \item PPTs have been shown to be effective in safeguarding the privacy of shared data in smart grids.  PPTs are utilized in the aforementioned research (e.g., TEEs, MPC) to mitigate the risk of unauthorized use of data when DRs receive data from DPs. This is because are unable to ascertain the specific purposes for which their data will be used after sharing it with DRs, who receive only processed results from the PPTs. 
\end{itemize}

\subsection{Data Marketplace}

With the proliferation of IoT and Internet of Everything, a massive amount of data generated and gathered from them not only adds value and insights to DPs, but also to third-party DRs when DPs resell their data. The data marketplace is a two-sided online market where DPs and DRs can sell and buy data personally\cite{pandey2022contribution}. DPs may manage, consider, and sell their available data in the data marketplace, whereas DRs can explore, choose, and acquire the data they require. Traditional data marketplaces, however, have challenges to their success. First, personal or sensitive data may be collected, publicized, and sold intentionally or unintentionally. Furthermore, the data may be falsified and used without the owners' knowledge or permission. Second, DPs do not have adequate motivation to participate in exchanging data in these centralized data marketplaces. They are concerned about the privacy and security of their shared data when participating in marketplaces managed by a central authority. Additionally, it is difficult for DPs to determine the price for the data, which necessitates economic understanding \cite{ha2019wts}. When it comes to DRs, they may have difficulty locating a high-quality and trustworthy source of data to purchase. Even though they are wealthy, it is uncertain whether they will be able to purchase quality datasets. Fortunately, blockchain technology can address the aforementioned concerns and challenges associated with traditional data marketplaces. By combining blockchain with the data marketplace, a new sort of data market, known as a decentralized data marketplace, is introduced. Decentralized data marketplaces with blockchain-based backends, similar to traditional data marketplaces, provide individuals, businesses, and organizations with a fantastic opportunity to reach an always-on and diversified pool of data. It improves privacy and security while also providing numerous benefits to DPs and DRs. In this part, we will explore research on blockchain-powered data marketplace applications.

\begin{figure}[t!]
	\centering
	\includegraphics[width=0.85\linewidth]{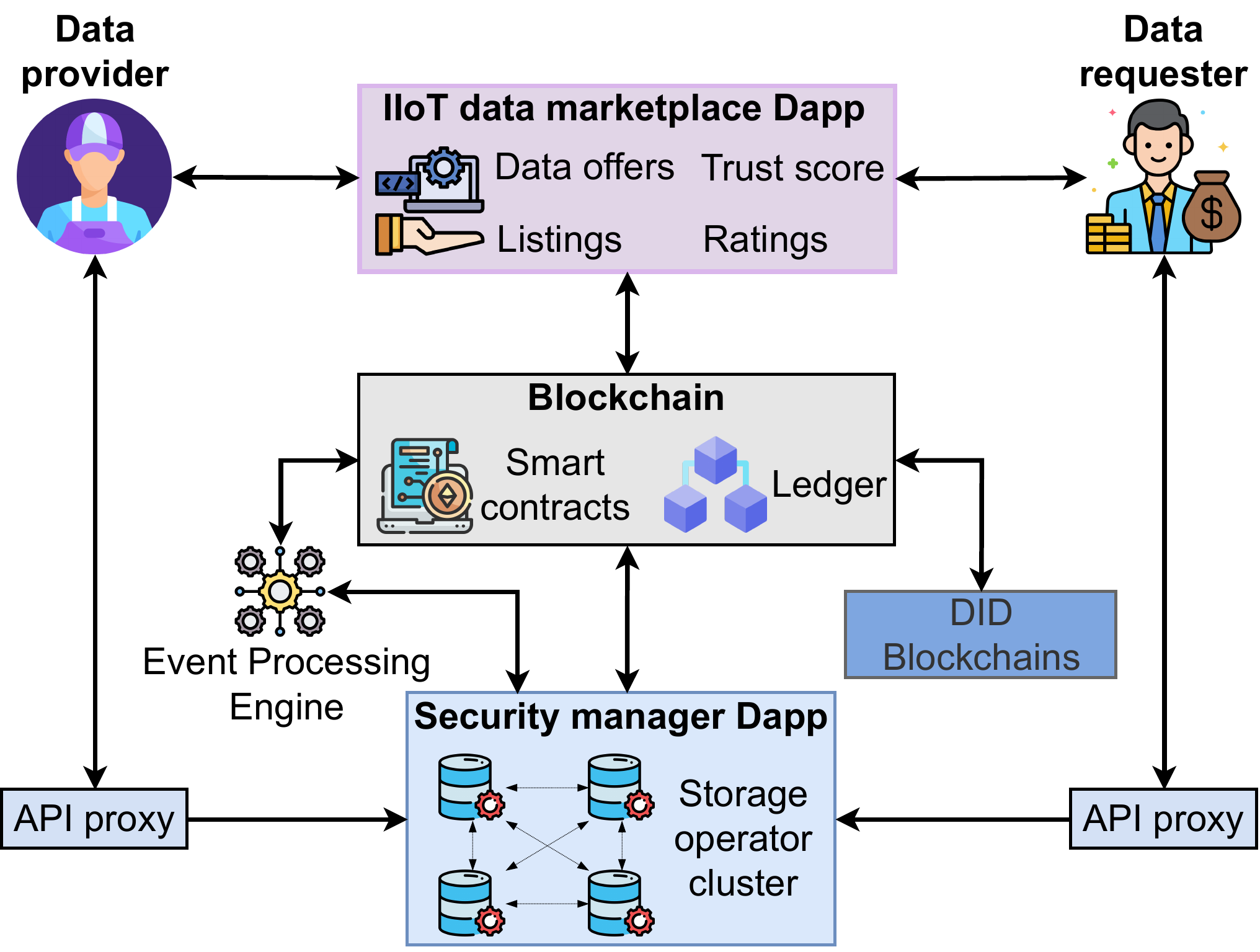}
	\caption{Blockchain-based data trading architecture in IIoT data marketplace \cite{dixit2021fast}}
	\label{fig:BC-datamarketplace}
	\vspace{-0.3cm}
\end{figure}

In \cite{nguyen2021marketplace}, Lam et al. proposed an ML trading marketplace using IoT data, blockchain technology, and FL. Via the system, initial ML models uploaded by buyers would be downloaded and trained by sellers who have data and want to sell their contribution to improve ML models without disclosing any information using FL. The authors also proposed a distributed data shapley value (DSV) algorithm for evaluating the quality of data and corresponding trained ML models as well. Therefore, participants may receive an equitable reward proportional to their contribution. All details about ML model trading are recorded in a transparent and immutable manner on the blockchain network.

The authors in  \cite{alsharif2020blockchain} presented a blockchain-based marketplace model for sharing and trading medical data through a smart contract. By using CP-ABE and developing the Zero-Knowledge Succinct Non-Interactive Arguments of Knowledge (zk-SNARK) protocol, the proposed model provides a flexible access control and verification scheme for encrypted shared data without disclosing any actual private information. Furthermore, a zero-knowledge contingent payment protocol was developed to realize a fair exchange.

The authors in  \cite{dixit2021fast} designed a secure blockchain-based IIoT data marketplace platform, ensuring trust and fairness. The proposed system has three key actors: sellers who possess and wish to sell their data, storage operators who store encrypted data submitted by sellers, and clients who want to buy data for a specified reason. The platform enables smooth and quite equitable payment for system actors using blockchain and smart contracts, with two payment modes described (batch and stream). Clients of the proposed marketplace have to go through decentralized identity verification using smart contracts/chaincode for traceability, trust, and reputable management. Finally, a proof-of-concept implementation using Hyperledger Fabric was created for feasibility testing.

In \cite{ramachandran2018towards}, authors presented a decentralized blockchain-based IoT data trading marketplace for smart cities. While the description of shared data is preserved in a distributed file storage, blockchain technology is used to store metadata about the storage identification and blockchain protocol type. Smart contracts are used to assist sellers to register their metadata to the blockchain networks and to allow both DPs and DRs to rate each other. The streaming data payment protocol (SDPP) is utilized to exchange data and make payments between buyers and sellers.

The authors in \cite{ozyilmaz2018idmob} proposed a blockchain-based IoT data marketplace, called IDMoB, allowing multiple parties to take part in data trading. The proposed marketplace provides a versatile filter for efficient data retrieval with a voting mechanism to rank data providers using smart contracts. Regarding storage, IDMoB uses Swarm \cite{tron2016swap} platform as an encrypted IoT data repository uploaded by data providers. Ethereum blockchain with payment channels is utilized for instant transfers and storing of data trading history.

In \cite{nguyen2019enabling}, Duy et al. utilized blockchain technology and crowdsensing techniques to develop a decentralized data collectability marketplace model, named DCDM. On the other hand, the proposed marketplace allows for on-demand data gathering and provisioning for future use (a service-based approach). The marketplace offers key functionalities such as recommendation and pricing service, which recommends the best providers to consumers, contract and transactions service, which is automatically executed by smart contracts for allocating sensing tasks to providers, and quality assurance service that assesses and rates data streams uploaded from consumers and providers.

The authors in  \cite{tang2022dmobas} designed a secure and efficient personal data marketplace with arbitrators' alliances using blockchain technology and the concept of side contracts, named dMOBAs. IPFS, which is a distributed storage system, is used to store encrypted data uploaded from sellers, and then it returns the hashes and multi-set hashes of the ciphertext to smart contracts. The entire trading process is executed by smart contracts on-chain. Besides, when receiving data from sellers, if buyers are not satisfied with the shared data, they are able to initiate an arbitration procedure using a side-contracts mechanism to resolve disputes and defend their rights. 

\smallskip
\noindent\textbf{Lessons learned.} The following are the key lessons from studies of blockchain-based data marketplace applications:

\begin{itemize}
    \item Blockchain-based data marketplaces bring trust, fairness, and democratization to DPs by allowing them to securely trade their data and use it as tradable assets that they are willing to share and find useful to providers in need. Furthermore, they also benefit from customized and enhanced services and applications that use their data for advancement. This also eliminates the reality of data monopolization by tech giants.

    \item The blockchain technology promotes the development of a decentralized, trusted, tamper-proof, fair, and transparent data marketplace for trading diverse types of data, such as ML models, IoT data, and personal data. Off-chain storage solutions, like other applications, assist the blockchain platforms in storing encrypted data, reducing the significant waste of storage and computing resources for peer nodes. Besides, smart contracts are used in data marketplace to automate trading procedures. They carry out buy/sell and rating requests between DPs and DRs.

    \item An efficient approach for data estimation, as well as a fair and transparent incentive solution, are mostly absent from the present blockchain-based data trade market. Giving set or ambiguous prices without taking the quality of the data into account results in issues like the sale of fraudulent data, which frustrates suppliers who sell high-quality data. Additionally, it is tough for buyers to buy sensitive data in data marketplaces to improve their businesses due to privacy regulations. For example, to achieve high accuracy on the face recognition feature of mobile applications, multiple small and medium ML companies are willing to purchase a huge amount of high-quality facial images that are very personal and sensitive to train their ML models. 
    
    \item Moreover, the widespread availability of online data marketplaces means that DPs face a significant risk of losing control over the data they share. This is because anyone can purchase data online, potentially leaving DPs without ownership or control over their shared data. Furthermore, exchanging processed data can result in high energy consumption and increased communication overhead. To mitigate these risks and enhance system performance, the use of DAC policies with PPTs in blockchain-based data marketplaces is worth considering. Such policies can help DPs maintain control over shared data, protect data privacy, and improve overall system efficiency.
\end{itemize}

\subsection{Others}

\input{Tables/4_Other_applications.tex}

A series of work has also been done by integrating blockchain technology into data sharing in other fields such as education, government, and decarbonization.

\emph{1) Education:} With the innovation of information technology, the education field has made strong shifts when paper physical records have been digitalized and stored on on-premise servers and cloud servers. Although the centralized management approach makes records, such as degrees, course grades, and awards, more accessible than traditional methods, it exposes educational records to a variety of hazards. For example, the test and student scores can be viewed and manipulated without permission. Furthermore, if the school server fails, there is a substantial danger of data loss. Furthermore, contemporary educational systems lack a trust-based data exchange environment. Each educational institution maintains a repository for students' academic records. When students want to transfer to a different institution, it takes a lot of effort and time for them to request and certify these records. Fortunately, salient features of blockchain technology can promote trusted data storage and sharing for academic institutions. Therefore, in this part, related research on blockchain-based educational data sharing will be reviewed and discussed. 

The authors in \cite{li2019edurss} proposed an educational record storage and sharing scheme based on consortium blockchain, storage servers, and smart contracts, named EduRSS. In the system, institutions are consensus nodes that form a consortium blockchain network that stores hash values of educational data, while the entirety of the encrypted original data is kept in the storage server, i.e., the MongoDB database. Moreover, three smart contracts are introduced to manage the joining, data storage, and data sharing processes of institutions. Via experimental results, the storage and encryption times of EduRSS prove to be lower than those of CP-ABE and multi-authority-based CP-ABE \cite{li2011multi} schemes. 

In \cite{bsssqs}, the authors presented a secure and smart scheme, called BSSSQS, for question sharing in education systems. In the scheme, four key entities participate in the data-sharing process: the question setter, the question cloud, the BSSSQS master, and the BSSSQS minion. First, the question setter prepares questions and sends them through an encrypted channel to the question cloud. Following that, the questions are double-checked and organized into question papers. The question papers are then automatically withdrawn from the question cloud and forwarded to the BSSSQS master. To prevent unauthorized access to question papers, the BSSSQS master uses a two-phase encryption technique and creates a smart contract that defines access rules for the BSSSQS minions. The BSSSQS minions, which are universities, exam centers, etc., can access question papers under BSSSQS permission and only decrypt question papers at a specific time. As a result, BSSSQS is safe from the threat of question paper leaks and provides an effective question-sharing scheme for future educational systems. 

The study in \cite{chou2020blockchain} proposed a blockchain-based work-sharing or collaboration system to ensure authorship when multiple contributors, i.e., teachers, participate in educational content creation. The author designs a new block structure for storing a work's authorship and work information, which can then be invoked by different teachers and exist in different blockchains of educational materials. When multiple teachers contribute to an educational resource, work blocks will be linked together to establish a blockchain. Besides, to evaluate the contribution of each contributor, a contribution share calculation function is given based on factors, e.g., file size. For system scalability, off-chain storage solutions are presented to keep the original work of contributors. 

\emph{2) Government:} To make wise decisions and create sustainable and far-reaching projects and services for the economy, society, and citizens, governments always collect, gather, analyze, and share data to capture and keep up with global and social information and events. With the brisk maturing of information communication technology, government systems have transformed paper-based data storage and sharing approaches into electronic counterparts, called electronic government (e-government), with goals of achieving fast and effective procedures, high transparency, ensuring accountability, and security and privacy preservation \cite{elisa2020consortium, zhang2019research}. Besides, data stored by governments include very sensitive information about citizens, healthcare, and politics, such as identity cards, demographic data, immigration, land registration, and so on. Moreover, current centralized government data storage and sharing solutions are in danger of cyberattacks and data breaches, thus designing and developing a secure, transparent, accountable, and privacy-preserving data storage and data sharing solution is mandatory and urgent \cite{elisa2020consortium, governmentattacks}. The essence of blockchain has assisted governments in moving closer to and reaching their objectives. Following that, we will discuss research on blockchain-based government data-sharing applications.

In \cite{wang2017blockchain}, the authors presented a blockchain-based information resource-sharing system among government departments to ensure data reliability, sharing efficiency, and accountability. A new block structure for storing government information resources is defined, where a block only associates with a transaction, keeping the owner's control over their shared data. Besides, a consensus mechanism is proposed for data verification and achieving common agreement on a single version of the truth among distributed nodes. With the testbed environment including two local networks comprising ten PCs and two routers, analysis and experiment results demonstrate the efficiency, reliability, and security of the proposed system.  

The work in \cite{chen2020blockchain} proposed a blockchain-based E-government data sharing framework to safeguard the authenticity and ownership of DPs, and data confidentiality, called GovChain. The GovChain framework is constituted of three layers: 1) application, 2) blockchain system and two function modules, i.e., identity \& key services module and CP-ABE, and 3) off-chain storage. In the application layer, application programming interfaces (APIs), command line interfaces, and software development kits are exposed for interaction between GovChain and clients with different data-sharing services. To improve the scalability of the data sharing system, only the hash of encrypted data is kept in blockchain, while original data are stored in the IPFS. Regarding the identity and key services module, it provides an unique identity and a digital certificate for each client and performs functions (e.g., client registration and attribute management) thanks to smart contracts. Finally, in the GovChain, every node is considered as an authority in decentralized CP-ABE scheme, so it can conduct off-chain encryption and decryption algorithm computation to reduce communication overhead.

The authors in \cite{shi2022secure} proposed a blockchain-based government data-sharing system with the support of a fine-grained access control scheme. To realize data confidentiality and flexible access management, the authors use the advanced encryption standard algorithm \cite{daemen1999aes} to encrypt data and CP-ABE algorithm for encrypting the hash of the data storage address returned from IPFS. Besides, a linear secret-sharing scheme is utilized to hide access control policies defined by DPs. During the data-sharing process, if a user has abnormal behaviors or malicious activities, its access privilege will be withdrawn by the key generation node, and a new attribute group will be established thanks to coverage technology.


\emph{3) Decarbonization:} 
Decarbonization is becoming the main priority for the majority of industries. Industrial ecosystems consist of multiple players from different sectors located in close proximity, using shared utilities and connected through physical networks. These ecosystems can contain over 20 production sites producing a range of products, such as chemicals, iron, steel, and petrochemicals. To optimize the energy and resource usage of these interconnected networks, secure and accurate information sharing is becoming crucial. The deployment of digital systems enables real-time interactions between all stakeholders.

An example of sharing for decarbonization is European Union Emissions Trading System (EU ETS) was completed in 2005-2007 and saw the allocation of 2.2 billion tons of CO2 emission allowances to over 12,000 installations across 27 EU member states. During this time, the trading of EU ETS allowances increased from 260 million tons to 1.44 billion tons. The authors of the research \cite{zhang2010overview} provide a comprehensive overview of existing studies on EU ETS from 2005 to 2009, focusing on its operational mechanisms and the economic benefits it generated. They summarize the evolution of the system's operating mechanisms, such as allowance allocation and pricing, and explain how these mechanisms have influenced the market's development and economic outcomes. The paper identifies five areas for future research on EU ETS: (1) its impact on energy industries, (2) the interaction between EU ETS and energy markets, (3) evaluation of its impact on socio-economic systems, (4) its impact on non-Annex I countries, and (5) its operating mechanisms. This paper fills a gap in the literature by providing a review of EU ETS after its first phase and offering ideas for future development.

Additionally, carbon pricing plays a crucial role in the global carbon emissions trading market. In the paper \cite{cao2019china}, the authors analyze the pricing decision principles of the Emissions Trading System (ETS), including the cap and trade principle and the baseline and credit approach. The EU ETS market \cite{brink2016carbon} provides various carbon pricing schemes with different national policies and compares their outcomes. Carbon pricing can be utilized as a means to control carbon emissions and steer the sustainable development of carbon trading \cite{narassimhan2017carbon}. The authors from \cite{tanaka2012comparison} examine a new energy-saving and emission reduction (ESER) system based on carbon price constraints in China. The authors \cite{nguyen2021b} introduce the concept of a trusted trading framework for emission allowance among vehicles. This research, for the first time, introduces the concept of emission trading for individual vehicles. 

Specifically, the Emission Allowance Balance (EAB) can be traded among vehicles based on predefined smart contracts. Whenever $B_i(t)<0$, there will be a red alert issued to $i$ for having a negative balance. This alert is in the form of penalties, or restricted road access to zero-balance vehicles. In this case, the vehicles can either wait until the next period for their EAB of to be reset or buy the EAB from other vehicles. We consider the case of vehicles exchanging EAB on-road via the execution of the smart contract and distributed ledger. For this, let $e_{i,j}(t)$ be the amount of allowances sold by vehicle $j$ from vehicle $i$ at time $t$. These operations are recorded on-chain.

\smallskip
\noindent\textbf{Remark}: \textit{The vehicle $j$ cannot sell more allowances $e_{ij}(t)$ than it actually owns. In other words, $i$ cannot buy more than is actually available. Hence,}
\begin{equation}
    e_{i,j}(t) \leq B_{j}(t), \text{for all } j\in\mathcal{V},\,t\in\left[0,T\right)
\end{equation}

\begin{figure}
    \centering
    \includegraphics[width=0.75\linewidth]{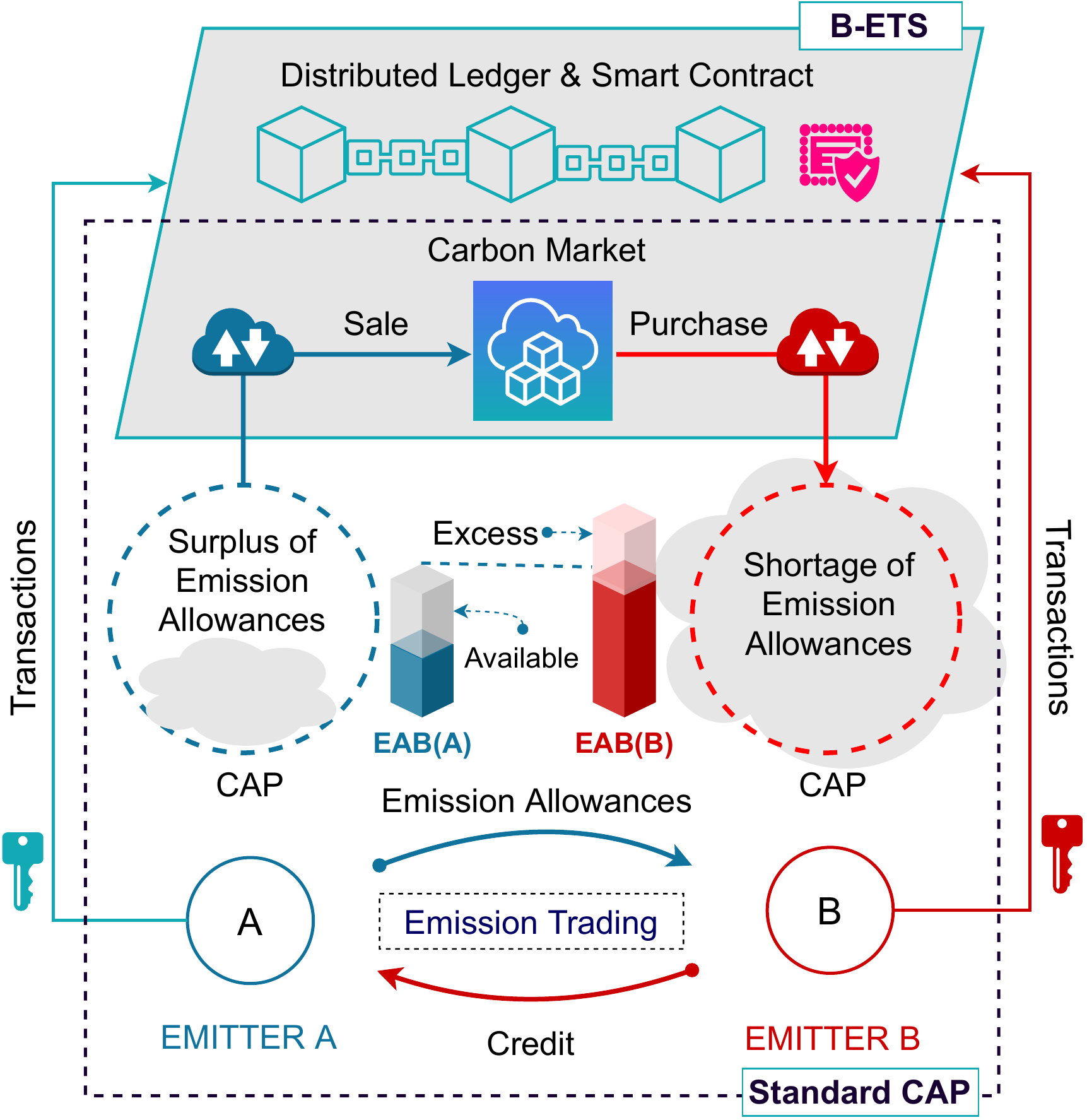}
    \caption{B-ETS Emission Trading architecture\cite{nguyen2021b}}
    \label{fig:my_label}
\end{figure}

\smallskip
\noindent\textbf{Lessons learned.} Then, we state the key lessons from leveraging blockchain-based data in the aforementioned sectors. 

\begin{itemize}

\item In the education area, blockchain technology provides a safe and transparent place for storing and sharing data in educational organizations. Students' data, e.g., school profiles, is kept authenticated, synchronized, immutable, and accessible to authorized users when needed. This eliminates a slow and complex paperwork process for both staff and students when students want to change schools, apply for jobs, or pursue higher degree programs. Besides, teachers can collaborate with others when writing material and have more freedom in building curricula and books without the distress of violating copyright ownership issues. This is because the contributions of authors, including cited works, are all recognized and recorded in the blockchain. Moreover, blockchain-based question sharing creates a fair competitive environment for students by avoiding the leakage of exam questions.

\item With regard to government data sharing between government-to-government, government-to-business, government-to-organization, and government-to-citizen \cite{chen2020blockchain, wang2017blockchain, shi2022secure}, applying blockchain technology enhances authenticity, confidentiality, responsibility, and security in government information exchange processes. Especially in cases where the data are sensitive, with the support of cryptographic techniques and blockchain, sharing of the data can be realized at a certain level of security and privacy. Therefore, e-government growth with effective public service delivery is fostered, and relationships and trust between governments, corporations, organizations, and people are strengthened. 

\item Blockchain-based data sharing can play a significant role in decarbonization efforts. By securely and transparently sharing data on carbon emissions, stakeholders such as governments, companies, and individuals can work together to reduce their carbon footprint. With a decentralized blockchain platform, each participant can maintain control over their data while still contributing to the larger effort. Smart contracts can be used to automatically enforce policies and trigger actions based on real-time data, providing a more efficient and effective means of achieving decarbonization goals, e.g., emission trading for vehicles. Furthermore, the immutable and auditable nature of blockchain records can ensure the integrity of emissions data, preventing fraud and promoting trust between stakeholders.

\end{itemize}

\subsection{Industrial Applications}

With the enormous potential and profits that data sharing provides, it is not unexpected that a plethora of technological startups, companies, and corporations are interested in and participating in fostering this engine for the maturation of digital innovation and transformation. Previously, many organizations were unwilling to share data or just a small fraction of it with others because they were concerned about and alerted to issues of privacy, intellectual property protection, benefits, and security. This is where blockchain technology comes in to alleviate the aforementioned data-sharing challenges by providing traceability, immutability, decentralization, and transparency. Blockchain-based data sharing enables companies to increase their partners across multiple disciplines, resulting in better product management and strategy. Businesses have greater chances to develop new categories of products and scale their markets. In the following, we review blockchain-enabled data-sharing applications launched by companies and corporations. 

\input{Tables/3_Taxonomy_of_applications.tex}

Nokia has created a data marketplace that promotes data sharing and trade, as well as ML orchestration for business-to-business (B2B) and business-to-business-to-customer transactions, based on private permissioned blockchains \cite{nokia}. The Nokia Data Marketplace platform is accessible, efficient, and scalable, enabling enterprises to share and sell data in a wide range of areas, from healthcare and transportation to telco, thanks to a Software-as-a-Service (SaaS) business model. It provides key features such as trusted data monetization, secure data exchange, and federated AI/ML orchestration with the use of blockchain technology, assisting organizations in facilitating digital transformations.

Transparent Supply \cite{ibm} is IBM's blockchain-based data-sharing platform for facilitating collaboration and visibility across a supply chain ecosystem. Transparent Supply, which uses private permissioned blockchain, is secure, fast, and scalable, with the ability to manage many transactions and deliver near real-time insights to stakeholders. Furthermore, it provides various features for industrial applications such as quality assurance, improved forecasting, decreased friction, and automation.

Mercedes-Benz has developed Acentrik, a decentralized data marketplace that promotes data exchange and monetization on blockchain with low fees for enterprises across industries while ensuring data sovereignty \cite{acentrik}. Currently, Acentrik allows customers to share and trade two types of assets: datasets and algorithms. In addition, through tokens such as ERC721 and ERC20, Acentrik makes access to data more secure and controlled. One notable feature of Acentrik is Compute-to-Data, in which data are not transferred to outside. However, consumers can still run compute jobs on it, thereby increasing the level of data privacy. 

Databroker \cite{dao} is one of the leading blockchain-based data marketplaces, allowing data producers and customers to exchange all types of data. Databroker has three key characteristics to promote data sharing and trade, such as: connecting and serving, personalized DataMatch service, and Platform-as-a-Service solution (PaaS). Especially, shared data are not stored on the Databroker platform but are securely transmitted between providers and buyers via Data eXchange Controller (DXC) software.

\smallskip
\noindent\textbf{Lessons learned.} Key lessons learned from the study of the blockchain-based data-sharing solutions in the industry discussed above are summarized below:

\begin{itemize}
    \item Companies and corporations are able to generate revenues and benefits from: 1) Building a blockchain-based trustworthy and transparent data marketplace where DPs (data sellers) and DRs (data buyers) can trade data. They then charge for transactions; 2) Building a blockchain-based data-sharing platform where a group of companies and corporations that agree to cooperate and participate in data sharing to develop their new business models, products, and services. 
\end{itemize}

In summary, we summarize technical aspects, key contributions, and limitations of blockchain-based data-sharing applications in the taxonomy~Table~\ref{tab:taxonomy1}, Table~\ref{tab:taxonomy2}, Table~\ref{tab:taxonomy3}, 
 and Table~\ref{tab:industrial_apps}. Then, key lessons learned from these applications are provided to give state-of-the-art and overall insights into blockchain-based data-sharing applications.

%% file: Tables/1_Taxonomy_of_applications.tex

\begin{table*}[t!]
\caption{TAXONOMY OF BLOCKCHAIN-BASED DATA SHARING SOLUTIONS} 
 \renewcommand{\arraystretch}{1.5}
 \label{tab:taxonomy1}
\centering
\small
\begin{tabular}
{c|
 P{0.05\linewidth} 
 P{0.1\linewidth} 
 P{0.025\linewidth} 
 P{0.025\linewidth} 
 P{0.025\linewidth} 
 P{0.025\linewidth} 
 P{0.05\linewidth}  
 p{0.28\linewidth}  
 p{0.14\linewidth}}
 \hline
\multicolumn{1}{c|}{{\textbf{Area}}} & 
 \multicolumn{1}{c}{\textbf{Ref.}} &
 \multicolumn{1}{c}{\textbf{Objective}} & 
 \multicolumn{4}{c}{{\textbf{Blockchain used}}} & 
 \multicolumn{1}{c}{\textbf{PPT}} & 
 \multicolumn{1}{c}{\textbf{Key contributions}} & 
 \multicolumn{1}{c}{\textbf{Limitations}}\\ \cline{4-7}
 
& {} & {} & {\textit{\textbf{Pub.}}} & {\textit{\textbf{Pri.}}}& {\textit{\textbf{Con.}}} & {\textit{\textbf{Hyb.}}} & {} & {} & \\\hline

 \multirow{18}{*}{\rotatebox[origin=c]{90}{\textbf{Healthcare}}} & \cite{zhang2018towards} & {Storage, search} & {-} & {\checkmark} & {\checkmark} & {-} & {$\times$} &{Introduces two blockchains for storage, and a keyword search algorithm.} & {Security risks on computer clients.} \\\cline{2-10}  
 
 &  \cite{chen2022fine} & {DAC, storage} & {-} & {-} & \checkmark & {-} & $\times$ & {Provides a fine-grained DAC using PRE and ABE with IPFS technology.} & {Computational overhead.}  \\\cline{2-10}
 
 &  \cite{7990130} & {DAC, auditability}& {-} & {\checkmark} & {-} & {-} & {$\times$} &{Proposes a four-layer EMR sharing system, data access revocation mechanism.} & {Data privacy protection.} \\ \cline{2-10}
 
& \cite{ZOU2021102604} & {DAC, data retrieval} & {\checkmark} & {-} & {-}  & {-} & {$\times$} & {Develops new block and chain structures with CHF.} & {Throughput, scalability.}   \\ \cline{2-10}

 &  \cite{fan2018medblock} &  {DAC, data retrieval} & {-} & {\checkmark} & {-} & {-} & {$\times$} & {Uses bread crumbs mechanism for finding encrypted EMR location.} & {Scalability, risks on CA.} \\\cline{2-10}

  & \cite{DAGHER2018283} & {DAC} & {-} & {\checkmark} & {-} & {-} & {$\times$} & {Uses a distributed PRE method with blinding.} & {Lacks performance evaluations.} \\\cline{2-10}

 & \cite{9831062} &  {DAC}  & {-} & {\checkmark} & {-} & {-} & {$\times$} &  {Develops hierarchical ABE, and adopts redactable BC.} & {Scalability and storage costs.}\\\cline{2-10}
 
  &  \cite{CHEN2021338} & {DAC, storage} & {-} & {\checkmark} & {-} & {-} & {$\times$} & {Implements IoT-based collection system and utilizes PRE algorithm.} & {Security on cloud servers.}\\\cline{2-10}
  
 & \cite{9547814} &{DAC, search} & {-} & {-} & {\checkmark} & {-} & {$\times$} &  {Develops constant-size ABE and on-chain boolean search schemes.} & {Management and security.} \\\cline{2-10}
 
  & \cite{9442539} & {Storage} & {-} & {\checkmark} & {-} & {-} & {$\times$} & {Proposes a cross-chain system, leverages Wanchain technology.} & {Security and privacy on CSPs.} \\
 
 \hline


\multirow{14}{*}{\rotatebox[origin=c]{90}{\textbf{Supply chain}}} & \cite{agrawal2021blockchain} & {Traceability, DAC} & {-} &{\checkmark}& {-} & {-} & {$\times$} & {A framework for textile \& clothing supply chains.} & {Security analysis.} \\\cline{2-10}

&  \cite{8780161} &  {DAC, IA} & {\checkmark} & {-} & {-} & {-} &{$\times$} &{Implements an IIoT collection system, utilizes ABE for DAC.} &  {Lacks performance evaluations.}\\ \cline{2-10}

& \cite{8010716} & {Traceability, IA} & {-} & {\checkmark}& {-} & {-} & {\checkmark} & {Combines BC with HE for privacy-preserving data sharing.} & {Lacks performance evaluations.} \\\cline{2-10}

& \cite{DWIVEDI2020102554} & {Traceability} & {-} & {\checkmark} & {-} &{-} &{$\times$} &{Proposes a key management scheme, a consensus protocol.}  & {Data gathering, data traceability.} \\\cline{2-10}

& \cite{casino2021blockchain} &{Traceability} & {-} & {\checkmark} & {-} & {-} & {$\times$} & {A data sharing framework for FSC traceability.} & {Scalability.} \\\cline{2-10}

& \cite{7996119} & {Traceability} & {-} & {-} & {-} & {-} & {$\times$} & {Combines BigchainDB, HACCP and IoT.} & {Lacks data privacy protection.} \\\cline{2-10}

& \cite{baralla2018blockchain} & {Transparency} & {\checkmark}& {-} & {-} & {-} & {$\times$} & {A FSC data sharing platform utilizing KSI, smart contracts.} & {Security on the central authority.} \\\cline{2-10}

& \cite{venkatesh2020system} & {Transparency} & {-} & {-} & {-} & {-} & {$\times$} & {A system architecture for SCSSM, theoretical contributions.} & {Research scope for SCSSM.} \\\hline

 \end{tabular}

\begin{tablenotes}
        \item * \textit{\textbf{Pub.}}, \textit{\textbf{Pri.}}, \textit{\textbf{Con.}}, and \textit{\textbf{Hyb.}} stand for \emph{Public}, \emph{Private}, \emph{Consortium}, and \emph{Hybrid} respectively.
        \item * \textit{\textbf{BC}} stands for blockchain.
\end{tablenotes}

 \end{table*}

%% file: Tables/2_Taxonomy_of_applications.tex

\begin{table*}[t!] 
\caption{TAXONOMY OF BLOCKCHAIN-BASED DATA SHARING SOLUTIONS (CONTINUED)}
 \label{tab:taxonomy2}
\renewcommand{\arraystretch}{1.5}
\centering
\small
\begin{tabular}
{c|
 P{0.05\linewidth} 
 P{0.1\linewidth} 
 P{0.025\linewidth} 
 P{0.025\linewidth} 
 P{0.025\linewidth} 
 P{0.025\linewidth} 
 P{0.05\linewidth}  
 p{0.28\linewidth}  
 p{0.14\linewidth}}
 \hline
\multicolumn{1}{c|}{{\textbf{Area}}} & 
 \multicolumn{1}{c}{\textbf{Ref.}} &
 \multicolumn{1}{c}{\textbf{Objective}} & 
 \multicolumn{4}{c}{{\textbf{Blockchain used}}} & 
 \multicolumn{1}{c}{\textbf{PPT}} & 
 \multicolumn{1}{c}{\textbf{Key contributions}} & 
 \multicolumn{1}{c}{\textbf{Limitations}}\\ \cline{4-7}
 
& {} & {} & {\textit{\textbf{Pub.}}} & {\textit{\textbf{Pri.}}}& {\textit{\textbf{Con.}}} & {\textit{\textbf{Hyb.}}} & {} & {} & \\\hline

 \multirow{16}{*}{\rotatebox[origin=l]{90}{\textbf{Transportation}}}   &  \cite{kang2018blockchain} & {Data reliability}& {-} & {-} & {\checkmark}  & {-} & {$\times$} & {A reputation-based data sharing scheme with TWSL model.} & {Lacks performance evaluations.} \\ \cline{2-10}
 
 &  \cite{9042215} & {DAC, services} & {-} & {-} & {\checkmark} & {-} & {$\times$} & {Proposes an ABPRE algorithm for data searching, data retrieval.} & {System performance.}  \\\cline{2-10}
 
 &  \cite{8746499} & {IV's privacy}& -& {\checkmark}  & {-} & {-} & {$\times$} & {Provides a cryptographic fingerprint and keys for each IV using PUF, public key infrastructure.} & {Lacks performance evaluations.} \\ \cline{2-10}

  &  \cite{zhang2019data} & {DUC, data reliability} & {-} & -& {\checkmark} & {-} &{$\times$} & {Provides a digital signature technique, an incentive mechanism.} & {Security and costs on RSUs.} \\ \cline{2-10}

 &  \cite{su2020lvbs} & {Storage}& {-} & {\checkmark} & {-} & {-} & {$\times$} & {A lightweight BC with a credit-based DPoS protocol for disaster rescue.} & {Scalability.} \\ \cline{2-10}
 
 &  \cite{9457110} & {Data reliability} & {-}& - & {\checkmark} & {-} & {$\times$} & {A high-speed data sharing system without RSUs, trust score management.} & {Scalability.} \\ \cline{2-10}

 &  \cite{lu2020blockchain} & {Data reliability}& {-} & {\checkmark} & {-}& {-} & {\checkmark} & {A BC-powered asynchronous FL architecture, two-stage verification.} & {Lacks of incentive mechanism.} \\ \cline{2-10}

 &  \cite{chen2019toward} & {Incentive}& {-}&- & {\checkmark} & {-} & {$\times$} & {A quality-focused assessment of the reward and off-chain data.} & {Risks on the TA.} \\ \cline{2-10}

 &  \cite{firdaus2021blockchain} & {Storage, incentive}& {-} &- & {\checkmark}  & {-} & {$\times$} & {A reputation-based data sharing scheme using smart contracts.} & {Lacks data privacy protection.} \\ \hline


\multirow{10}{*}{\rotatebox[origin=l]{90}{\textbf{Smart grid}}}  &  \cite{wang2020spds} & {DAC, DUC} & -& -& {\checkmark} & {-} & {\checkmark} & {Avails DPaaS, develops a TEE-enabled off-chain smart contract.} & {Identities are not guarded.} \\ \cline{2-10}

& \cite{samuel2019blockchain} & {Incentive, DAC} & {-} & {\checkmark} &{-}  & {-} & {\checkmark} & {A PoA protocol, a privacy-based incentive mechanism, using differential privacy.} & {Communication overhead.} \\\cline{2-10}

 &  \cite{yang2020secure} & {DAC} & {-} & {-} & {\checkmark} & {-} & {$\times$} & {On-chain/off-chain solutions with edge computing, a distributed authority.} & {Lacks data privacy protection.} \\ \cline{2-10}

 &  \cite{li2021lightweight} & {DAC} & {-} & {-} & {-} & {\checkmark} & {$\times$} & {A dual BC with cloud storage for intelligent pricing system.} & {Security, incentive. } \\ \cline{2-10}
 
  &  \cite{reijsbergen2022securing} & {Incentive} & {-} & {\checkmark}& {-} & {-} & {$\times$} & {A realistic FDI attack model, an incentive-compatible mechanism.} & {Lacks data privacy protection.} \\ \cline{2-10}
  
 &  \cite{guan2021blockchain} & {Dual-side privacy} & {-} & {-}& {\checkmark} & {-} & {\checkmark} & {Presents a data obfuscation method.} & {Computational overhead.} \\ \hline

 \multirow{14}{*}{\rotatebox[origin=l]{90}{\textbf{Data marketplace}}}  &  \cite{nguyen2021marketplace} & {ML models, IoT data} & {-} & {\checkmark}& {-} & {-} & {\checkmark} & {Proposes an AI model marketplace, a data estimation algorithm.} & {Time for training on IoT devices.} \\ \cline{2-10}

 &  \cite{alsharif2020blockchain} & {DAC, health data} &  & {\checkmark}& {-} & {-} & {$\times$} & {Proposes a medical data marketplace model, a zk-SNARK protocol.} & {Lacks incentive mechanism.}  \\\cline{2-10}
 
 &  \cite{dixit2021fast} & {IoT data} & {-} & {\checkmark} & {-} & {-} & {$\times$} & {Presents an IIoT data martketplace, trust score calculation, DID verification.} & {Lacks shared data quality evaluation.} \\ \cline{2-10}

  &  \cite{ramachandran2018towards} & {IoT data} & {\checkmark} & {-} & {-} & {-} & {$\times$} & {Proposes an IoT data marketplace for smart cities using SC, SDPP.} & {Scalability, data privacy.} \\ \cline{2-10}

 &  \cite{ozyilmaz2018idmob} & {IoT data}& {\checkmark} & -& - & {-} & {$\times$} & {Presents an IoT data marketplace with querying and voting mechanism.} & {Lacks data privacy protection.} \\ \cline{2-10}

  &  \cite{nguyen2019enabling} & {IoT data} & {-} & {-} & - & {-} & {$\times$} & {Proposes a collectability-based IoT data marketplace model.} & {Scalability, data privacy.} \\ \cline{2-10}

  &  \cite{tang2022dmobas} & {Personal data} & {-} & {-} & - & {\checkmark} & {$\times$} & {A data marketplace with arbitration using a side-contracts mechanism.} & {Time consumption for encryption.} \\ \hline
  
\end{tabular}
\begin{tablenotes}
        \item * \textit{\textbf{Pub.}}, \textit{\textbf{Pri.}}, \textit{\textbf{Con.}}, and \textit{\textbf{Hyb.}} stand for \emph{Public}, \emph{Private}, \emph{Consortium}, and \emph{Hybrid}, respectively.
        \item * \textit{\textbf{BC}} stands for blockchain.
\end{tablenotes}

 \end{table*}

%% file: Tables/4_Other_applications.tex

\begin{table*}[t!]
\caption{TAXONOMY OF BLOCKCHAIN-BASED DATA SHARING SOLUTIONS (CONTINUED)}
 \renewcommand{\arraystretch}{1.5}
  \label{tab:taxonomy3}
\centering
\small
\begin{tabular}
{c|
 P{0.05\linewidth} 
 P{0.1\linewidth} 
 P{0.025\linewidth} 
 P{0.025\linewidth} 
 P{0.025\linewidth} 
 P{0.025\linewidth} 
 P{0.05\linewidth}  
 p{0.28\linewidth}  
 p{0.14\linewidth}}
 \hline
\multicolumn{1}{c|}{{\textbf{Area}}} & 
 \multicolumn{1}{c}{\textbf{Ref.}} &
 \multicolumn{1}{c}{\textbf{Objective}} & 
 \multicolumn{4}{c}{{\textbf{Blockchain used}}} & 
 \multicolumn{1}{c}{\textbf{PPT}} & 
 \multicolumn{1}{c}{\textbf{Key contributions}} & 
 \multicolumn{1}{c}{\textbf{Limitations}}\\ \cline{4-7}
 
& {} & {} & {\textit{\textbf{Pub.}}} & {\textit{\textbf{Pri.}}}& {\textit{\textbf{Con.}}} & {\textit{\textbf{Hyb.}}} & {} & {} & \\\hline

 \multirow{14}{*}{\rotatebox[origin=c]{90}{\textbf{Others}}} &  \cite{li2019edurss} & {DAC, storage} & {-} & {-} & \checkmark & {-} & {$\times$} & {Develops smart contracts for educational record storage and sharing.} & {Scalability, centralized storage solution.}  \\\cline{2-10}
 
 &  \cite{bsssqs} & {Security}& {-} & {\checkmark} & {-} & {-} & {{$\times$}} &{Addresses question paper leaking using smart contracts and two-phase encryption.} & {Lacks performance evaluations.} \\ \cline{2-10}

  & \cite{chou2020blockchain} & {Storage, authorship} & {-} & {\checkmark} & {-} & {-} & {{$\times$}} & {Proposes a \textit{work} block structure, contribution calculation, and utilizes off-chain solutions.} & {Lacks performance evaluations.} \\\cline{2-10}
  
  & \cite{wang2017blockchain} & {Data reliability} & {-} & {\checkmark} & {-} & {-} & {{$\times$}} & {Designs BC structure, network sharing model and consensus algorithms.} & {Packet dropping and disordering.} \\\cline{2-10}
  
  & \cite{chen2020blockchain} & {DAC, storage} & {-} & {\checkmark} & {-} & {-} & {{$\times$}} & {Implements cross domain authentication, decentralized CP-ABE scheme and IPFS.} & {Computation overhead on each node.} \\\cline{2-10}

    & \cite{shi2022secure} & {DAC, storage} & {-} & {-} & {\checkmark} & {-} & {{$\times$}} & {Proposes an encryption based revocable attribute and a hybrid data storage model.} & {Scalability.} \\\hline

\end{tabular}
\begin{tablenotes}
        \item * \textit{\textbf{Pub.}}, \textit{\textbf{Pri.}}, \textit{\textbf{Con.}}, and \textit{\textbf{Hyb.}} stand for \emph{Public}, \emph{Private}, \emph{Consortium}, and \emph{Hybrid} respectively.
        \item * \textit{\textbf{BC}} stands for blockchain.
\end{tablenotes}

 \end{table*}

%% file: Tables/3_Taxonomy_of_applications.tex
\begin{table*}[t!]
\small
\caption{COMPARISON OF BLOCKCHAIN-BASED DATA SHARING SOLUTIONS IN INDUSTRY}
\renewcommand{\arraystretch}{1.5}
\centering
\label{tab:industrial_apps}
\begin{tabular}
{P{0.05\linewidth} 
 P{0.26\linewidth} 
 P{0.12\linewidth} 
 P{0.11\linewidth} 
 P{0.12\linewidth} 
 P{0.12\linewidth}}
\hline
\rowcolor[HTML]{EFEFEF} 
\multicolumn{1}{c}{\textbf{Ref.}} &
  \multicolumn{1}{c}{\textbf{Application}} &
  \multicolumn{1}{c}{\textbf{Company}} &
  \multicolumn{1}{c}{\textbf{Domain}} &
  \multicolumn{1}{c}{\textbf{DLT type}} &
  \multicolumn{1}{c}{\textbf{Cryptocurrency}}
\\\hline
\cite{nokia} &  B2B data marketplace & Nokia & Varies & Private permissioned & -  \\ \hline
\cite{ibm} & B2B data sharing platform & IBM & Supply chain & Private  permissioned &  - \\ \hline

\cite{acentrik} & B2B data marketplace  &  Mercedes-Benz & Varies & Public & Matic and USDC  \\ \hline
\cite{dao} & Databroker: Data marketplace & SettleMint NV & Varies & Public & DTX \\ \hline
\cite{streamr} & Streamr: Data marketplace & Streamr Network AG & Real-time data  & Public/Private & Ether \\ \hline

\cite{popov2018tangle} & IOTA: Data marketplace & The IOTA Foundation & IoT & Tangle (based on DAG) &  MIOTA  \\ \hline

\cite{Advaneo} & Advaneo: Data marketplace & Advaneo GmbH & Varies & Private &   - \\ \hline

\end{tabular}

\end{table*}

%% file: Sections/6-Open-Research-Issues.tex
\section{Research challenges and future directions}
\label{sec-issuesDirections}
As discussed, blockchain is playing an increasingly important role in enabling real-world data-sharing applications. Despite its enormous promise, the extensive survey highlights a number of significant research difficulties that must be addressed before future blockchain-based data-sharing system adoption. We examine numerous critical difficulties in blockchain-based data-sharing systems, including security, privacy, scalability, storage, and redactable blockchains. Several research avenues for these difficulties are also presented.

\subsection{Connectivity}
There is a missing topic in current blockchain studies about communication aspects\cite{nguyen2022blockchain}. The communication between DPs and DRs could be wired or wireless communication. Initiated by \cite{nguyen2020trusted, danzi2020communication}, the authors introduced a trusted monitoring network based on Blockchain and narrowband IoT protocol to share air pollution data among companies at the city level and analyze the trade-off between latency and security. The model and analysis of Blockchain-based narrowband IoT network to share IoT sensing data are firstly provided in \cite{nguyen2021modeling}. The first three IoT data-sharing protocol via Ethereum smart contract was introduced. Then, in order to guarantee the privacy of sensitive data, the authors introduced a new concept of model trading with a focus on the quality of data and common communication efficiency of exchanging ML models via IoT networks automated by Ethereum smart contracts\cite{nguyen2021marketplace}. However, the performance of Blockchain-based data-sharing protocols has not been analyzed and studied in various communication protocols, e.g., Bluetooth, LPWAN, etc. Transferring data over wireless IoT networks faces many challenges regarding security, medium loss, and resource constraints. Therefore, communication aspects of Blockchain-based systems are still open challenges for both academic and industrial communities.

\subsection{Security} 

Although integrating blockchain technology into data-sharing solutions can enhance the security in terms of integrity and availability of them by utilizing properties of blockchain, the security of existing blockchain-based data-sharing systems is still a major concern due to the rapid development of quantum computing technology and the technical vulnerabilities of blockchain platforms.


In recent years, the monumental breakthroughs of quantum computing have shown its capabilities to break encryption and digital signature algorithms used to protect the confidentiality, authenticity, and integrity of shared data in blockchain-based data-sharing solutions \cite{guo2022survey}. In general, the security of conventional cryptography algorithms is guaranteed thanks to the hardness of solving mathematical problems. However, in the coming years, powerful quantum computers, which have extraordinary computing capability, can be predicted to pose a threat to these cryptography systems if malicious parties own them. Particularly, by Google's estimation in 2019, their 54-qubit Sycamore processor performed a computational task in 200 seconds, while the most powerful supercomputer took nearly 10,000 years to solve the same task \cite{arute2019quantum}. In addition, how to manage cryptographic keys securely remains a challenge in those blockchain-based data-sharing systems.

Besides, blockchain systems are also susceptible to attacks and security vulnerabilities. For example, due to deficiencies in programming languages, code bugs, and execution environments, the number of attacks on blockchain systems targeted at smart contracts is growing rapidly such as well-known Ethereum smart contract attacks, including re-entrancy attacks \cite{rodler2018sereum}, MitM attacks \cite{qin2020cecoin}, replay attacks \cite{sonnino2020replay}, short address attacks  \cite{saad2020exploring}, and re-ordering attacks \cite{wang2022exploring}. For example, the DAO \cite{DAOattacks} has lost \$50 million USD and Ethereum network suffered a hard fork by separating into two chains, leading to a huge amount of monetary loss of the investors and users. 


In these circumstances, researching and developing new methods is critical for secure and sustainable blockchain-based data-sharing systems. As a leading blockchain platform, Ethereum has been testing post-quantum cryptography algorithms whose security is considered to resist quantum attacks \cite{fernandez2020towards}. Particularly, some new post-quantum signature algorithms such as eXtended
Merkle Signature Scheme \cite{buchmann2011xmss} and SPHINCS \cite{bernstein2015sphincs}, are contemplated to replace the Elliptic Curve Digital Signature Algorithm in the Ethereum 2.0 Serenity \cite{guo2022survey}. The security of these post-quantum signature algorithms is based upon the security of the hash function instead of the difficulty of mathematical problems. Regarding smart contract security issues, four smart contract development phases, i.e., security design, security implementation, testing before deployment, and monitoring and analysis, need to be carefully designed, developed, and inspected \cite{huang2019smart}.

\subsection{Privacy}

Another issue raised by blockchain-based data-sharing systems is privacy. Public blockchains hide a user's true identity by utilizing pseudonymous identifiers. In Bitcoin, for example, addresses that users use to communicate with the network without disclosing their true identity are hashes of public keys. However, there are still risks associated with revealing users' identities in blockchain platforms \cite{conti2018survey}. Moreover, in data-sharing applications, it is a really challenging undertaking for DPs to maintain control over shared data. As a result, shared data may be used for malicious purposes, which violates the terms of the agreement between DPs and DRs. Without a protection mechanism, blockchain-based data-sharing systems will lose the trust of their users for data contribution. 

There are several feasible solutions to protect the privacy of users and their data from threats. The study in \cite{guan2021blockchain} not only safeguards the identity of DP but also conceals the identity of DR by proposing a data obfuscation method based on the ring signatures and a new one-time address scheme. Furthermore, the authors solved the problem of DUC by combining MPC and edge computing for collaborative computation without learning the original data of any DPs. Only the final processed results are shared with DRs and publicly exposed on the blockchain network. Besides, an effective solution to protect data privacy and address the distress of DUC for users is presented in \cite{wang2020spds}. While the entire smart contract execution is outsourced to TEE, i.e., Intel SGX, raw data are stored in off-chain storage. Similar to the above MPC solution, only computed results are shared instead of transferring raw data. These PPTs appear promising for use in blockchain-based data-sharing systems, ensuring participant data privacy and compliance with privacy regulations. 

\subsection{Incentive/Punishment Mechanisms}

Incentive mechanisms are essential in encouraging industrial actors to participate and contribute their data to solve the DOP\cite{nguyen2021marketplace}. Without proper incentives, actors may be hesitant to share data due to data privacy and misuse concerns. When actors can better relate to incentives, they are more committed to contributing high fidelity, volume, and velocity data over a long time. This enables the ecosystem to establish a more accurate picture of the current system design. This, in turn, could improve the accuracy and robustness of the sharing outputs \cite{pandey2022fedtoken}. There are some incentive mechanisms as, monetary compensation, data ownership, reputation systems, access permissions, social benefits, etc. 
When the number of active users is vast and diversified, DRs will have more opportunities to exchange relevant and high-quality data for product development. Also, there remains a significant obstacle in that DPs are not compensated transparently and fairly for their contributions to data collection, data processing, data privacy, and data sharing in current blockchain-based data-sharing solutions. Meanwhile, as the benefits of data-sharing, such as incentives and price, increase, more malicious entities will engage in unethical behavior to maximize their earnings. Nevertheless, designing a detection method for malicious operations and a deterrent punishment solution for offenders is a tough challenge.

Several ways have been proposed to facilitate user enrollment in blockchain-based data-sharing systems. In \cite{nguyen2021marketplace}, authors presented a federated DSV mechanism for evaluating DP contributions for the ML model marketplace. The suggested system can determine the quality of the data and the local training model provided by each DP based on improvements in loss function metrics such as accuracy or mean squared error. As a result, DPs are fairly rewarded in proportion to their contributions. Another strategy is to create reputation-based incentive mechanisms. DPs who share a significant amount of high-quality data and obtain verified positive feedback will be rated at the highest level. Based on their reputation, DPs stand a good chance of increasing revenue from their data, attracting customers to use their services and products, and receiving more privileges from the data-sharing system. On the other hand, malicious DPs who dishonestly exchange data with blockchain-based data-sharing systems will be penalized based on the gravity of the breach. The authors of \cite{7990130} employed smart contracts to track actions on shared data; if they identified any aberrant behaviors, data access permission was revoked and DRs could no longer access the shared data. Furthermore, in \cite{reijsbergen2022securing}, a fair incentive mechanism is proposed to reward DPs who are honest and share high-quality data and to punish abnormal DPs who fail to share or share fake and malicious information. These malicious DPs could be banished from the blockchain network and lose their entire deposit. 

Further, game theory is also a potential approach and can be considered to be widely adopted in designing better incentive mechanisms in blockchain-based data-sharing applications. By utilizing game theory, players, e.g., DPs and DRs, can choose their strategies with the aim of maximizing their utility, i.e., rewards \cite{liu2019survey}. In \cite{xuan2020incentive}, evolutionary game theory is utilized to propose a smart contract-based incentive model dynamically adjusting incentives and participation costs to keep the blockchain-based data-sharing system active and improve users' (i.e., DPs and DRs) revenue via sharing of data. With this proposed model, when the number of DPs and DRs shows a sign of decreasing, incentives are increased to motivate DPs and DRs to join and trade data. 

However, designing a fair incentive mechanism is challenging due to some open problems. 
\begin{itemize}
    \item \textit{Data heterogeneity}: data sharing systems often involve data from multiple sources that may be different in terms of quality, quantity, and relevance. Designing an incentive mechanism that fairly rewards all data contributors while accounting for differences in the data, can be challenging.
   
    \item \textit{Privacy concerns}: Many data providers may be hesitant to share their data due to privacy concerns. Designing an incentive mechanism that addresses these concerns and assures data contributors that their data will be kept private and secure is essential.
    
    \item \textit{Fairness and equity:} An incentive mechanism that is perceived as unfair or inequitable may lead to distrust among DPs and undermine the effectiveness e.g., with FL.

    \item \textit{Trust and transparency:} DPs need to trust that the incentive mechanism is designed and implemented fairly and transparently. The incentive mechanism should be transparent in terms of the rewards and the criteria used to determine them.

    \item \textit{Limited information:} The incentive mechanism should be designed with the limited information that is available about the data and the DPs. It may be difficult to accurately measure the value of each data contribution, and DPs may not be able to provide complete information about their data.

    \item \textit{Dynamic environments:} Sharing data usually takes place in dynamic environments, where the data and the needs of the stakeholders may change over time. Designing an incentive mechanism that is flexible and adaptive to these changes is important.
\end{itemize}

\subsection{Scalability}

As previously stated, the volume of data created and consumed is increasing exponentially due to the rapid growth in the number of DPs and DRs participating in and exchanging data in blockchain-based data-sharing applications. As a result, the question of how to enhance systems in terms of scalability to respond to massive changes in workloads and user demands is still open and attractive to researchers and industries. Two popular and important performance metrics directly affect the quality of service and the user's quality of experience, i.e., transaction throughput and transaction verification latency in blockchain-based data-sharing systems \cite{zheng2018detailed}. Regarding transaction throughput, current major public blockchains, e.g., Bitcoin and Ethereum, have low throughput that accounts for 7 TPS and 20 TPS, respectively \cite{khan2021systematic}. Therefore, these public blockchains are unsuitable for applications requiring a high volume of transactions. To increase TPS for exchanging more data, the size of each block should be expanded. However, when disseminating the block to blockchain peer nodes, a larger block results in a longer block propagation time, which may produce a fork problem. Besides, to keep users using services and applications, transaction verification latency, which is the time between sending a transaction to the blockchain network and including that transaction in a block, should be as low as possible. Nevertheless, miners cannot verify every transaction due to the restricted size of each block, preferring to confirm transactions with high fees. Lower-fee transactions are forced to wait in the large line, resulting in increased transaction verification latency.

Additionally, miners may lack high computing, networking, and storage capacities when blockchain is applied to real-world business applications. Safekeeping and processing a large volume of data on a distributed ledger is a difficult task. Thus, it is paramount to design schemes to increase transaction throughput and decrease transaction verification latency to make blockchain-based data-sharing applications more scalable. 

Several potential solutions should be considered to improve the scalability of blockchain-based data-sharing systems. One direction is to implement separate permissioned blockchains, i.e., private or consortium blockchains, for each data-sharing application. The studies in \cite{9042215, 8746499, zhang2019data, su2020lvbs, 9457110, chen2019toward, kang2018blockchain, firdaus2021blockchain} build blockchain networks for storing and sharing data where RSUs or even resource-constrained vehicles play a role as miners in the transportation field. Because of the partially decentralized, fully controlled capabilities and restricted number of approved validators, deploying permissioned blockchains for data-sharing speeds up verification procedures far faster than public ones. 

Moreover, an efficient technique for horizontally scaling blockchain-based data-sharing systems is sharding, where transactions are processed in parallel \cite{dang2019towards}. Sharding divides the blockchain network into multiple groups of nodes (or multiple shards) while running the consensus protocol within each shard, and each node only keeps data about its own shard. It means that each node in blockchain sharding does not manage and process the entire transaction volume like the traditional approach. The study in \cite{hashim2021medshard} proposed a transaction-based sharding strategy for scaling blockchain-based EHR sharing systems. Nodes in one shard DP do not need to communicate with nodes in other shards since shards are formed for each appointment, decreasing cross-shard communication overhead.
Additionally, consensus nodes that are previous caregivers run the PoA protocol to achieve a consensus inside a shard and reduce consensus latency. The simulation results suggest that the proposed model performs better regarding scalability metrics such as consensus latency and throughput. Furthermore, in \cite{wang2021privacy}, the authors presented a vehicular data-sharing framework on top of a multi-sharding blockchain network. The key point is that using the multi-sharding protocol allows blockchain nodes to process transactions in other shards directly without incurring cross-shared communication overhead. Multi-sharding blockchain outperforms Bitcoin and standard sharding protocols in terms of throughput, bandwidth consumption, and storage redundancy, as demonstrated by simulation experiments.

\subsection{Redactable Blockchains}

Although blockchain technology brings many useful features for efficient data-sharing in blockchain-based data-sharing systems, it also faces several hurdles, such as unintentionally or maliciously injecting sensitive and harmful data into blockchain and storage limitations even with off-chain solutions \cite{tziakouris2018cryptocurrencies}. For fake and malicious data, storing them in the blockchain network causes very serious consequences for users. For example, false information leads to misdiagnoses and inappropriate treatment for patients in the healthcare field, as well as accidents for drivers in the transportation sector. Moreover, with the explosion of data volume, the storage space of DPs is relatively constrained compared to the rapid extension of the blockchain ledger in the long term. Particularly, the amount of irrelevant and obsolete data kept in the blockchain grows daily. For example, weather forecasts and traffic jam information kept on the vehicular blockchain network five years ago are no longer useful for drivers choosing an appropriate route for travel today. Besides, great effort is being expended today to support the ``Right to be Forgotten,'' which gives individuals the right to seek the deletion of personal data without undue delay if certain conditions are met. Based on the observations above, personal, fictitious, malicious, and nonsensical data should be modified and eliminated from the blockchain. But this need is contrary to the philosophy of blockchain, where the data stored is immutable and cannot be deleted. Therefore, the question of how to alter and remove data from blockchain while ensuring data authenticity, regulations, security, and privacy is posed.

One possible solution to address these issues is using a redactable blockchain that allows authorized entities to alter and redact blocks of transactions under restricted constraints \cite{ateniese2017redactable}. The research in \cite{9831062} proposed a hierarchical access control redactable blockchain model that allows data updates on the blockchain while limiting the power of modifiers to prevent abuse of power in medical data-sharing. These properties can be achieved by developing a hierarchical ABE scheme for granting access permission to modifiers and using CHF to keep the hash value untouched when rewritable operations on the blockchain appear. Furthermore, for particular applications, such as healthcare, where the possibility of storing personal and sensitive data is considerable, permissioned blockchains are preferable. When the number of consensus nodes is restricted, and there is any information that needs to be removed from the blockchain, this can be accomplished with the majority's agreement, which is practically impossible in permissionless blockchains \cite{politou2019blockchain}.

%% file: Sections/7-Conclusion.tex
\section{Conclusion}
\label{sec-conclusion}
Blockchain-based data sharing provides a trustworthy method for sharing data among individuals, businesses, and organizations. In this comprehensive survey, we presented an in-depth survey of the fundamentals, techniques, and applications of distributed data sharing. Specifically, we introduced a novel Blockchain-base data sharing solution and discussed its key characteristics, enabling technologies, and deployment tutorial. Then, we analyzed the applicability of Blockchain-based data sharing techniques in various applications and gave valuable lessons learned. Afterward, we discussed the  research challenges e.g, connectivity, security, privacy, incentive/punishment mechanisms, design bottlenecks, and future directions of Blockchain-based data sharing. We expect that this survey can shed light on designing trusted data-sharing applications for both enterprises and inspire more pioneering research in this emerging area.